\documentclass[11pt]{article}
\usepackage{amssymb}
\usepackage{cite}
\usepackage{mathtools}
\usepackage{graphicx}
\usepackage{caption}
\usepackage{subcaption}
\usepackage{amsmath}
\usepackage{amsthm}
\usepackage{hyperref}
\usepackage{tablefootnote}
\usepackage[bottom]{footmisc}

\usepackage{siunitx}
\usepackage{url}

\DeclareMathOperator{\Harmonic}{Harmonic}
\DeclareMathOperator{\Polynom}{Polynom}

\begin{document}
\thispagestyle{empty}
\begin{center}
{\Large\bf Data-driven chimney fire risk prediction using machine learning and point process tools.}\\[.4in]

\noindent
{\large C.~Lu$^{1}$, M.N.M.~van~Lieshout$^{2,1}$, M.~de~Graaf$^{3,1}$ and P.~Visscher$^{4}$}\\[.1in]
\noindent
{\em $^{1}$Department of Applied Mathematics, University of Twente, Enschede, The Netherlands\\
 $^{2}$Centrum Wiskunde \& Informatica, Amsterdam, The Netherlands\\
 $^{3}$Thales Nederland B.V., Huizen, The Netherlands\\
 $^{4}$Brandweer Twente, Enschede, The Netherlands
}\\[.1in]
\end{center}
\begin{verse}
{\footnotesize
\noindent
{\bf Abstract}\\
\noindent
Chimney fires constitute one of the most commonly occurring fire types. Precise prediction and prompt prevention are crucial in reducing the harm they cause. In this paper, we develop a combined machine learning and statistical modeling process to predict chimney fires. Firstly, we use random forests and permutation importance techniques to identify the most informative explanatory variables. Secondly, we design a Poisson point process model and apply associated logistic regression estimation to estimate the parameters. Moreover, we validate the Poisson model assumption using second-order summary statistics and residuals. We implement the modeling process on data collected by the Twente Fire Brigade and obtain plausible predictions. Compared to similar studies, our approach has two advantages: i) with random forests, we can select explanatory variables non-parametrically considering variable dependence; ii) using logistic regression estimation, we can fit the statistical model efficiently by tuning it to focus on important regions and times of the fire data.\\[0.1in]

\noindent
{\em Keywords \& Phrases:}
fire prediction, spatio-temporal point pattern, Poisson point process, variable importance, logistic regression, pair correlation function, $K$-function\\[0.1in]
}
\end{verse}

\bibliographystyle{plain}

\section{Introduction}
\label{sec:intr}
During the last decade, the Dutch fire and rescue services are developing an interest in applying business intelligence to improve their strategy of fire prediction and prevention \cite{Brandweer2010Strategy}. In order to prepare for risk reducing measures, such as essential public awareness campaigns and proper fire staffing and equipment arrangements, accurate predictions are required. In this paper, we focus on chimney fires, as they occur frequently, rely heavily on environmental factors and impact people's daily life. Collaborating with the Twente Fire Brigade, we conduct a complete risk prediction study for chimney fires. We define the risk prediction as an occurrence modeling problem, analyze underlying patterns and design appropriate prediction models. Our approach for chimney fire prediction is quite general and can be transferred to other similar fire types, such as kitchen fires.

The literature for fire risk prediction is mostly concerned with wildfires.  Overall, the prediction methods can be divided into two categories: machine learning based approaches (e.g.\ \cite{Rodrigues2014MLreview,Jain2020MLreview,Malik2021MLclassification}) and statistical approaches (e.g.\ \cite{Brandweer2010Strategy,Xu2011basicModeling,Turner2009wholeProcess}). Usually, machine learning based approaches do not require prior knowledge of a fire type and its corresponding dependent variables, and they can detect the dependence between the fire risk and a large number of environmental candidate variables automatically using specialised learning algorithms, such as logistic regression~\cite{Preisler2004ProbabilisticLR}, support vector machine~\cite{Sakr2010SVMclassification}, decision trees~\cite{Stojanova2012DT}, random forests~\cite{Rodrigues2014MLreview} and neural networks~\cite{Satir2016NNdiscretized}. Most of machine learning algorithms are applicable for discrete data, whereas continuous hazard maps of fire risk are expected. Moreover, fire occurrences are usually recorded as spatio-temporal point patterns. To solve these gaps, certain studies (e.g.\ \cite{Sakr2010SVMclassification,Stojanova2012DT,Satir2016NNdiscretized}) discretized fire incidents into areal unit data and sometimes transformed the risk prediction from an occurrence modeling problem to a risk scale classification problem (i.e.\ label different amounts of fires to corresponding risk scales), however, complete and real data inference may be lost. In addition, machine learning approaches often require relatively large amounts of data in order to obtain a satisfactory performance, and due to their `black box' behaviour, it is difficult to specify the influence of an explanatory variable on fire occurrences.

In contrast, statistical approaches are able to learn spatio-temporal point patterns directly using point process tools, and are more interpretable as they are based on a specific, parametric, mathematical model which allows for theoretical confidence intervals as well. For instance, Hering et al.~\cite{Hering2009Kfunction} and Costafreda-Aumedes et al.~\cite{Costafreda-Aumedes2016Kfunction} used the $K$-function to analyze the clustering patterns of wildfire data. Ye~\cite{Ye2011Poisson} and Boubeta et al.~\cite{Boubeta2015Poison} utilized Poisson structures to generate hazard maps of fire occurrences. As for complex models, M{\o}ller and D\'{i}az-Avalos~\cite{Moller2010shotNoiseCoxprocess} considered spatial and temporal explanatory variables in a shot-noise Cox process and fitted it with minimum contrast techniques. Pereira et al.~\cite{Pereira2013LGCP} and Serra et al.~\cite{Serra2014LGCP} modeled the spatio-temporal point patterns in a log-Gaussian Cox process, which was particularly designed to simulate latent phenomena. Recently, a Bayesian framework was developed and suggested to improve fire prediction \cite{Juan2019BayesianNetwork}. Pimont et al.~\cite{Pimont2020BayesianFramework} employed a spatio-temporal log-Gaussian Cox process model (`Firelihood') to predict wildfires and established Bayesian inference for model components using integrated nested Laplace approximation \cite{Rue2009INLA}. Koh et al.~\cite{Koh2021BayesianNetwork} developed a joint hierarchical model framework by combining extreme-value theory and point processes and studied summer wildfire data for the French Mediterranean basin. For chimney fires, School~\cite{School2018Thesis} also used a log-Gaussian Cox process to predict fire risk based on the explanatory variables selected by Pearson correlation coefficients and the random effects simulated by Gaussian random fields.

Although various point processes models have been proposed to improve the performance of fire prediction, many studies (e.g.\ \cite{Moller2010shotNoiseCoxprocess,Pereira2013LGCP}) determine the model structures firstly  and select and fit explanatory variables afterwards. This procedure does not fully exploit information contained in variables and sometimes requires complex modeling of residuals. In addition, the selection of explanatory variables is conducted either manually \cite{Turner2009wholeProcess} or using basic statistical methods \cite{Yang2015modelBasedvariableImportance,School2018Thesis}. Several studies \cite{Thurman2014variableSelection,Yue2015variableSelection,Choiruddin2018increasingDomainCI} proposed regularized penalty functions in the estimation of point processes for variable selection. However, a parametric form of the intensity function including all variables needs to be specified in advance. 

To conduct a data-driven fire risk prediction study, we combine machine learning and point process tools in our modeling process to leverage the advantages of both types of approaches. In an initial study~\cite{Lu2021ISI}, we used the permutation importance techniques of random forests to select important variables, as random forests are the most accurate machine learning method for fire prediction \cite{Rodrigues2014MLreview}. Accordingly, we designed a generalized linear Poisson model and fitted it on areal unit data to predict chimney fire risk in Twente. In this paper, we refine this model to a spatio-temporal Poisson point process as it targets at point patterns (cf.\ Figure~\ref{fig:maps}(b)) directly, enabling more detailed analysis. Specifically, our contributions are: i) we use machine learning algorithms to select explanatory variables of chimney fires non-parametrically while taking variable dependence into account, so that the utilization of variable information can be maximized; ii) we adhere to a statistical Poisson point process model with an interpretable parametric structure designed on explanatory variables for fire occurrence modeling, so that complete inference on the data can be carried out; iii) we fit the model parameters using the efficient logistic regression approach by tuning it to focus on important parts of point patterns and validate the model using second-order summary statistics and residuals.

The remainder of the paper is organized as follows. Section~\ref{sec:data} introduces the fire data we use as well as some data pre-processing steps. Section~\ref{sec:slev} discusses the selection of important variables to be used in the chimney fire risk prediction. In Section~\ref{sec:psml}, we motivate and fit a Poisson point process model. Section~\ref{sec:mdvd}  tests for point interactions in our fire data and validates the Poisson model assumption of absence of such interactions. Section~\ref{sec:disc} discusses the modeling process, compares the point process model with an areal unit model~\cite{Lu2021ISI} and explores the role of the dummy intensity in logistic regression estimation. Section~\ref{sec:conclusion} provides a conclusion and ideas to further refine the model.

\section{Data}
\label{sec:data}
In this section, we introduce the chimney fire data, the relevant environmental variables and the data pre-processing operations.

\subsection{Chimney Fires and Environmental Variables}
\label{subsec:data_collection}

\begin{figure}[h]
    \centering
    \begin{subfigure}[t]{0.48\textwidth}
        \centering
        \includegraphics[scale=0.34]{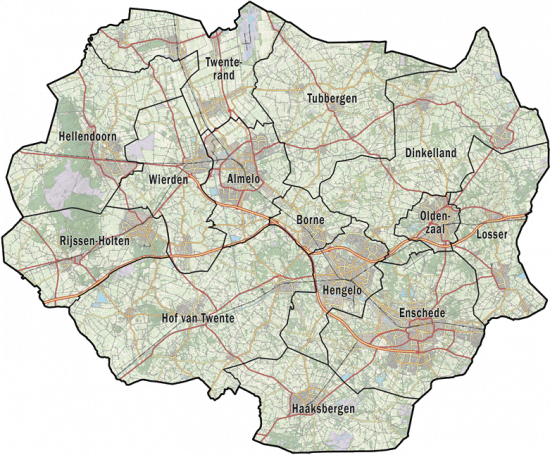}
        \caption{Twente municipalities}
        \label{subfig:Twente}
    \end{subfigure}%
    \begin{subfigure}[t]{0.48\textwidth}
        \centering
        \includegraphics[scale=0.234]{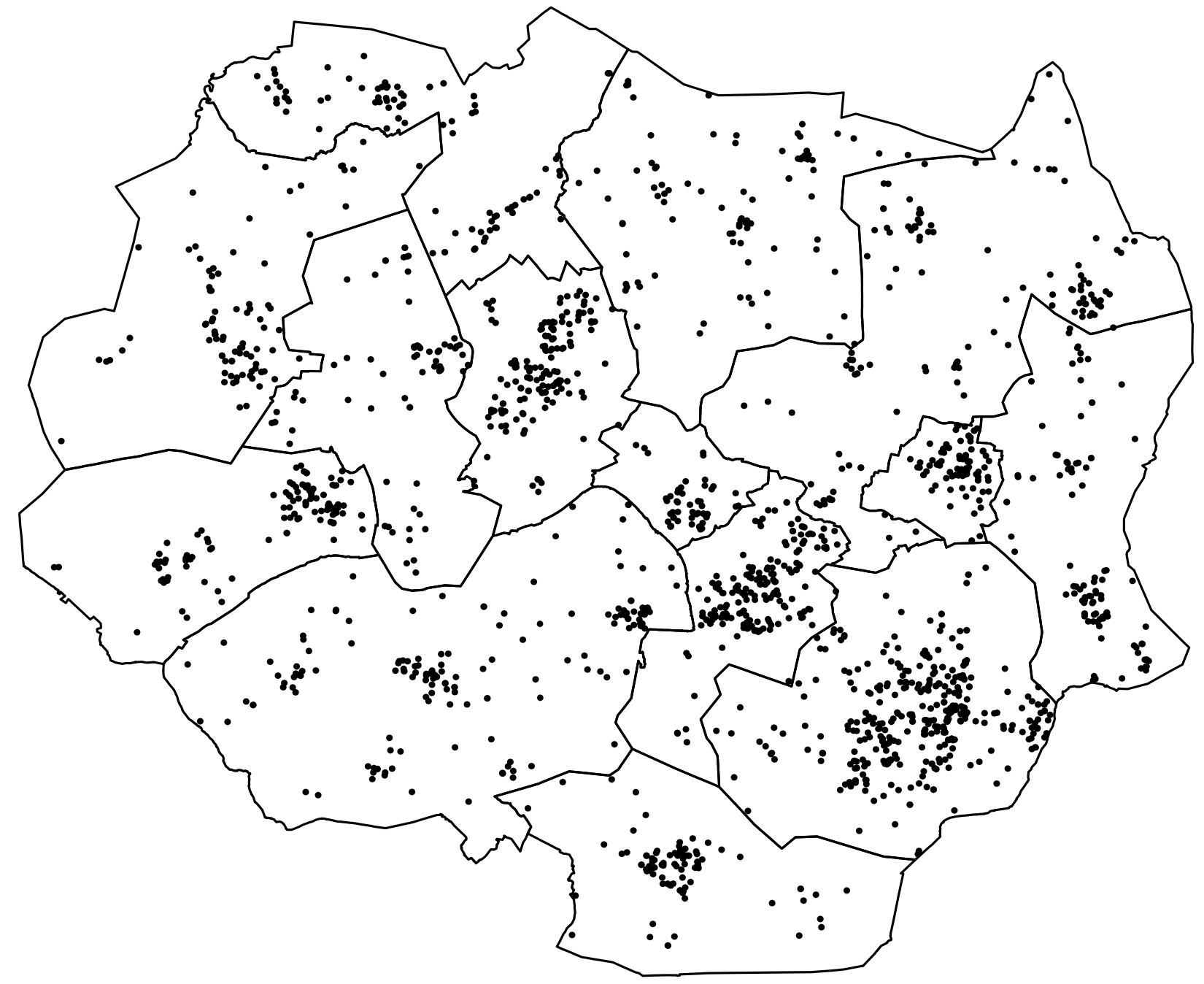}
        \caption{spatial pattern of fires}
        \label{subfig:spatial_distribution}
    \end{subfigure}%
    \caption{Map of Twente municipalities (a) and spatial projection of the chimney fire incidents during 2004--2020 (b).}
    \label{fig:maps}
\end{figure}

\begin{figure}[h]
    \centering
    \includegraphics[scale=0.4]{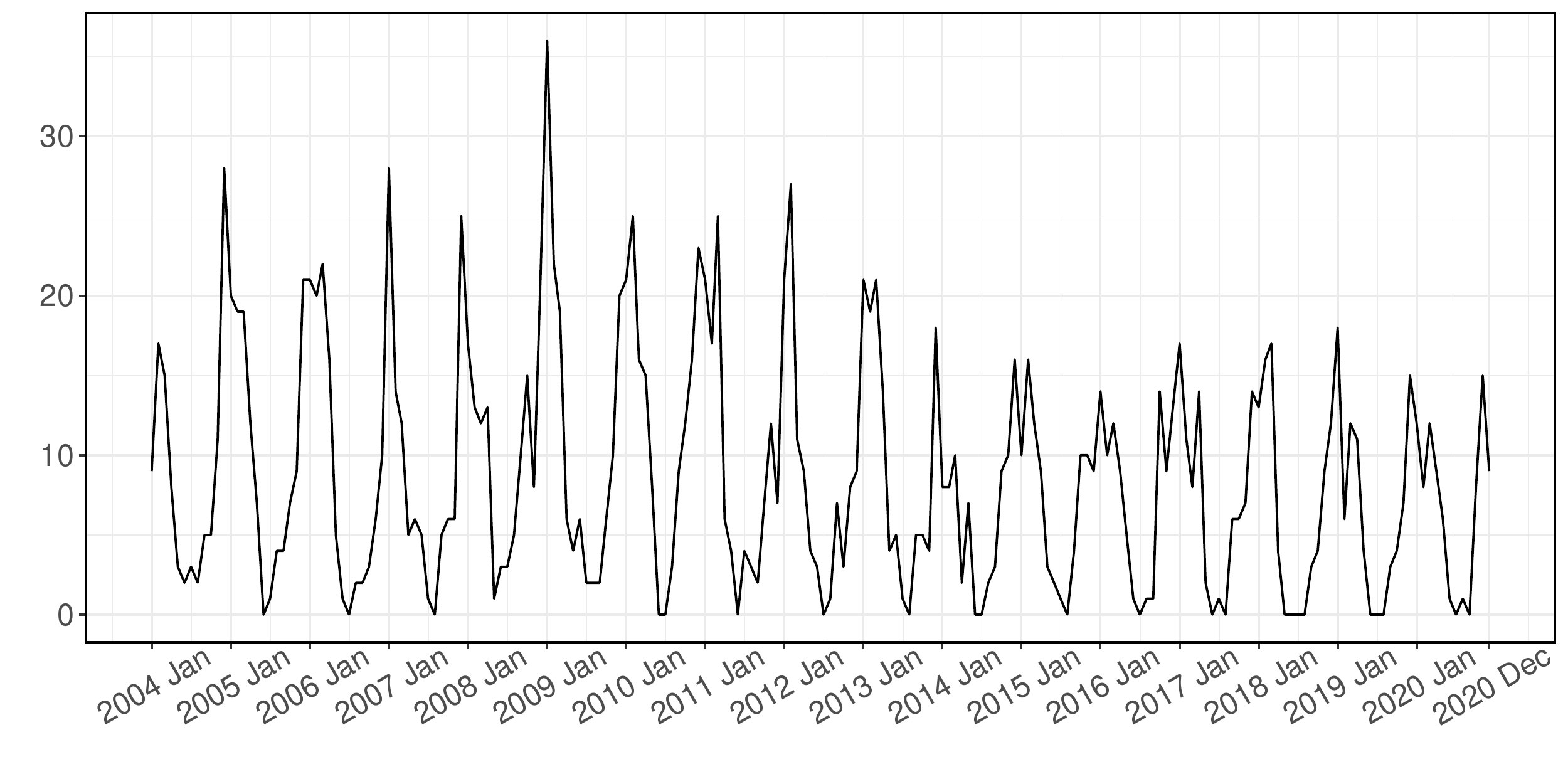}
    \caption{Temporal projection (monthly counts) of the chimney fire incidents ($y$-axis) during 2004--2020.}
    \label{fig:temporal_distribution}
\end{figure}

Collaborating with the Twente Fire Brigade, we collected the data of all reported chimney fire incidents occurring between January 1, 2004 and December 31, 2020, in the Twente region, in the eastern part of the Netherlands (map shown in Figure~\ref{fig:maps}(a)). After a manual check that removes obvious mistakes, the dataset consists of 1759 incidents. Each incident is reported individually with its ID number, location, time and a brief description of the circumstances of the fire and the rescue process\footnote{Location is recorded in the form of Dutch RD coordinates and in the unit of metre; Time is recorded in the unit of day to match weather data.}. The spatial and temporal projections\footnote{Fire incidents on leap days have already been excluded, as discussed in Section~\ref{subsec:data_pre-processing}.} are plotted, respectively, in Figure~\ref{fig:maps}(b) and \ref{fig:temporal_distribution}. It is clearly visible in Figure~\ref{fig:maps}(b) that the spatial distribution of chimney fires is heterogeneous in the sense that most incidents occur in urban areas, especially in cities with a higher population, such as Almelo, Hengelo and Enschede (cf.\ Figure~\ref{fig:maps}(a)). Apparent clustering may also arise, because neighbouring buildings tend to have identical chimney types. Moreover, locations neighbouring in space and time share similar weather conditions. In Figure~\ref{fig:temporal_distribution}, the temporal distribution of chimney fires is clearly periodic: chimney fires occur more frequently in winter than in summer.

According to the experts of the Twente Fire Brigade, various environmental factors could influence the risk of chimney fires. Spatially, building types, population density and composition and urbanity degrees in an area may determine the baseline risk of chimney fires. Temporally, season and weather conditions may also influence chimney fire risk. Table \ref{tab:influencing variables} lists 27 putative explanatory variables with specific abbreviations, descriptions and sources. Among them, population and urbanity information are recorded over 6291 pre-defined $500m \times 500m$ area boxes in Twente, whereas building information contains the precise locations, ages and functions of the houses. Temporal data derives from the daily weather information observed at the weather station at Twenthe airport. To assess the influence of small variations in weather among different parts of the Twente region, we also collected the data from two neighbouring weather stations (i.e.\ Heino, Hupsel) outside Twente for analysis. Their locations relative to Twente can be found on \url{https://www.knmi.nl}.

\begin{table}[tbh]
    \scriptsize
    \centering
    \begin{tabular}{|c|l|l|l|}
        \hline
        \textbf{Variable} & \textbf{Abbreviation} & \textbf{Description} & \textbf{Source\tablefootnote{IFV: \textit{Instituut Fysieke Veiligheid}, CBS: \textit{Centraal Bureau voor de Statistiek}, KNMI: \textit{Koninklijk Nederlands Meteorologisch Instituut}. All data is provided by the Twente Fire Brigade especially for this research.}}\\
        \hline
        $V_{\sigma,1}$ & House & The total number of houses & IFV\\
        \hline
        $V_{\sigma,2}$ & House\_indu & The number of houses with an industrial function & IFV\\
        \hline
        $V_{\sigma,3}$ & House\_hotl & The number of houses with a hotel function & IFV\\
        \hline
        $V_{\sigma,4}$ & House\_resi & The number of houses with a residential function & IFV\\
        \hline
        $V_{\sigma,5}$ & House\_20 & The number of houses constructed before 1920 & IFV\\
        \hline
        $V_{\sigma,6}$ & House\_2045 & The number of houses constructed between 1920 and 1945 & IFV\\
        \hline
        $V_{\sigma,7}$ & House\_4570 & The number of houses constructed between 1945 and 1970 & IFV\\
        \hline
        $V_{\sigma,8}$ & House\_7080 & The number of houses constructed between 1970 and 1980 & IFV\\
        \hline
        $V_{\sigma,9}$ & House\_8090 & The number of houses constructed between 1980 and 1990 & IFV\\
        \hline
        $V_{\sigma,10}$ & House\_90 & The number of houses constructed after 1990 & IFV\\
        \hline
        $V_{\sigma,11}$ & House\_frsd & The number of free standing (detached or semi-detached) houses & IFV\\
        \hline
        $V_{\sigma,12}$ & Resid & The total number of residents & CBS\\
        \hline
        $V_{\sigma,13}$ & Resid\_14 & The number of residents with an age in the range of 0 till 14 & CBS\\
        \hline
        $V_{\sigma,14}$ & Resid\_1524 & The number of residents with an age in the range of 15 till 24 & CBS\\
        \hline
        $V_{\sigma,15}$ & Resid\_2544 & The number of residents with an age in the range of 25 till 44 & CBS\\
        \hline
        $V_{\sigma,16}$ & Resid\_4564 & The number of residents with an age in the range of 45 till 64 & CBS\\
        \hline
        $V_{\sigma,17}$ & Resid\_65 & The number of residents with an age of 65 or higher & CBS\\
        \hline
        $V_{\sigma,18}$ & Man & The number of male residents & CBS\\
        \hline
        $V_{\sigma,19}$ & Woman & The number of female residents & CBS\\
        \hline
        $V_{\sigma,20}$ & Address & The number of addresses in the neighbourhood & CBS\\
        \hline
        $V_{\sigma,21}$ & Urbanity\tablefootnote{Categorical, where urbanity of a neighbourhood is evaluated into levels varying from 1 to 5 (urban-not urban), thus is treated as numerical in random forests.} & The urbanity of the neighbourhood & CBS\\
        \hline
        $V_{\sigma,22}$ & Town & Boolean variable indicating the presence of a town & CBS\\
        \hline
        $V_{\tau,1}$ & WindSpeed & Daily mean wind speed (km/h) & KNMI\\
        \hline
        $V_{\tau,2}$ & Temperature & Daily mean temperature (\SI{}{\degreeCelsius}) & KNMI\\
        \hline
        $V_{\tau,3}$ & WindChill & Daily mean wind chill (\SI{}{\degreeCelsius}) (calculated from $V_{\tau,1}, V_{\tau,2}$) &\\
        \hline
        $V_{\tau,4}$ & Sunshine & Daily sunshine duration (h) & KNMI\\
        \hline
        $V_{\tau,5}$ & Visibility\tablefootnote{Categorical, where minimum visibility distances (0–$\infty$ km) are defined to levels varying from 1 to 80, thus is treated as numerical in random forests.} & Daily minimum visibility & KNMI\\
        \hline
    \end{tabular}
    \caption{Putative explanatory variables, with their abbreviations, descriptions and sources. $\sigma$: spatial variables, $\tau$: temporal variables.} \label{tab:influencing variables}
\end{table}

\subsection{Data Pre-processing}
\label{subsec:data_pre-processing}
To unify the data to be used in the modeling process, we perform several data pre-processing operations on environmental variables. Firstly, we collect building information at the level of living units, as it may reflect the geographical information of chimneys in a more representative way. For instance, two families living in a semi-detached house actually use their individual chimneys. Secondly, some spatial variables (i.e.\ population and urbanity information) consist of a list of historical data, whereas others (i.e.\ building information) are only accessed in a single actual value. To enable similar treatment of the data, we use the averaged value over the time period of interest for the variables consisting of a list of historical data, so that for all spatial variables, a single value is accessible. Moreover, we only consider the buildings that are currently in use and assume that buildings keep their functions and types during the time period of interest. Observing that the age and type information of some buildings are missing, we assign them the label `extra' during the counting process. Additionally, we exclude the fire incidents and weather data on the leap day in leap years, while in the application of future fire risk prediction, we will use the prediction of February~28 to compensate for the missing prediction of the leap day, February~29. Finally, to obtain spatial environmental variables for every location in Twente from areal unit data (i.e.\ population and urbanity information) or data with precise coordinates (i.e.\ building information), we employ kernel smoothing. For temporal variables, we assume their values at different times of a day remain invariant and equal to the daily mean values. 

Based on the pre-processing steps above, we introduce the following notation for environmental variables:
\begin{itemize}
\item[-] $V_{\sigma, i}(u)$: the smoothed value of the $i$-th spatial variable at location $u$,
\item[-] $V_{\tau, i}(t)$: the value of the $i$-th temporal variable at time $t$,
\end{itemize}
where $(u,t)$ denotes any location and time combination in the spatio-temporal domain.

\section{Selection of Explanatory Variables}
\label{sec:slev}
Obviously, not all putative environmental variables listed in Table~\ref{tab:influencing variables} are required to explain the incidence of chimney fires and some of them are mutually dependent. To select the most important variables, we perform a non-parametric variable importance analysis using the permutation importance techniques of random forests~\cite{leo2001randomForest}. 

\subsection{Random Forests and Permutation Importance}
\label{subsec:random_forest_permutation_importance}
Random forests \cite{leo2001randomForest} are widely used as robust classification and regression methods in many applications, such as risk prediction \cite{Wongvibulsin2019randomForestapp} and data mining \cite{Schonlau2020randomForestapp}. A random forest is usually composed of hundreds or thousands of decision trees, where each tree is generated on a sampled subset of the data by repeated bagging (i.e.\ bootstrap sampling with replacement) and trained to fit the explanatory variables to the response variable. Afterwards, a combined result over all trees will be reported as the final output. In addition, random forests can be used to assess the importance of a variable by means of permutation importance techniques \cite{leo2001randomForest, Strobl2008permutationImportance,Altmann2010permutationImportance}. Through randomly permuting the values of a variable over the observations, the importance of a variable is defined as the mean increase of the prediction error over all trees computed on the permuted data compared to that computed on the original data.

More formally, consider a regression problem on a dataset $\mathcal{D}$ with $n$ observations, and suppose that for each observation of the dataset, there are $m$ explanatory variables. Let $x_{i}^{j}$ and $y_{i}$, with $i=1,...,n$ and $j=1,...,m$, denote the $j$-th explanatory variable and the response variable respectively for the observation $i$; $x_{i}$ is the vector that collects all $x_i^j$. The construction of the random forest consists of the generation of $T$ decision trees. To generate a tree $t$, a subset of $\mathcal{D}$, denoted as $\mathcal{B}(t)$, is sampled by bagging. At each node of the tree, a number of explanatory variables are selected randomly from all variables as the candidates. Then, one of the candidates, say $x^{j}$, is used to split the node into two subsets, $\mathcal{B}_{L}(t)$ and $\mathcal{B}_{R}(t)$ (e.g.\ for a numerical $x^{j}$, $\mathcal{B}_{L}(t)=\{(x_{i},y_{i}):x_{i}^{j}\leq c\}$ and $\mathcal{B}_{R}(t)=\{(x_{i},y_{i}):x_{i}^{j}> c\}$), in a way that the residual sum of squares 
\begin{equation}
    \sum_{(x_{i},y_{i})\in \mathcal{B}_{L}(t)}(y_{i}-\Bar{y}_{\mathcal{B}_{L}(t)})^{2}+\sum_{(x_{i},y_{i})\in \mathcal{B}_{R}(t)}(y_{i}-\Bar{y}_{\mathcal{B}_{R}(t)})^{2}
    \label{e:RF_sum}
\end{equation}
is minimized. Here, $\Bar{y}_{\mathcal{B}_{L}(t)}$ and $\Bar{y}_{\mathcal{B}_{R}(t)}$ denote the mean of the response variables in the corresponding subsets. Such node splitting procedure is continued until the tree satisfies certain preconditions (e.g. a maximum number of levels of the tree). Similar tree generation processes are employed to construct all other trees in the forest as well. To measure the importance of an explanatory variable, say $x^{j}$, in tree $t$, we randomly permute its values over the out-of-bag observations for tree $t$, denoted as $o\mathcal{B}(t)$ (i.e.\ $o\mathcal{B}(t)=\mathcal{D}(t) \setminus \mathcal{B}(t)$). Writing $\pi_{j}$ for the
permutation, the value of the of the $j$-th explanatory variable for response $y_i$ changes from $x_{i}^{j}$ to $x_{\pi_{j}(i)}^{j}$ when $i\in o\mathcal{B}(t)$. The values of other explanatory variables are left unchanged. Then, the increase of the prediction error is 
\begin{equation}
    I(x^{j};t)=\frac{\sum_{i\in o\mathcal{B}(t)}(y_{i}-\hat{y}_{i,\pi_{j}})^{2}}{|o\mathcal{B}(t)|}-\frac{\sum_{i\in o\mathcal{B}(t)}(y_{i}-\hat{y}_{i})^{2}}{|o\mathcal{B}(t)|},
    \label{e:RF_error}
\end{equation}
where $\hat{y}_{i}$ denotes the prediction in tree $t$ for observation $i$ using unpermuted explanatory variables, $\hat{y}_{i,\pi_{j}}$ denotes the corresponding prediction using explanatory variables with the $j$-th variable permuted and $|o\mathcal{B}(t)|$ denotes the number of out-of-bag observations for tree $t$. Finally, the mean increase of the prediction error over all trees, $\sum_{t=1}^{T}I(x^{j};t)/T$, is used to illustrate the importance of the $j$-th variable in the forest.

In practice, the original construction algorithms of random forests tend to bias the variable selection at tree nodes to factorial variables with many categories or continuous variables with many cut points. In addition, traditional permutation importance techniques can sometimes be misled by  correlated explanatory variables. To address these problems, unbiased random forests \cite{Hothorn2006recursivePartitioning,Strobl2007unbiasedRandomForest} and conditional permutation importance techniques \cite{strobl2008conditionalVariableImportance} were proposed based on a conditional inference framework of recursive partitioning. Compared to earlier methods, the idea is to partition the variable space in order to obtain groups of observations with similar association patterns instead of groups of observations with merely similar values of the response variable. For instance, to measure the importance of variable $x^{j}$ under conditional permutation in tree $t$, the set of out-of-bag observations, $o\mathcal{B}(t)$, is partitioned into a grid where each block of it shares the same information on the remaining variables, $x\setminus x^{j}$. Then, the permutation is applied to each block and the increase of the prediction error is computed and summed up over all blocks as the importance score of $x^{j}$. With these significant increments and optimizations, the random forest approach has proved very useful to measure variable importance non-parametrically considering variable dependence \cite{strobl2009partyOn}. Note that the notation used in this section is only valid here to explain random forests and permutation importance statistically.

\subsection{Variable Importance Analysis}
\label{subsec:variable_importance_analysis}
In our study, we measure the importance of spatial and temporal variables separately on areal unit data \cite{Lu2021ISI}. The main reason for this is that spatial variables are accessed in a single actual value while temporal variables are accessed as historical data. Specifically, for spatial variables, we group the incidents in 6291 pre-defined $500m \times 500m$ area boxes and merge them into a large data table consisting of 6291 rows and 23 columns. Each row indicates an area box, and the 23 columns refer to the number of incidents occurring in that box as well as the values of 22 spatial variables. In practice, we find that some boundary area boxes lie only partially in Twente. However, the population statistics for such boxes do not distinguish between Twente and the neighbouring regions. To avoid bias, we filter out the data of these boundary boxes in our variable importance analysis. For temporal variables, we group the incidents in 6205 days and merge them into a data table consisting of 6205 rows and 6 columns. Each row indicates a specific day, and the 6 columns refer to the number of incidents on that day as well as the values of 5 temporal variables. In the analysis, we construct unbiased random forests of 2000 trees and set the proportion of the number of input variables that are randomly sampled as candidates at a tree node to approximately $1/3$ (i.e.\ spatial: 7 candidates, temporal: 2 candidates). Considering the high correlation between certain variables (e.g.\ the urbanity and the population), we use conditional permutation importance techniques \cite{strobl2008conditionalVariableImportance} instead of the traditional ones \cite{leo2001randomForest} to suppress the importance scores biasing towards correlated variables. Our implementation of variable importance analysis is based on the R-package \textit{party}  \cite{Hothorn2006party,Strobl2007unbiasedRandomForest}.  We use the method proposed in \cite{Debeer2020conditionalVI} to compute the conditional permutation importance which provides a faster computation and shows more stable results than the original implementation \cite{strobl2008conditionalVariableImportance}.

\begin{figure}[h]
    \centering
    \begin{subfigure}[]{0.48\textwidth}
        \centering
        \includegraphics[scale=0.32]{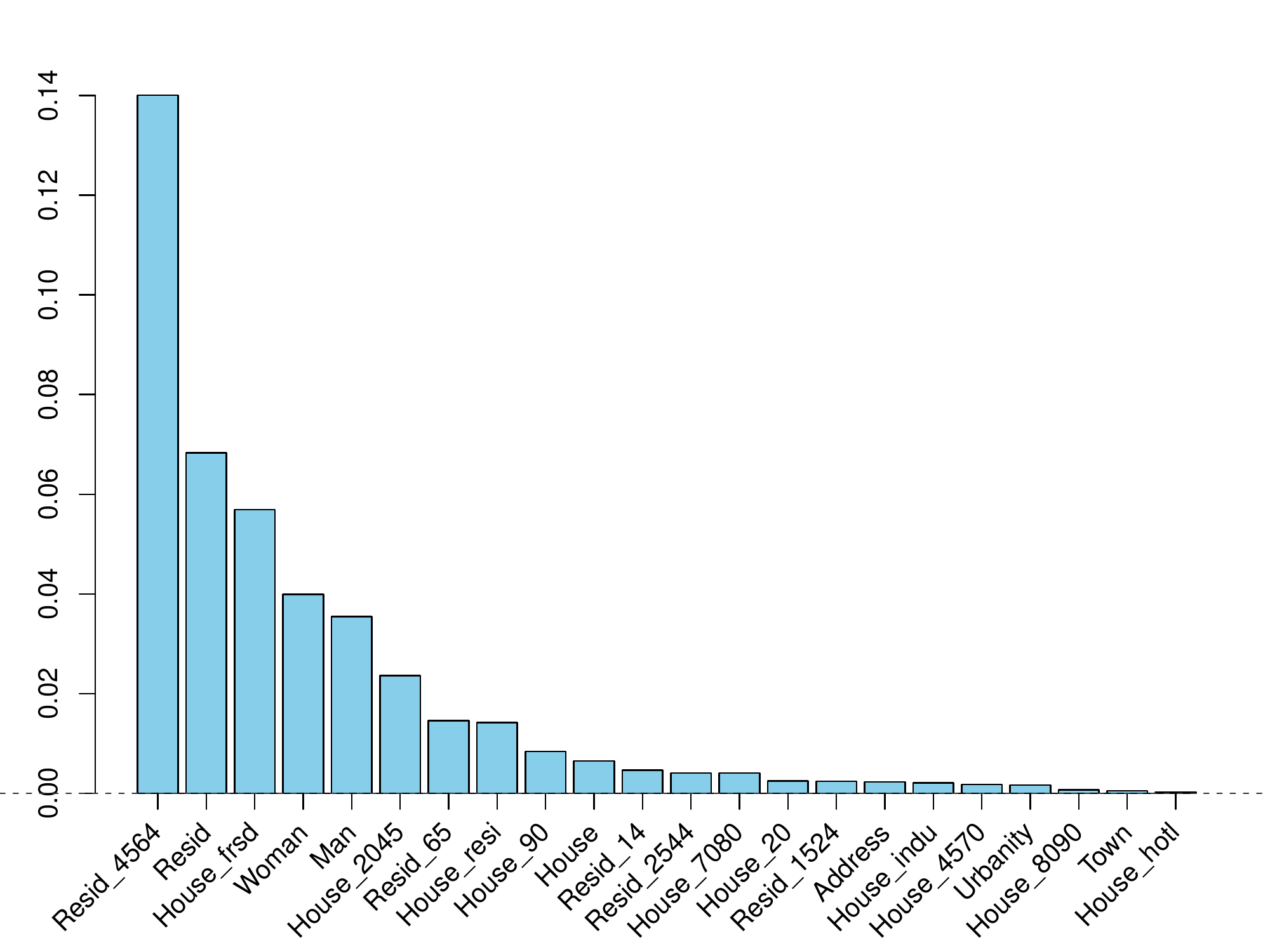}
        \caption{traditional permutation}
        \label{subfig:VI_spatial_traditional}
    \end{subfigure}%
    \begin{subfigure}[]{0.48\textwidth}
        \centering
        \includegraphics[scale=0.32]{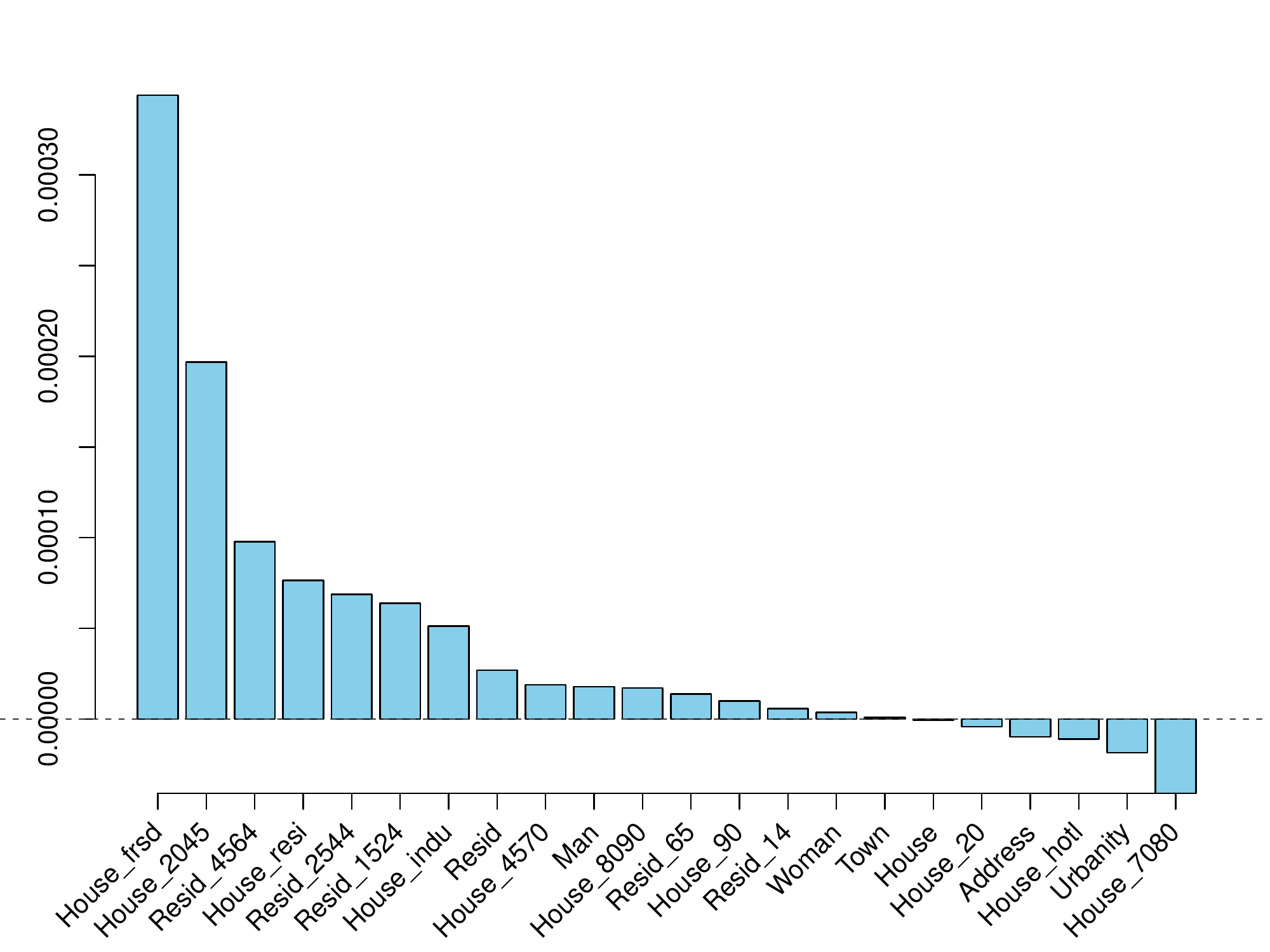}
        \caption{conditional permutation}
        \label{subfig:VI_spatial_conditional}
    \end{subfigure}%
    \caption{Importance ($y$-axis) obtained for spatial variables using traditional (a) and conditional (b) permutation techniques.}
    \label{fig:VI_spatial}
\end{figure}

\begin{figure}[h]
    \centering
    \begin{subfigure}[]{0.48\textwidth}
        \centering
        \includegraphics[scale=0.32]{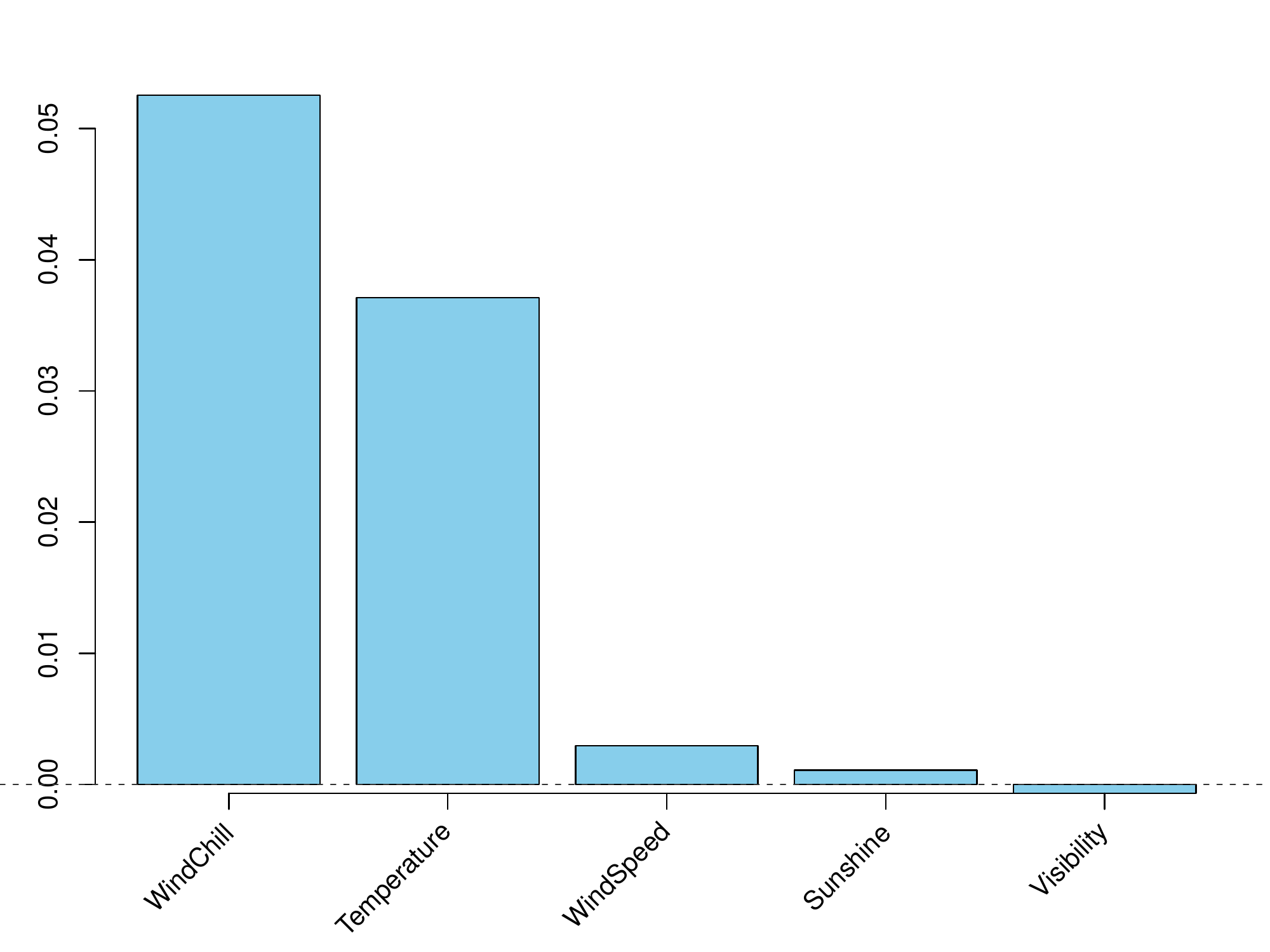}
        \caption{traditional permutation}
        \label{subfig:VI_temporal_traditional}
    \end{subfigure}%
    \begin{subfigure}[]{0.48\textwidth}
        \centering
        \includegraphics[scale=0.32]{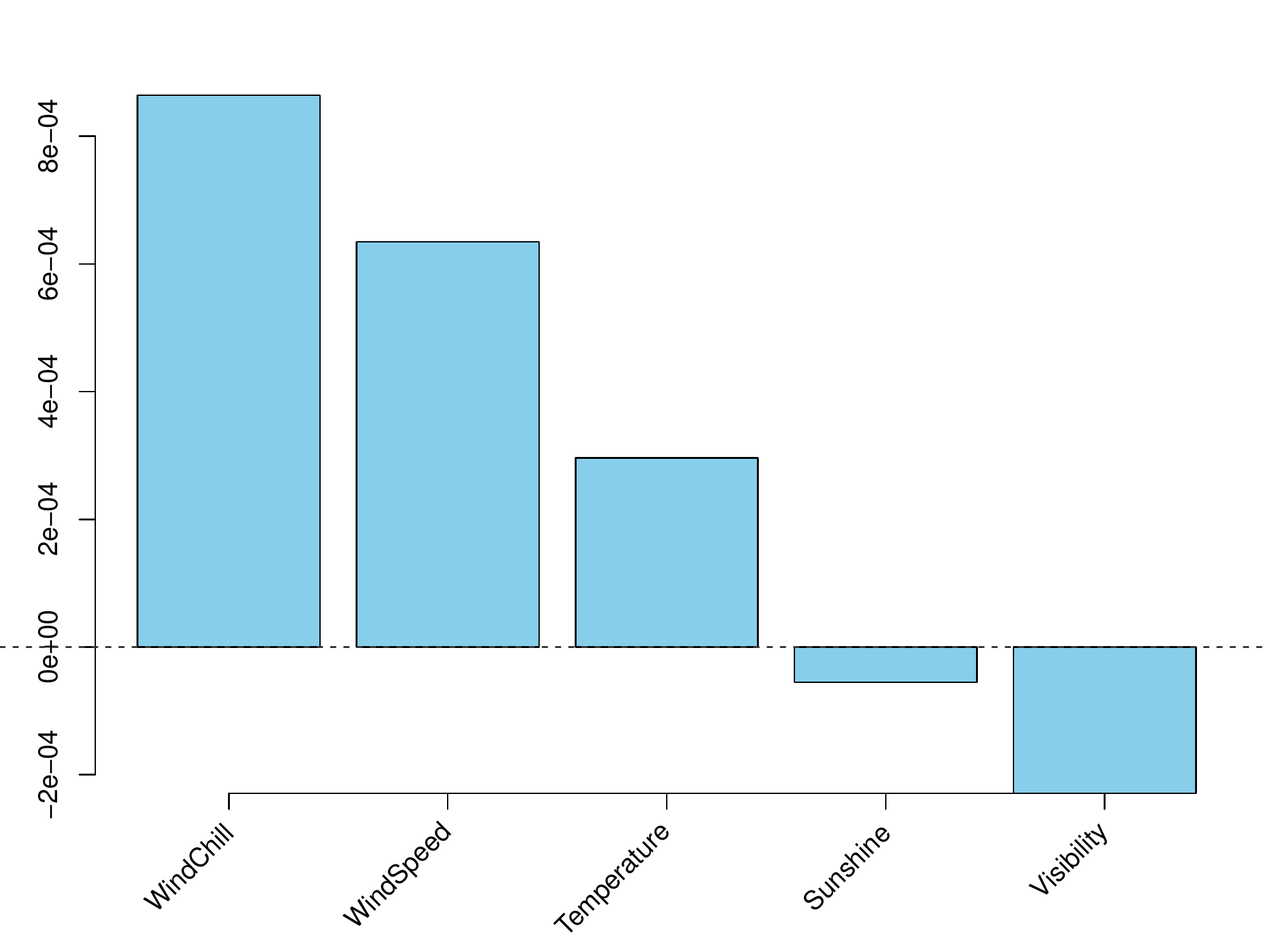}
        \caption{conditional permutation}
        \label{subfig:VI_temporal_conditional}
    \end{subfigure}%
    \caption{Importance ($y$-axis) obtained for temporal variables using traditional (a) and conditional (b) permutation techniques.}
    \label{fig:VI_temporal}
\end{figure}

The conditional permutation variable importance results for spatial and temporal variables are plotted in Figure~\ref{fig:VI_spatial} and \ref{fig:VI_temporal} respectively. For comparison, we plot the traditional permutation results as well. In all plots, the $y$-axis refers to the increase of the squared prediction error when traditional or conditional permutation on a variable is applied. A large increase indicates that the variable is very important for correct predictions, while a decrease indicates that the variable has no influence on or even hampers the prediction. If we compare the traditional and conditional permutation results, we see that the bias due to correlated variables is well suppressed. For instance, the number of residents aged between 45 and 64 and the total number of residents have the largest importance scores under traditional permutation, with the number of free standing houses coming the third. However, the latter has the highest variable importance score under conditional permutation. A possible explanation could be that free standing houses are mostly occupied by families whose adult members aged between $45$ and $64$, which implies a strong positive correlation among all three categories. Moreover, other age groups such as the elderly or families with young children may be less inclined to use their chimney. As a result, the importance scores of the two variables concerning residents decrease a lot under conditional permutation. Similar observations hold for temporal variables as well. Both wind chill and temperature obtain high importance scores under traditional permutation although they are correlated. Conditional permutation detects the underlying correlation and suppresses temperature, as wind chill is defined in terms of both temperature and wind speed information. Since wind speed still obtains a relatively large importance score even under conditional permutation, we consider it in further modeling. Additional analysis to assess the influence of small weather variations among different parts in Twente indicates that these variations can be ignored.

Overall, according to conditional permutation variable importance, \textbf{the number of buildings constructed between 1920 and 1945} and \textbf{the number of free standing houses} are the most important spatial variables, and \textbf{wind chill} and \textbf{wind speed} are the most important temporal variables. The results could have the following explanations: i) most free standing houses contain chimneys whereas other types do not; ii) chimney pipes in old buildings tend to be made of brick rather than metal, which increases the risk to catch fires; iii) strong wind can fuel a fire; iv) wind chill reflects people feeling cold thus inducing them to use their fires and chimneys.

\section{Poisson Point Process Model}
\label{sec:psml}
In this section, we motivate and propose a nested Poisson point process model for chimney fire prediction. We also present the model fitting and selection procedure and propose confidence intervals for both model parameters and predicted fire intensities.

\subsection{Motivations}
\label{subsec:motivations}
To design an appropriate prediction model structure, we investigate the relations between the selected explanatory variables and chimney fire occurrences. 

\begin{figure}[h]
    \centering
    \begin{subfigure}[]{0.48\textwidth}
        \centering
        \includegraphics[scale=0.28]{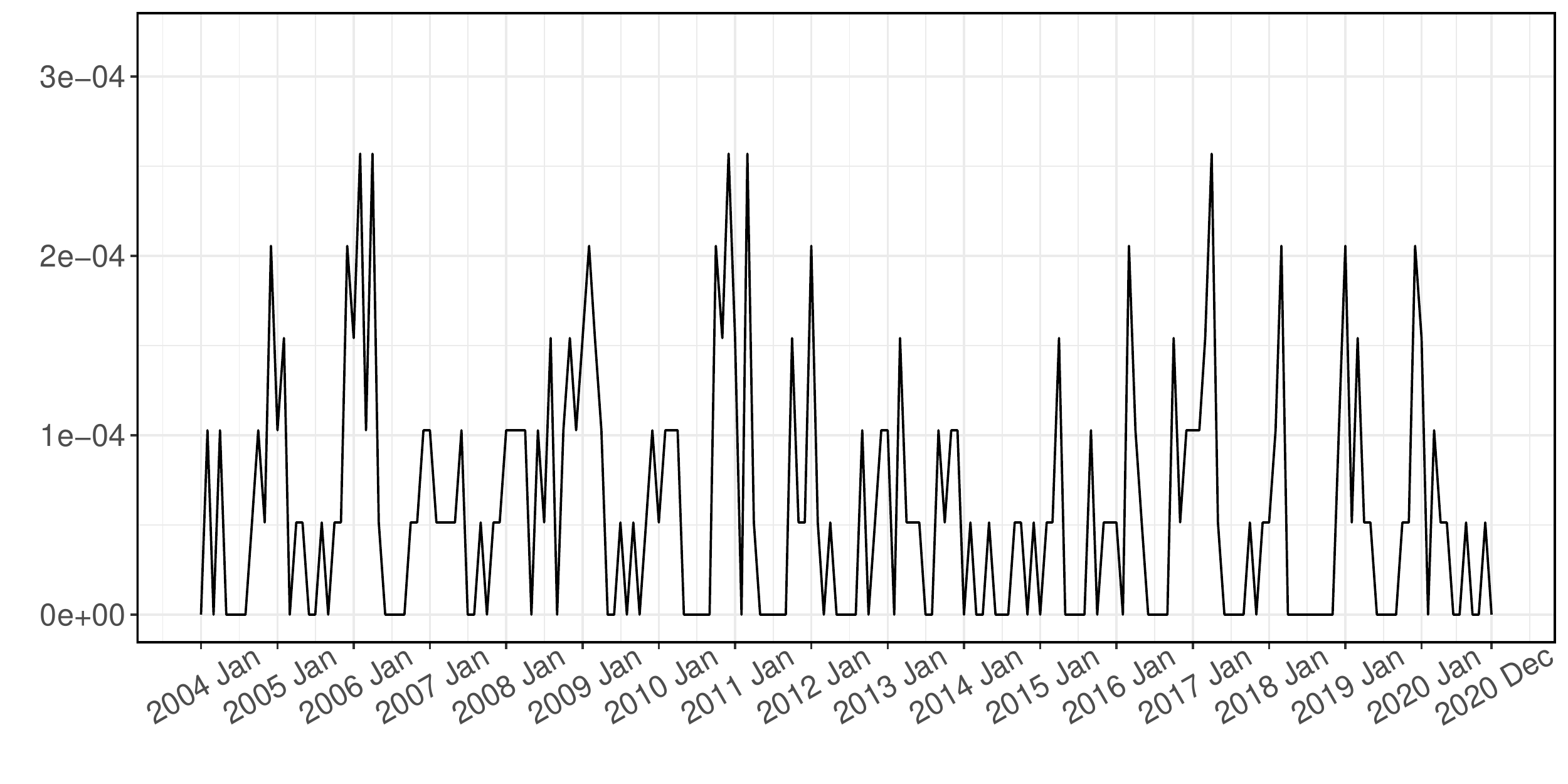}
        \caption{house type 1 (2045 -- freestanding)}
        \label{subfig:counts_house_type1}
    \end{subfigure}%
    \hfill
    \begin{subfigure}[]{0.48\textwidth}
        \centering
        \includegraphics[scale=0.28]{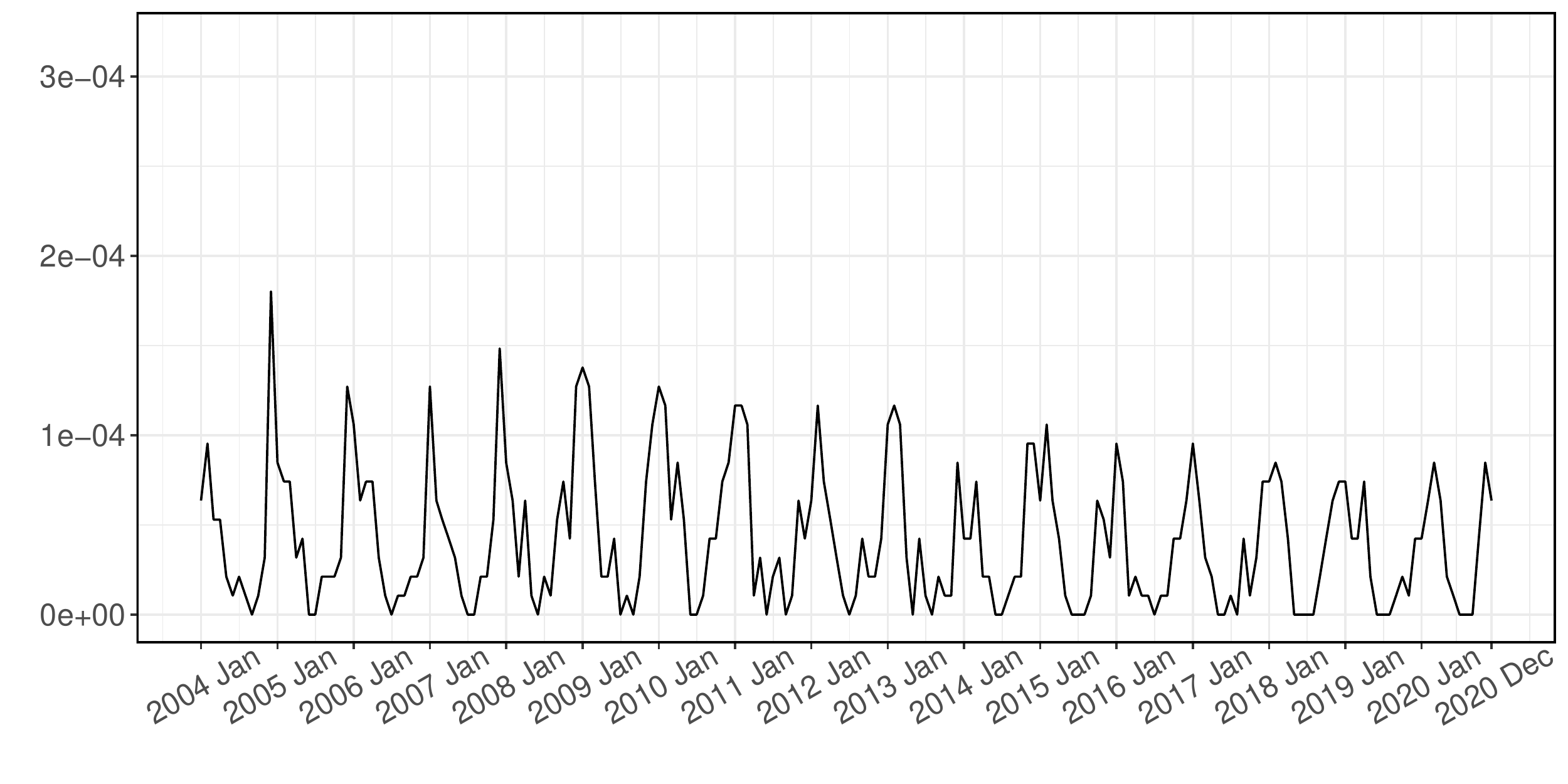}
        \caption{house type 2 (not 2045 -- freestanding)}
        \label{subfig:counts_house_type2}
    \end{subfigure}%
    \vskip\baselineskip
    \begin{subfigure}[]{0.48\textwidth}
        \centering
        \includegraphics[scale=0.28]{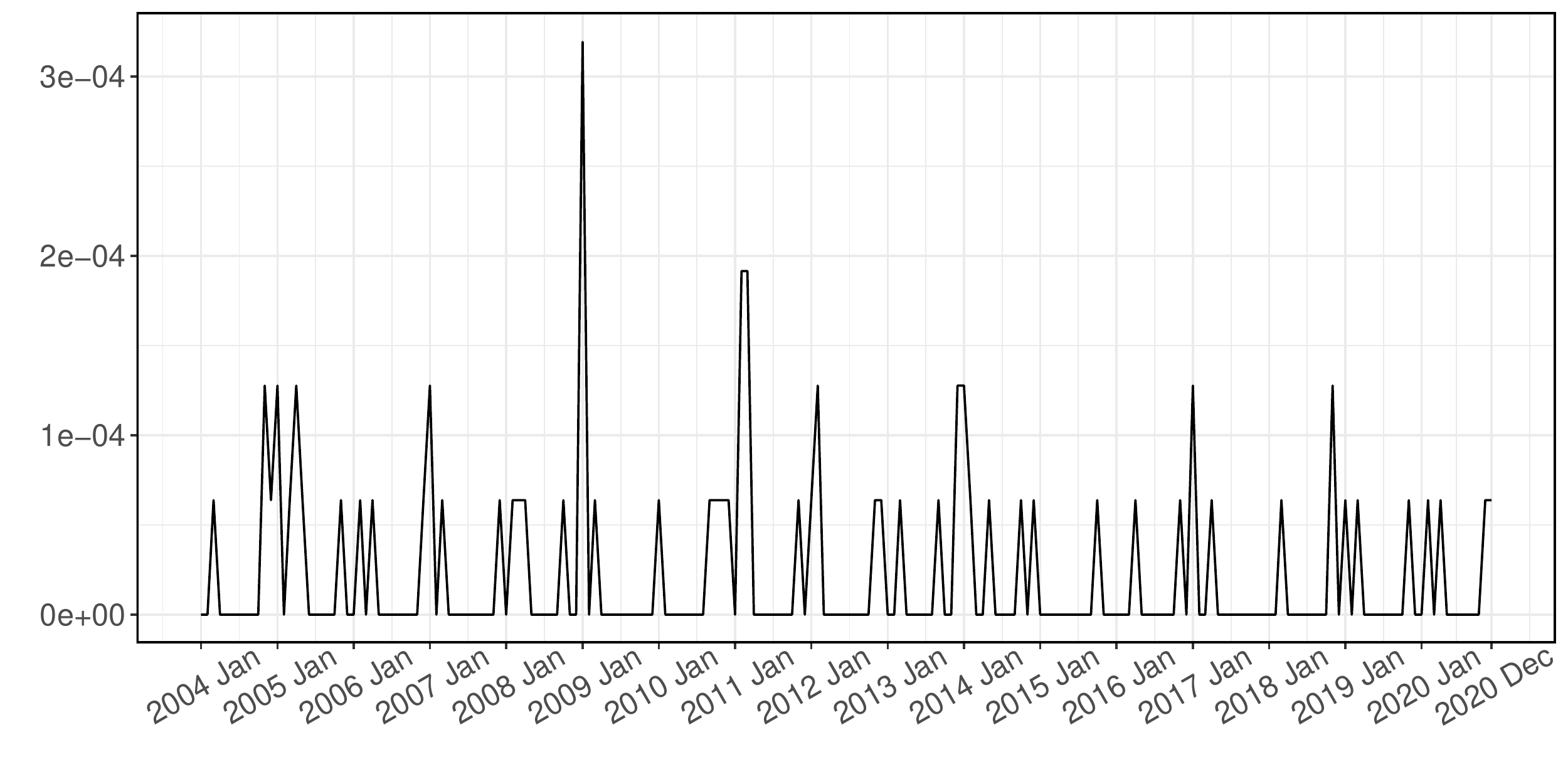}
        \caption{house type 3 (2045 -- not freestanding)}
        \label{subfig:counts_house_type3}
    \end{subfigure}%
    \hfill
    \begin{subfigure}[]{0.48\textwidth}
        \centering
        \includegraphics[scale=0.28]{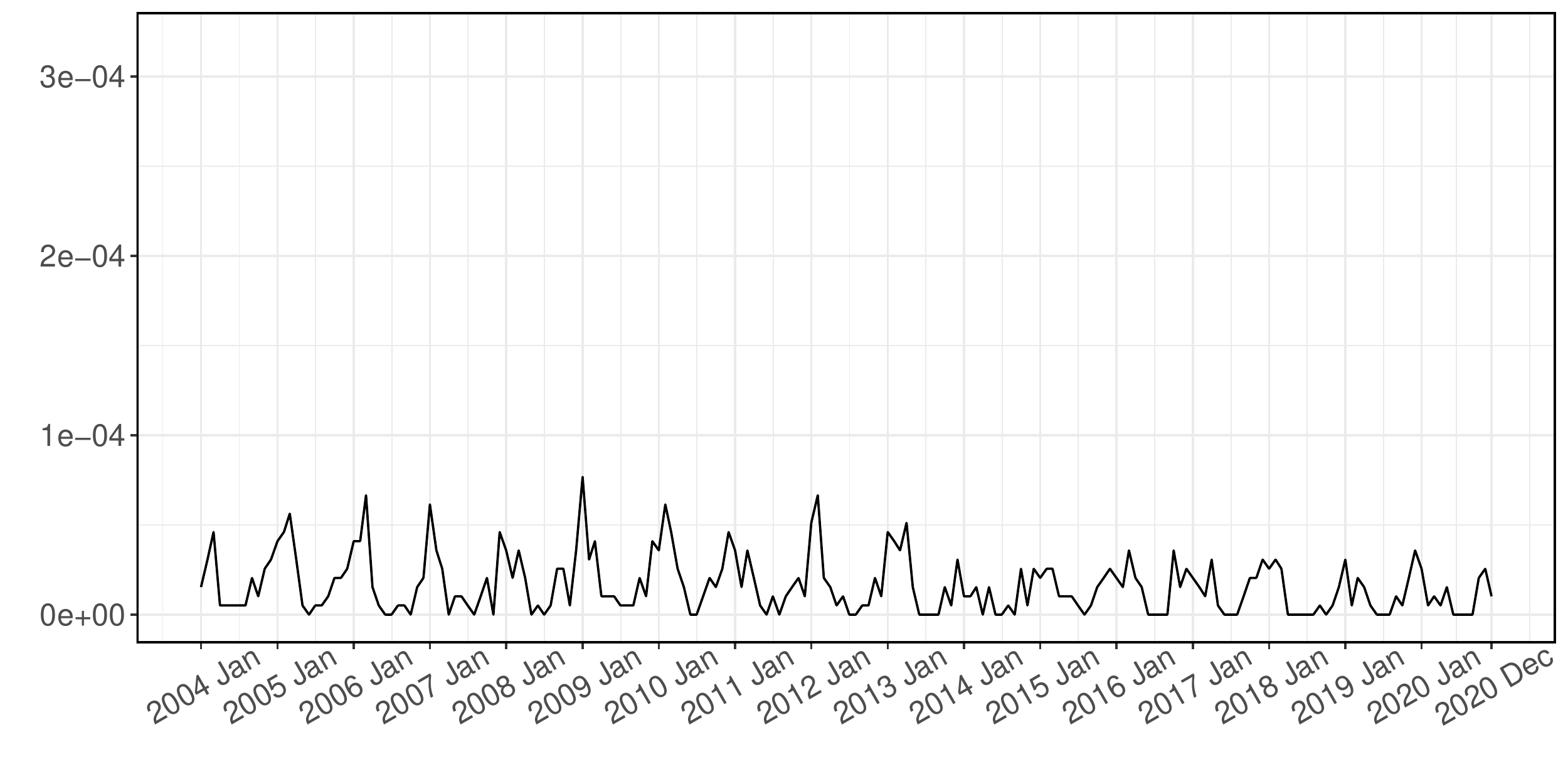}
        \caption{house type 4 (not 2045 -- not freestanding)}
        \label{subfig:counts_house_type4}
    \end{subfigure}%
    \caption{Monthly intensities of chimney fires per house ($y$-axis) for different house types during 2004--2020.}
    \label{fig:house_types}
\end{figure}

Firstly, we divide the houses into four house types depending on the age (whether they have been constructed between 1920 and 1945 or not) and on whether they are free standing or not, and plot the monthly intensities of chimney fires per house for different house types separately. Figure \ref{fig:house_types} shows that, generally, chimney fire occurrences in all house types are periodic, with incidents concentrated in the colder seasons. However, different house types run different risks of chimney fires: the intensities for old houses (i.e.\ House\_2045) and free standing houses are higher than others. Moreover, note that the patterns are not perfectly periodic; the amplitudes of the peaks vary per year. To model such varying patterns, we take wind chill and wind speed into account.

\begin{figure}[h]
    \centering
    \begin{subfigure}[]{0.48\textwidth}
        \centering
        \includegraphics[scale=0.27]{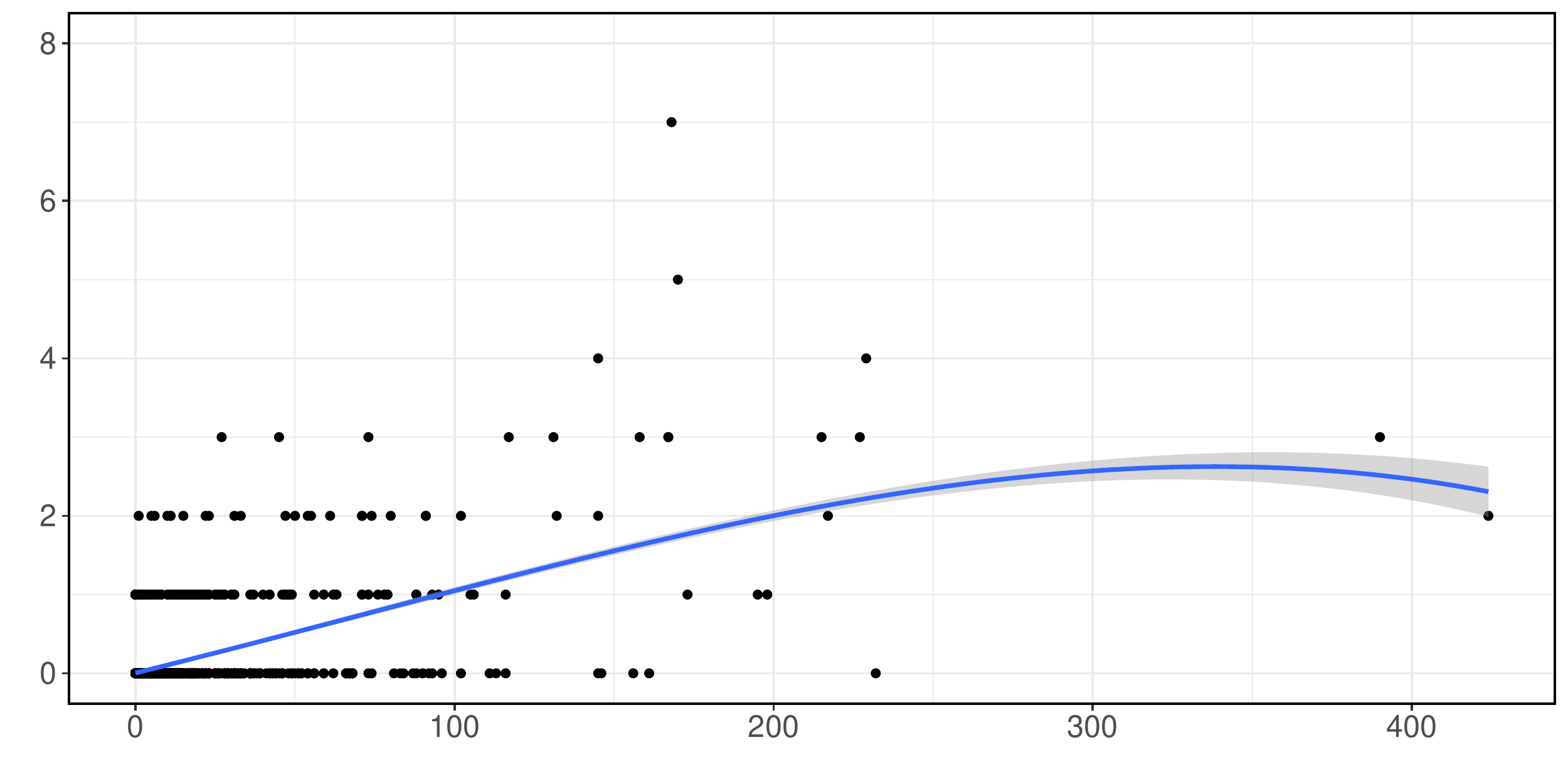}
        \caption{house type 1 (2045 -- freestanding)}
        \label{subfig:relation_house_type1}
    \end{subfigure}%
    \hfill
    \begin{subfigure}[]{0.48\textwidth}
        \centering
        \includegraphics[scale=0.27]{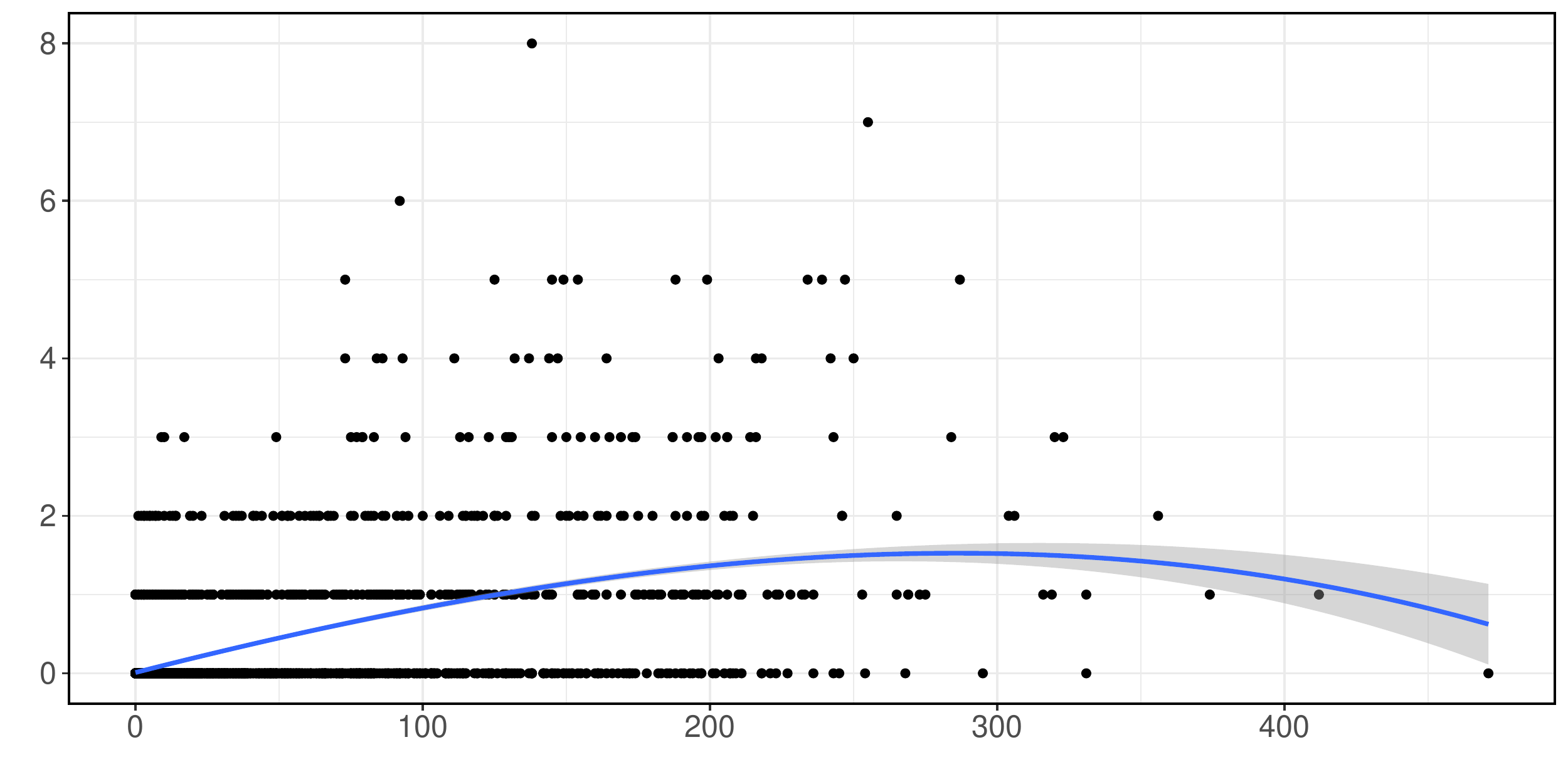}
        \caption{house type 2 (not 2045 -- freestanding)}
        \label{subfig:relation_house_type2}
    \end{subfigure}%
    \vskip\baselineskip
    \begin{subfigure}[]{0.48\textwidth}
        \centering
        \includegraphics[scale=0.27]{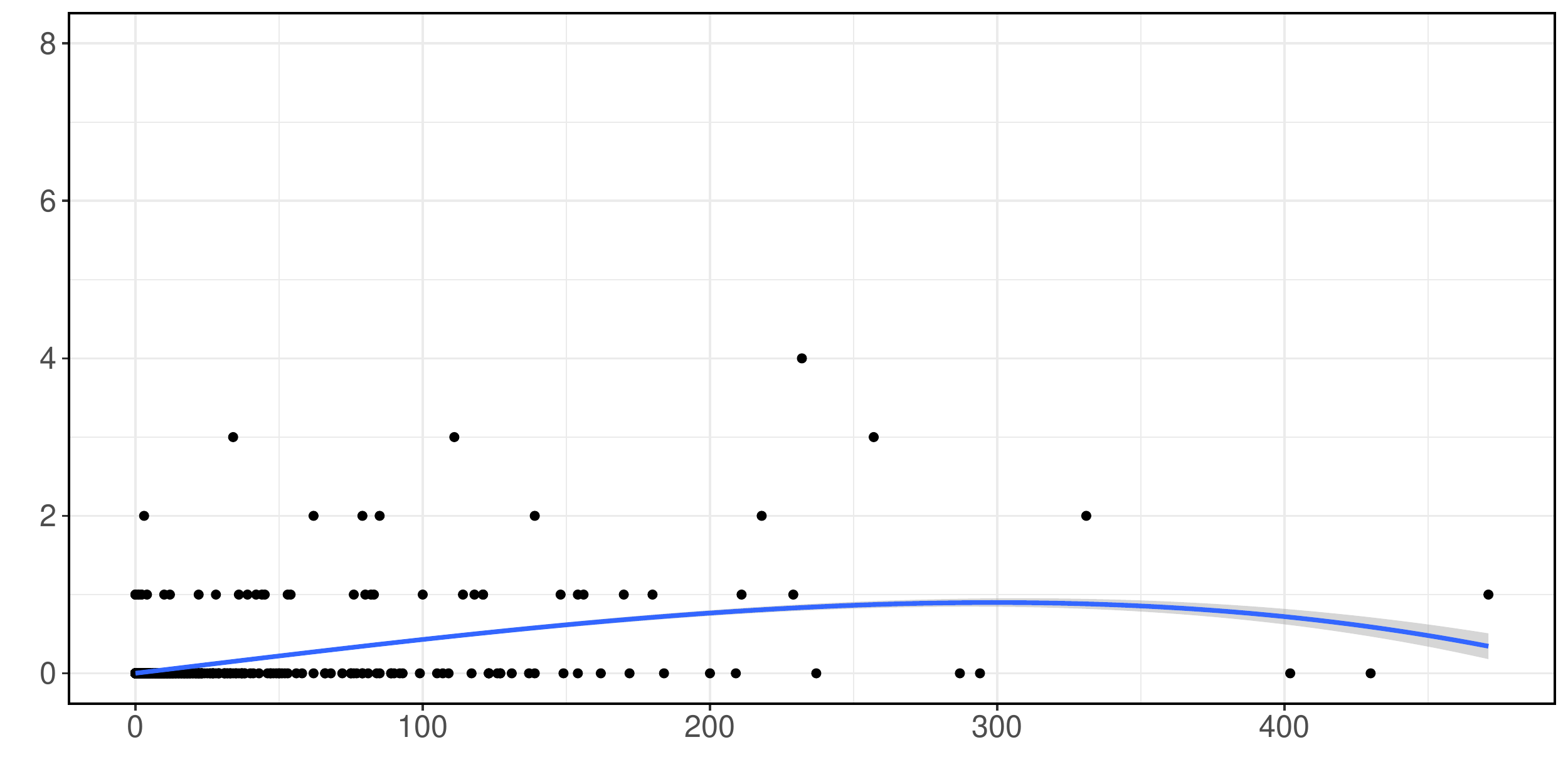}
        \caption{house type 3 (2045 -- not freestanding)}
        \label{subfig:relation_house_type3}
    \end{subfigure}%
    \hfill
    \begin{subfigure}[]{0.48\textwidth}
        \centering
        \includegraphics[scale=0.27]{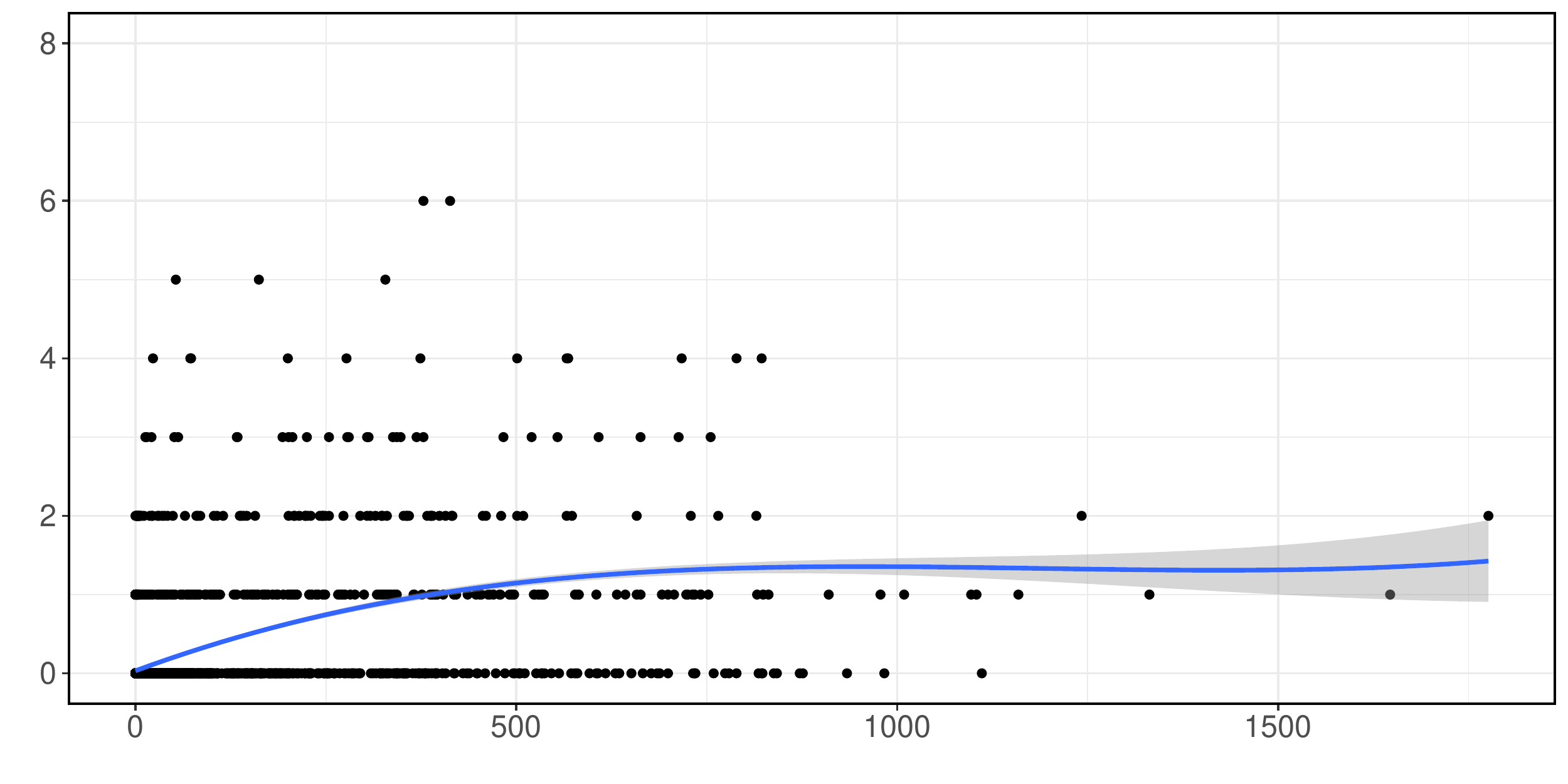}
        \caption{house type 4 (not 2045 -- not freestanding)}
        \label{subfig:relation_house_type4}
    \end{subfigure}%
    \caption{Numerical relations between chimney fire occurrences ($y$-axis) and the number of houses of different types ($x$-axis). The blue curves are the estimated trends using the locally estimated scatterplot smoothing method with parameter $span=1$. The $95\%$ envelopes are also included to show the confidence of such estimates. }
    \label{fig:variable_relations}
\end{figure}

In addition, we plot the numerical relations between chimney fire occurrences and the number of houses of different types in Figure \ref{fig:variable_relations} based on areal unit data \cite{Lu2021ISI}. To approximate the underlying trends in scatter plots, we apply the locally estimated scatter plot smoothing method \cite{Cleveland1992loess} and plot the estimated trends. It is interesting to find that, for each house type, chimney fire occurrences increase approximately linearly in the number of houses of that type. However, this observation only holds when the number of houses is not very large. If it is too large, the number of chimney fires tends to a saturated value. 

To summarize, we draw the following conclusions: i) chimneys in different house types catch fires at a type-dependent rate, and the rate is influenced by both seasonal information and temporal explanatory variables; ii) chimney fire occurrences in houses of a specific type are approximately proportional to the number of houses of that type.

\subsection{Model Structure}
\label{subsec:nested_model}
Based on the motivations above, we design the following chimney fire risk prediction model. We suppose that each house catches a chimney fire independently at a type-dependent rate that varies in time. Under this assumption, the model structure is that of a Poisson point process defined on the Twente region and on the period of year 2004--2020, with the intensity function of the form $\lambda(u,t)=\sum_{k}\lambda_{k}(u,t)$ where $\lambda(u,t)$ indicates the risk intensity at location $u$ at time $t$ and the subscript $k$ indicates for houses of type $k$. Moreover, $\lambda_{k}(u,t)= h_{k}(u)\lambda_{k}(t)$, where $h_{k}(u)$ indicates the density of house type $k$ at location $u$ and $\lambda_{k}(t)$ indicates the risk intensity for a house of type $k$ at time $t$. Such a structure reflects the conclusions reached in Section~\ref{subsec:motivations}, although it relies on the condition that the density of houses at a location is not too large.

\begin{figure}[h]
    \centering
    \begin{subfigure}[t]{0.45\textwidth}
        \centering
        \includegraphics[scale=0.286]{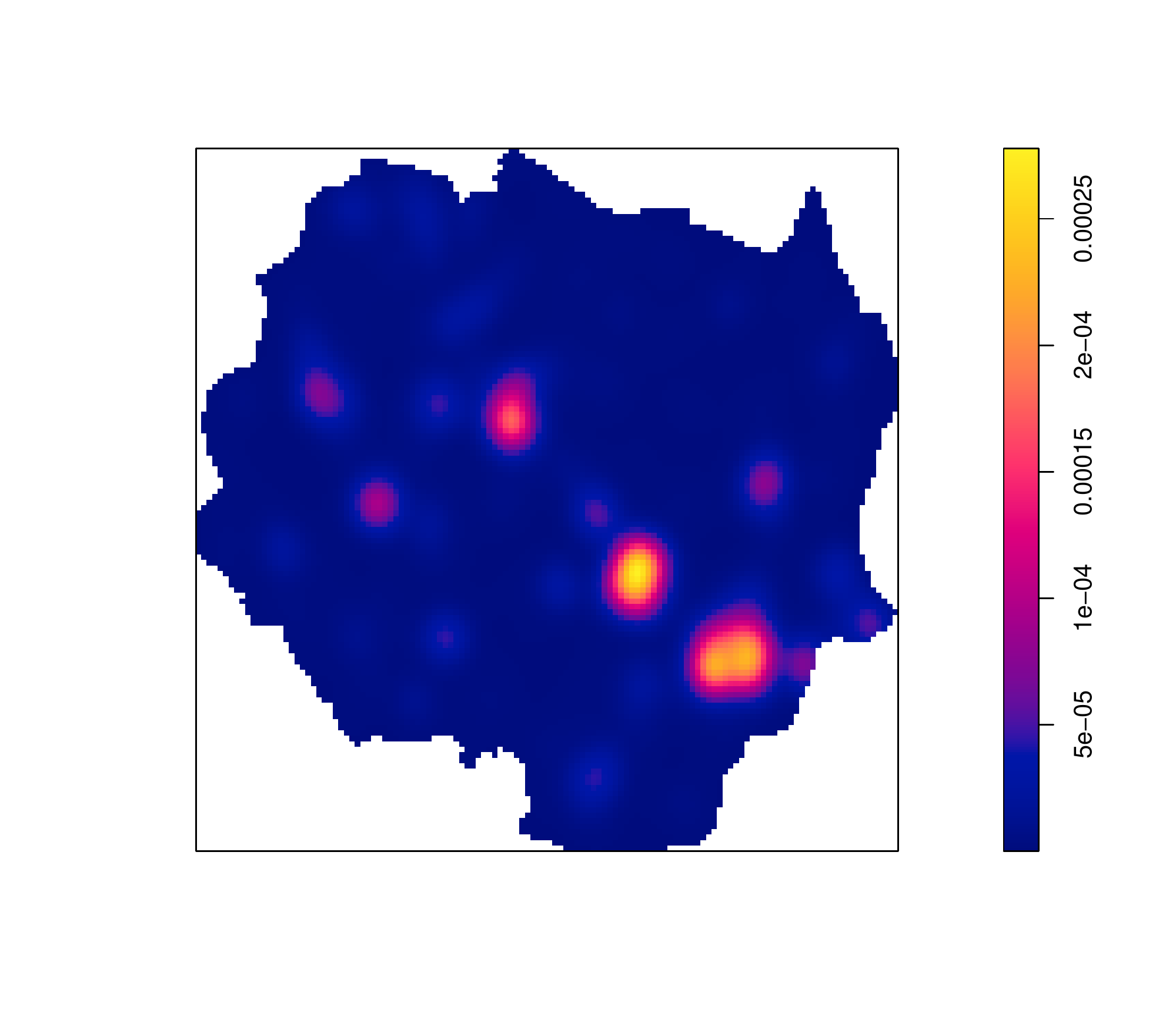}
        \caption{house type 1 (2045 -- freestanding)}
        \label{subfig:house_type_1_density}
    \end{subfigure}%
    \hfill
    \begin{subfigure}[t]{0.45\textwidth}
        \centering
        \includegraphics[scale=0.286]{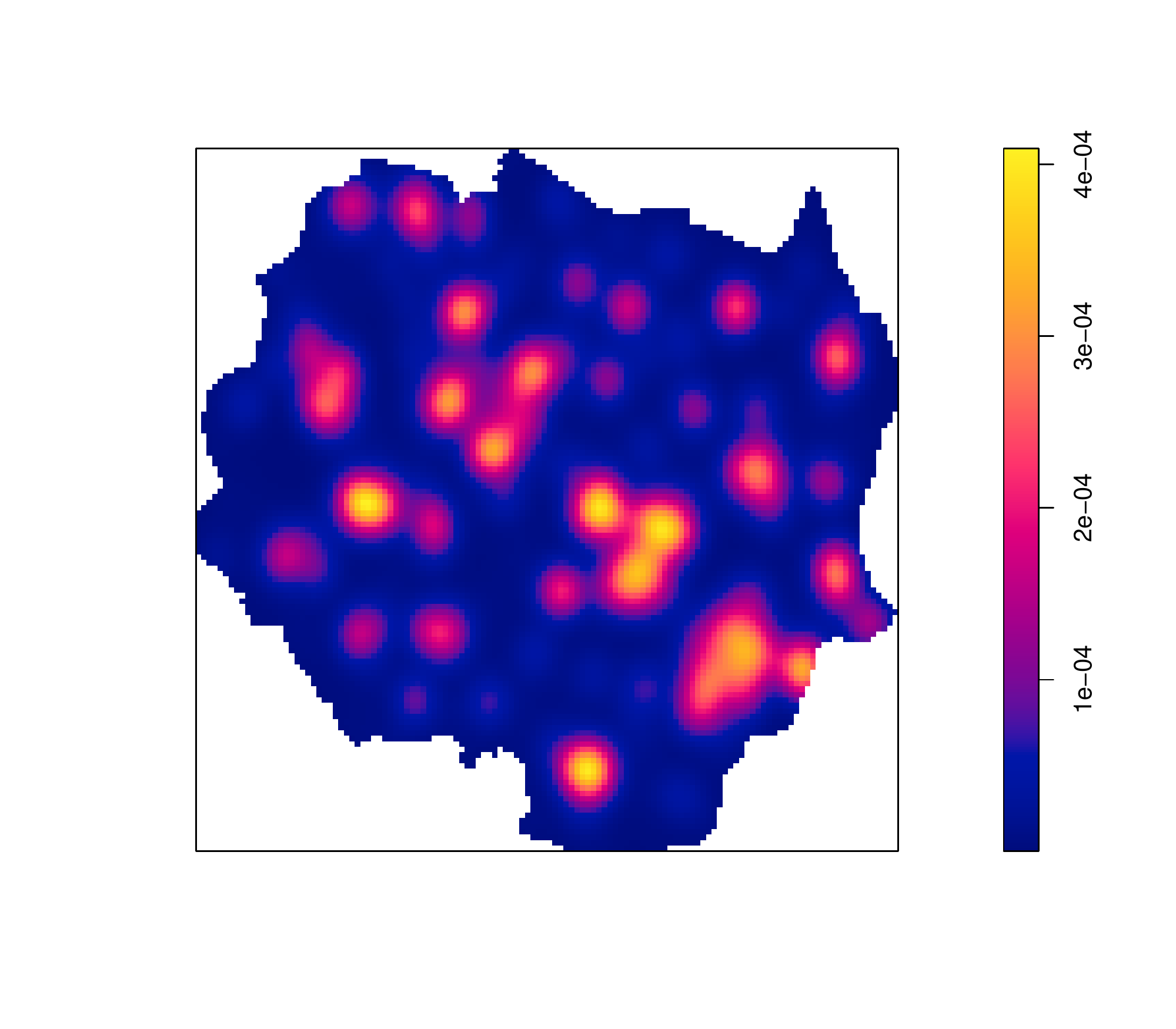}
        \caption{house type 2 (not 2045 -- freestanding)}
        \label{subfig:house_type_2_density}
    \end{subfigure}%
    \hfill
    \begin{subfigure}[t]{0.45\textwidth}
        \centering
        \includegraphics[scale=0.286]{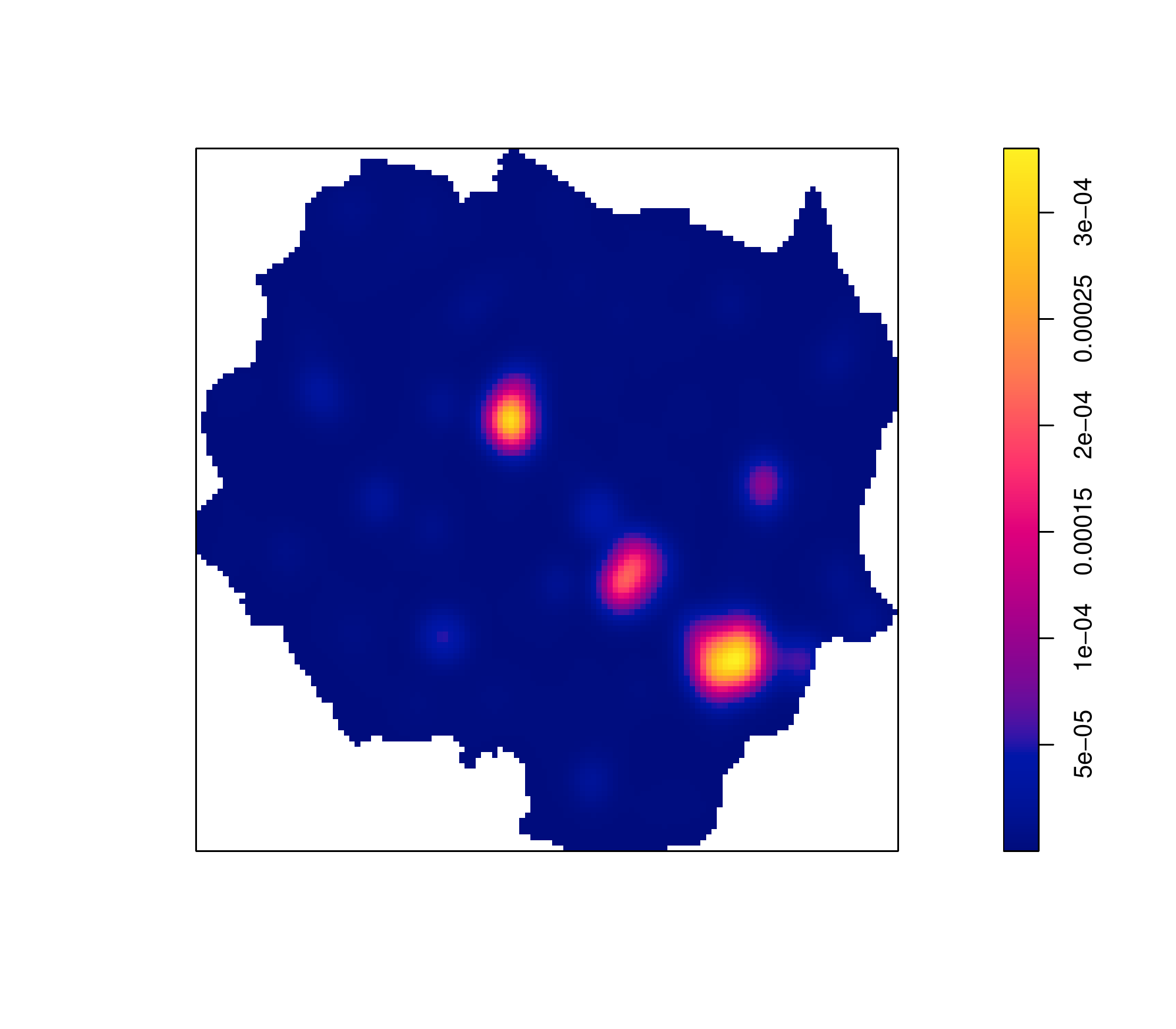}
        \caption{house type 3 (2045 -- not freestanding)}
        \label{subfig:house_type_3_density}
    \end{subfigure}%
    \hfill
    \begin{subfigure}[t]{0.45\textwidth}
        \centering
        \includegraphics[scale=0.286]{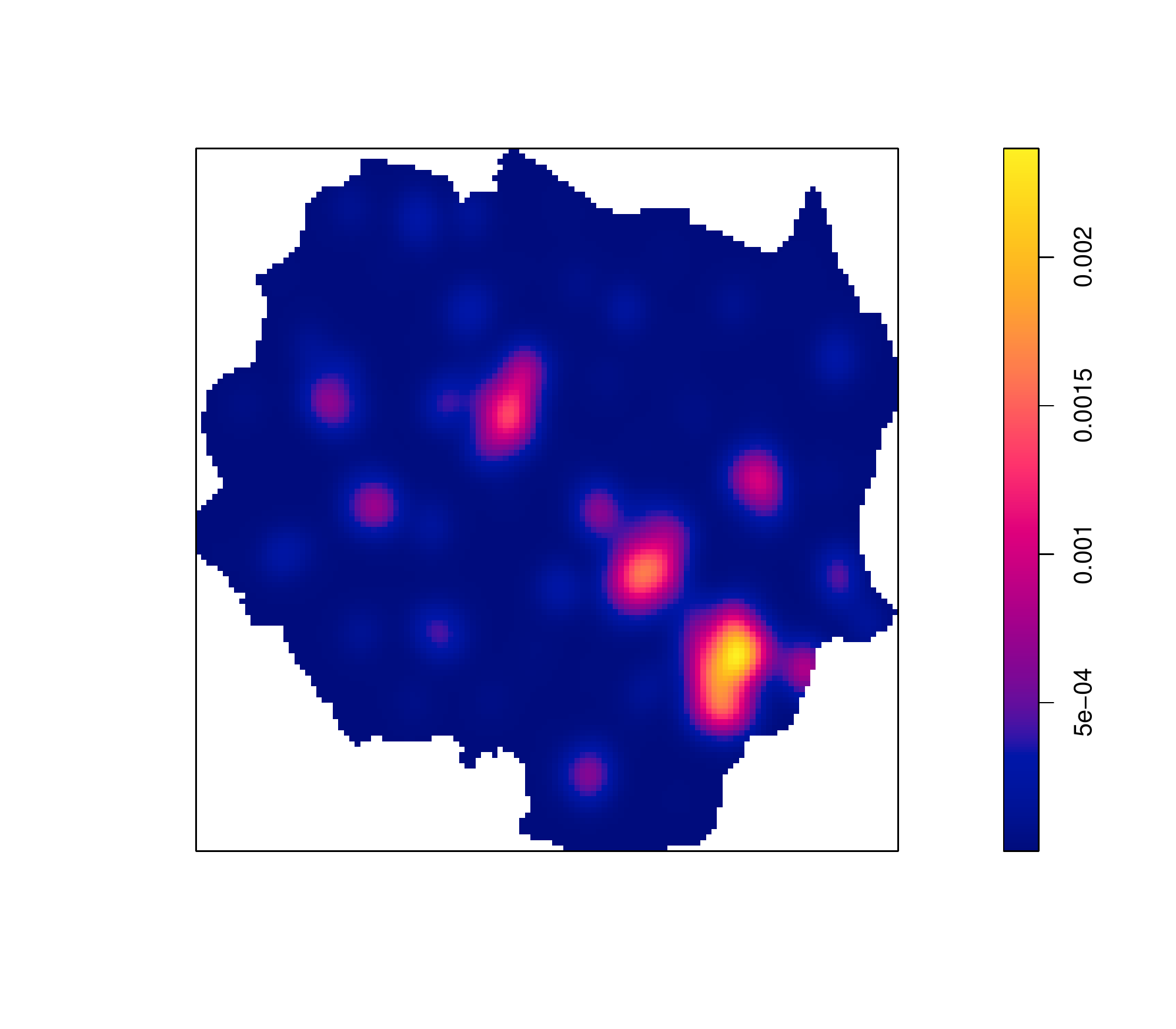}
        \caption{house type 4 (not 2045 -- not freestanding)}
        \label{subfig:house_type_4_density}
    \end{subfigure}%
    \caption{Density maps of houses of four types. The unit in all plots is $\text{metre}^{-2}$. Note that the intensity bars have different scales.}
    \label{fig:building_density}
\end{figure}

Formally, our nested Poisson point process model reads
\begin{equation}
    \lambda(u,t)=\sum_{k=1}^{4}\lambda_{k}(u,t)=\sum_{k=1}^{4}h_{k}(u)\lambda_{k}(t).
    \label{e:intensity}
\end{equation}
The density maps of houses of four types (i.e.\ $h_{k}(u)$), as shown in Figure~\ref{fig:building_density}, are derived from $V_{\sigma,6}(u)$ and $V_{\sigma,11}(u)$ in Table \ref{tab:influencing variables} by smoothing, using a Gaussian kernel with a standard deviation of 1,000 metres. Since both wind speed, $V_{\tau,1}(t)$, and wind chill, $V_{\tau,3}(t)$, were selected as important temporal variables in Section~\ref{subsec:variable_importance_analysis}, we propose a temporal intensity function of the form 
\begin{equation}
\begin{split}
    \lambda_{k}(t)=&\exp(\Harmonic(t;o_{k,1})+\Polynom(V_{\tau,1}(t);o_{k,2})+\\
    &\Polynom(V_{\tau,3}(t);o_{k,3})+\Polynom(V_{\tau,1}(t)V_{\tau,3}(t);o_{k,4})),
\end{split}
\label{e:lambda_k}
\end{equation}
where a harmonic function\footnote{Harmonic function here refers to cosine and sine functions.} with order $o_{k,1}$ is employed to model seasonal variations and polynomial functions with order $o_{k,2}$ and $o_{k,3}$ are used to model information of wind speed and wind chill.  In addition, we employ another polynomial function of $V_{\tau,1}(t)V_{\tau,3}(t)$ with order $o_{k,4}$ to allow for interactions between wind speed and wind chill. The exponential function is applied to guarantee that the risk intensity function stays positive. For purposes of parametric representation, the temporal intensity functions $\lambda_{k}(t)$ can also be described as $\lambda_{\theta_{k}}(u,t)$, where $\theta_{k}$ indicates the vector of coefficients in the harmonic and polynomial functions.

\subsection{Model Fitting}
\label{subsec:model_fitting}
In \cite{Lu2021ISI}, we fitted a fire prediction model similar to (\ref{e:intensity}) and (\ref{e:lambda_k}) on areal unit data using maximum likelihood estimation. We established theoretical confidence intervals for predicted intensities based on classical theories: the Fisher information for asymptotic normality of generalized linear models \cite{McCullagh2019GLM, Fahrmeir1985AsymptoticsGLM} and the Delta method \cite{VerHoef2012deltaMethod}. In principle, we could also apply this approach here to maximize the log-likelihood function of our Poisson point process  
\begin{equation}
    \sum_{k=1}^{4}\left\{\sum_{x_{k}}\log\lambda_{\theta_{k}}(x_{k})-\int_{\text{Twente}}\int_{\text{Year 2004--2020}}\lambda_{\theta_{k}}(u,t)dudt\right\},
    \label{e:log-likelihood}
\end{equation}
where $x_{k}$ runs through chimney fire incidents occurring in houses of type $k$. However, since we have a continuous space-time domain, the integral in the log-likelihood function must be approximated numerically, e.g.\ using quadrature points \cite{Baddeley2000PLE}. In our case, the number of quadrature points would have to be rather large, thus, we use the logistic regression estimation \cite{Baddeley2014LRPP} to fit our Poisson point process model efficiently. 

The idea behind the logistic regression estimation approach is the well-known Campbell-Mecke theorem (see, e.g. \cite{Daley2009CampbellMecke}). Consider a point process $X$ that is defined on a space-time domain $\mathbb{R}^{2} \times \mathbb{R}$ and is locally finite, which means $X$ is almost surely finite for every bounded subset of $\mathbb{R}^{2} \times \mathbb{R}$. We assume that $X$ is confined to a bounded domain $W \times T \subseteq \mathbb{R}^{2} \times \mathbb{R}$ and has intensity function $\lambda(u,t)$, where $(u,t)$ denotes a location and a time in the domain. For any real-valued non-negative or integrable function $f(u,t)$ defined on $W \times T$, the Campbell-Mecke theorem reads 
\begin{equation}
    E\left\{\sum_{x\in X}f(x)\right\}=\int_{W }\int_{T}f(u,t)\lambda(u,t)dudt,
    \label{e:CM}
\end{equation}
where $x$ runs through the points of $X$. Suppose the intensity function is parametric and depends on a parameter vector $\theta$, thus can be described as $\lambda_{\theta}(u,t)$, one might estimate both sides of the Campbell--Mecke formula and equate them to obtain an estimating equation for $\theta$.

The logistic regression estimation is based on the function
\begin{equation}
    f(u,t)=\frac{\partial}{\partial \theta}\log \left[\frac{\lambda_{\theta}(u,t)}{\lambda_{\theta}(u,t)+\rho(u,t)}\right]=\frac{\rho(u,t)/\lambda_{\theta}(u,t)}{\lambda_{\theta}(u,t)+\rho(u,t)}\nabla \lambda_{\theta}(u,t),
    \label{e:CM-LR}
\end{equation}
where $\nabla \lambda_{\theta}(u,t)$ denotes the gradient of $\lambda_{\theta}(u,t)$ over $\theta$ and $\rho$ is a positive-valued function defined on $W\times T$. Here, one needs to assume that $\lambda_\theta > 0$.  To estimate the integral in 
(\ref{e:CM}) with $f(u,t)$ defined in (\ref{e:CM-LR}), we use a `dummy' point process $D$ on $W\times T$
that is independent of $X$ and has intensity function of $\rho$. Apply the Campbell-Mecke theorem (\ref{e:CM}) to $D$, one has 
\begin{equation}
    E\left\{\sum_{x\in D}\frac{1}{\lambda_{\theta}(x)+\rho(x)}\nabla \lambda_{\theta}(x)\right\}
    =\int_{W}\int_{T}\frac{\rho(u,t)}{\lambda_{\theta}(u,t)+\rho(u,t)}\nabla \lambda_{\theta}(u,t)dudt.
    \label{e:CM-dummy}
\end{equation}
Having obtained unbiased estimators for both the left and right hand side of equation (\ref{e:CM}) with
$f$ given by (\ref{e:CM-LR}), we plug them in to obtain the estimating equation. It solves
\begin{equation}
    s(X,D;\theta)=\sum_{x\in X}\frac{\rho(x)/\lambda_{\theta}(x)}{\lambda_{\theta}(x)+\rho(x)}\nabla \lambda_{\theta}(x)-\sum_{x\in D}\frac{1}{\lambda_{\theta}(x)+\rho(x)}\nabla \lambda_{\theta}(x)=0
    \label{e:sLR}
\end{equation}
over the parameter $\theta$. Note that the subscript $x$ in two sums runs through the points of $X$ and $D$ respectively. It is interesting to observe that (\ref{e:sLR}) is precisely the derivative of a logistic log-likelihood function
\begin{equation}
    l(X,D;\theta)=\sum_{x\in X}\log\left[\frac{\lambda_{\theta}(x)}{\lambda_{\theta}(x)+\rho(x)}\right]+\sum_{x\in D}\log\left[\frac{\rho(x)}{\lambda_{\theta}(x)+\rho(x)}\right].
    \label{e:likelihood}
\end{equation}
In this way, the fitting of Poisson point processes can be transformed into a maximum likelihood estimation problem of logistic regression. It can be easily implemented based on the R-package \textit{stats} \cite{Venables2002stats} and treated as the fitting of generalized linear models with a \textit{logit} link function. Since the log-likelihood function is concave, the existence and uniqueness of parameter $\theta$ that maximizes the log-likelihood are guaranteed under certain conditions \cite{Silvapulle1981LRestimates}. 

\subsection{Modeling Chimney Fire Data}
We use logistic regression estimation to fit the intensity functions of our four house types separately. Consider (\ref{e:lambda_k}) for fixed $k$ and write $\theta_{k}$ for the vector of coefficients in the harmonic and polynomial functions. Also consider the chimney fires occurring in houses of type $k$ as a point process $X_{k}$ with intensity function $\lambda_{\theta_{k}}(u,t)$. Denote  the corresponding dummy point process by $D_{k}$ with intensity function $\rho_{k}(u,t)$. Assuming we have $m$ parameters in $\theta_{k}$, the estimating equations read
\begin{equation}
    s_{p}(X_{k},D_{k};\theta_{k})=\sum_{x\in X_{k}}\frac{\rho_{k}(x)C_{p}(x)}{\lambda_{\theta_{k}}(x)+\rho_{k}(x)}-\sum_{x\in D_{k}}\frac{\lambda_{\theta_{k}}(x)C_{p}(x)}{\lambda_{\theta_{k}}(x)+\rho_{k}(x)}=0, \qquad p=1,...,m,
    \label{e:logistic}
\end{equation}
where $C_{p}(x)$ denotes the $p$-th temporal covariate (i.e.\ the harmonic components, wind chill terms with different orders and the interaction term of wind chill and wind speed in (\ref{e:lambda_k})) at point $x$. Furthermore, we perform a three-step implementation to estimate $\theta_{k}$. Firstly, we specify the intensity function $\rho_{k}(u,t)$ of the dummy point process $D_{k}$. Considering that, spatially, most chimney fires occur in urban areas, and temporally, there are more chimney fires in winter than in summer, we tune $\rho_{k}(u,t)$ to concentrate on important regions and seasons so as to fit the model efficiently. Specifically, we use the density map of houses of the given type $k$ (cf.\ Figure~\ref{fig:building_density}) to distinguish urban areas from rural areas and design a sine function to assign higher $\rho_{k}(u,t)$ to winter seasons so as to comply with the observations in Figure~\ref{fig:temporal_distribution}. Thus, the intensity function $\rho_{k}(u,t)$ of $D_{k}$ is given by
\begin{equation}
    \rho_{k}(u,t)=r_{k} h_{k}(u) (0.5+0.25 ( \sin(\frac{2\pi}{365}t+\frac{\pi}{2})+1)),
    \label{e:rho}
\end{equation}
where $r_{k}$ is a multiplication factor\footnote{Specifically, we set $r_{k}$ to 60, 20, 20 and 8 for the four house types respectively.} used to guarantee that $\rho_{k}(u,t)$ is at least four times $\lambda_{\theta_{k}}(u,t)$ as suggested in \cite{Baddeley2014LRPP} and $h_{k}(u)$ is the density of houses of type $k$ at location $u$. Secondly, we generate a realisation from the dummy point process $D_{k}$ using  the R-package \textit{spatstat} \cite{Baddeley2015spatstat}. Thirdly, based on the observation of the target point process $X_{k}$ and the realisation of the dummy point process $D_{k}$, we use the R-package \textit{stats} \cite{Venables2002stats}
to estimate the model parameters $\theta_{k}$ for house type $k$. In addition, to determine the optimal function orders (i.e.\ $o_{k,1}$, $o_{k,2}$, $o_{k,3}$, $o_{k,4}$ in (\ref{e:lambda_k})) simultaneously, following \cite{Lu2021ISI}, we apply a grid search to select the combination that yields the smallest Akaike information criterion \cite{Choiruddin2020AIC} given a set of ranges\footnote{$o_{k,1}$: 1--4, $o_{k,2}$: 1--5, $o_{k,3}$: 1--5, $o_{k,4}$: 1--5.} for them. Considering that wind chill already contains some information of wind speed, we also perform likelihood ratio tests over the best models with and without wind speed and the interaction term.

\begin{figure}[tbh]
    \centering
    \begin{subfigure}[t]{0.32\textwidth}
        \centering
        \includegraphics[scale=0.276]{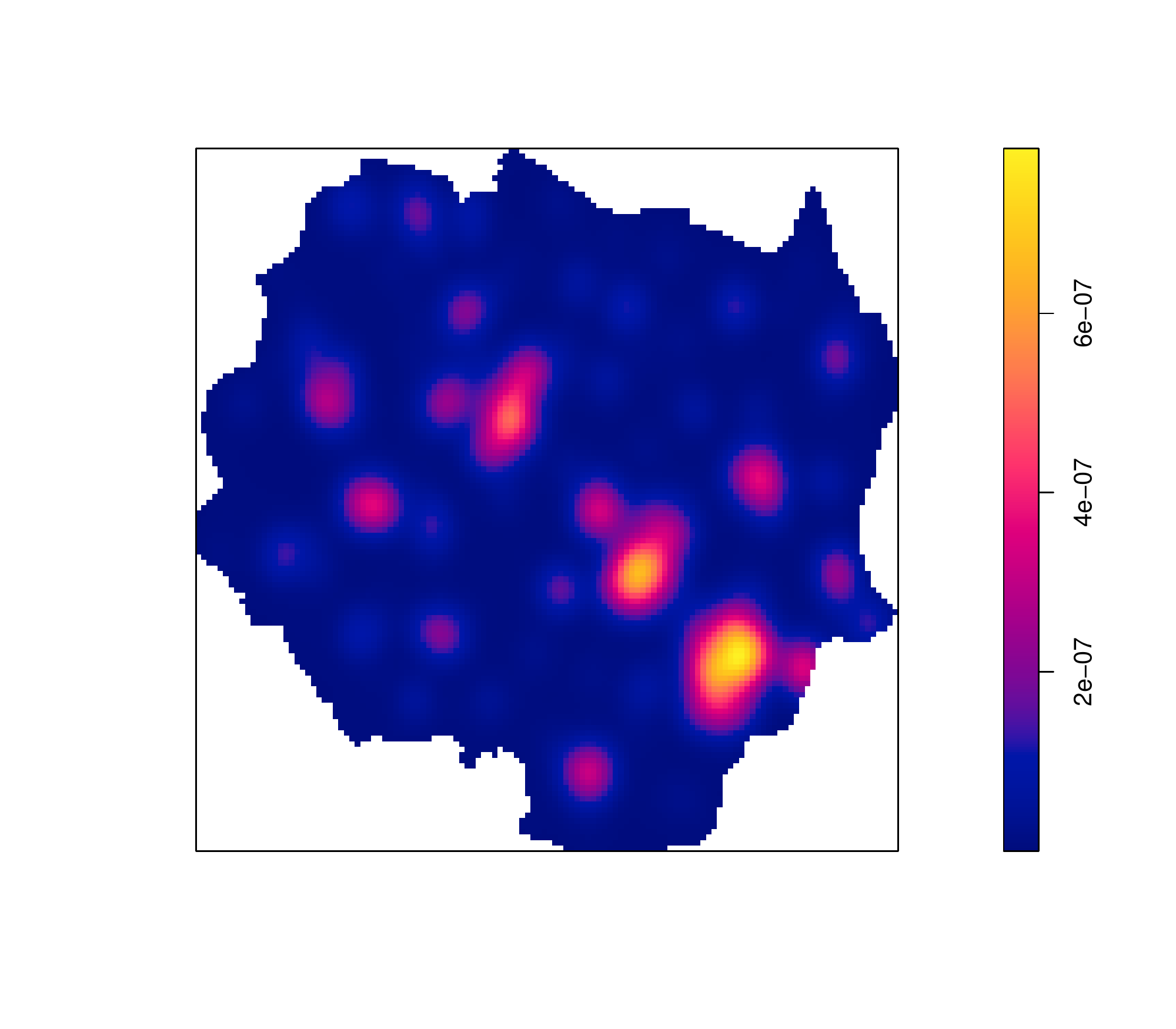}
        \caption{spatial prediction}
        \label{subfig:PP_pred_spatial_2020}
    \end{subfigure}%
    \hfill
    \begin{subfigure}[t]{0.66\textwidth}
        \centering
        \includegraphics[scale=0.34]{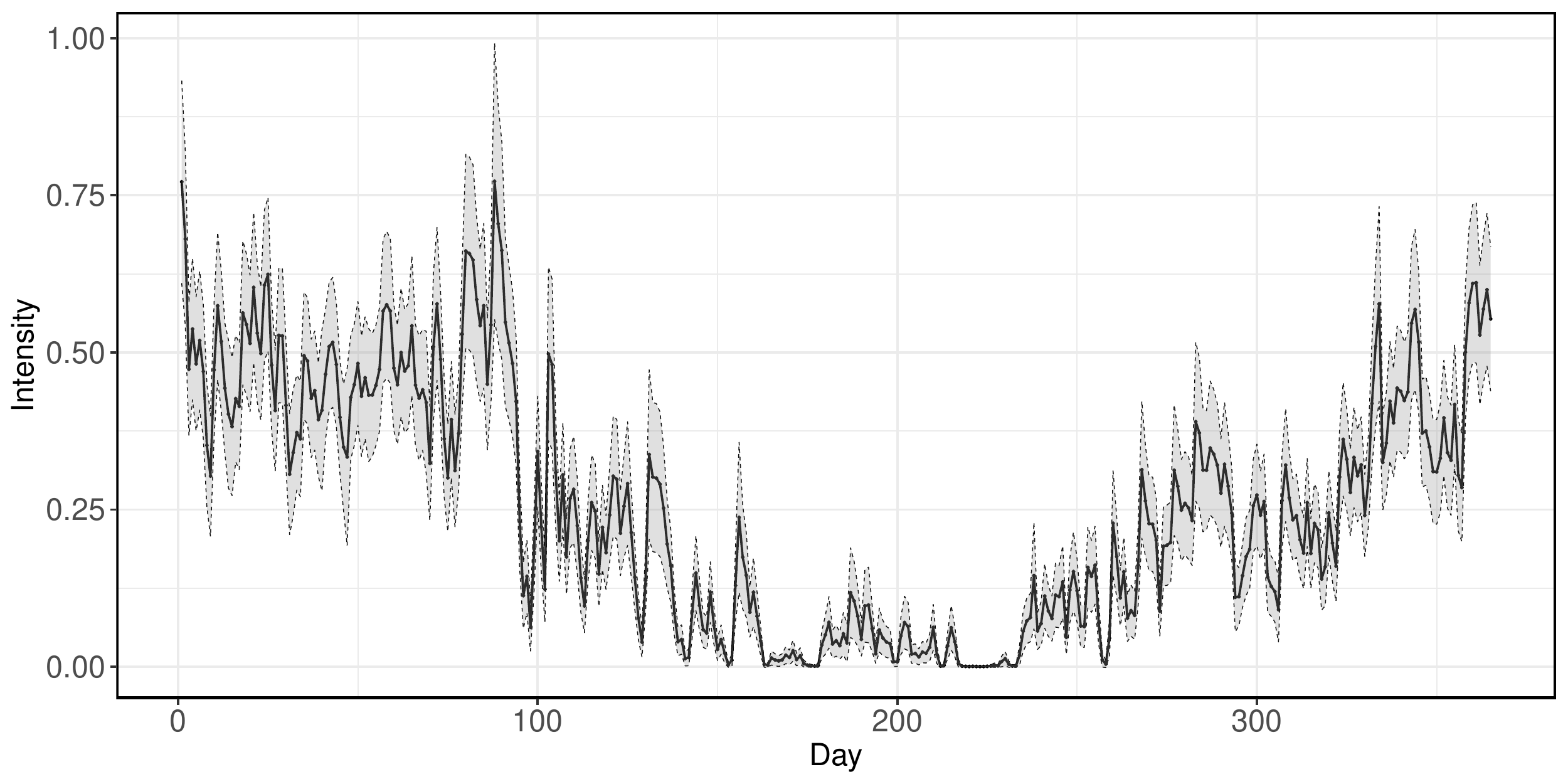}
        \caption{temporal prediction}
        \label{subfig:PP_pred_temporal_2020}
    \end{subfigure}%
    \hfill
    \begin{subfigure}[t]{0.32\textwidth}
        \centering
        \includegraphics[scale=0.276]{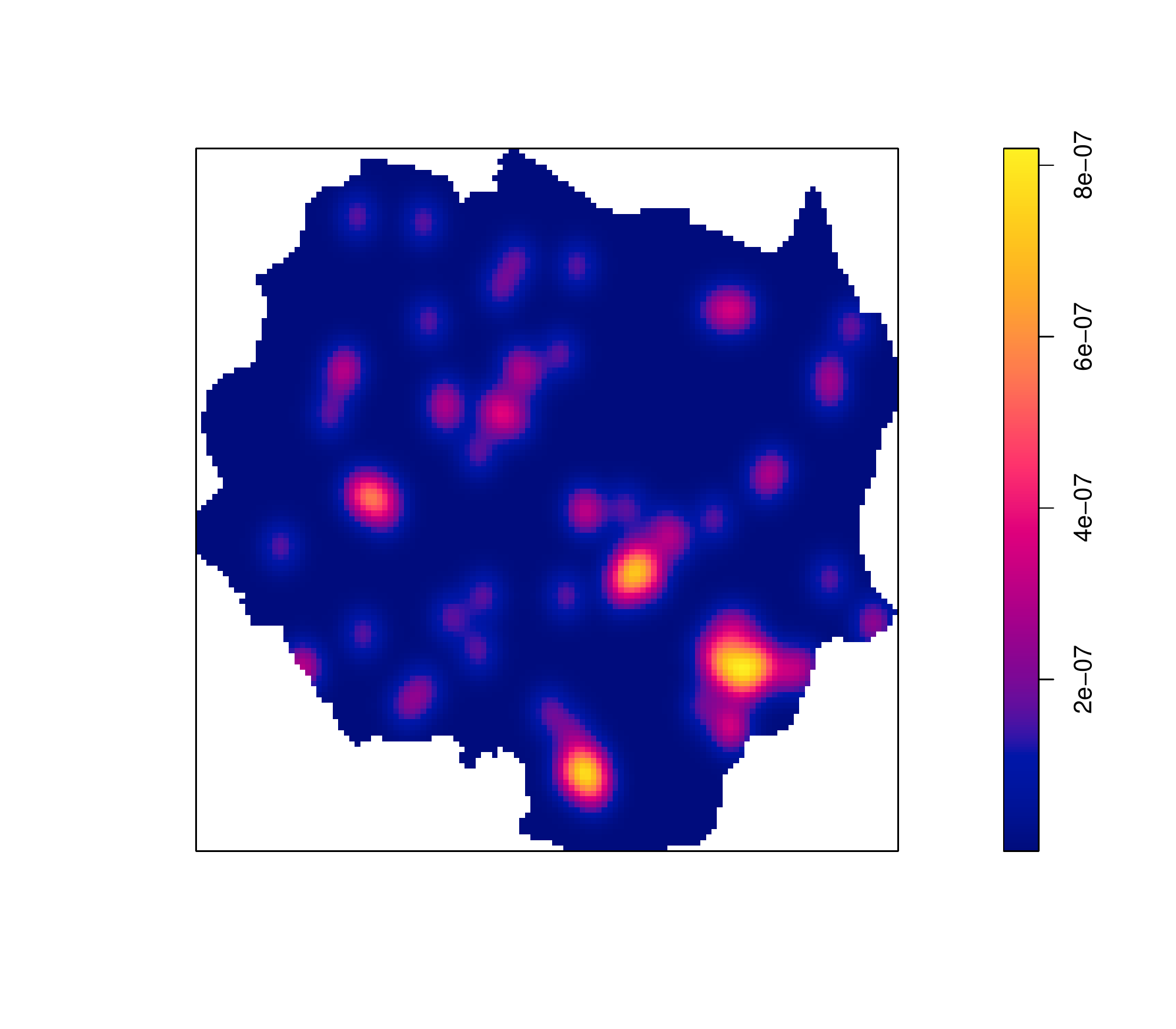}
        \caption{spatially smoothed\protect\footnotemark\ pattern}
        \label{subfig:PP_actual_spatial_2020}
    \end{subfigure}%
    \hfill
    \begin{subfigure}[t]{0.66\textwidth}
        \centering
        \includegraphics[scale=0.34]{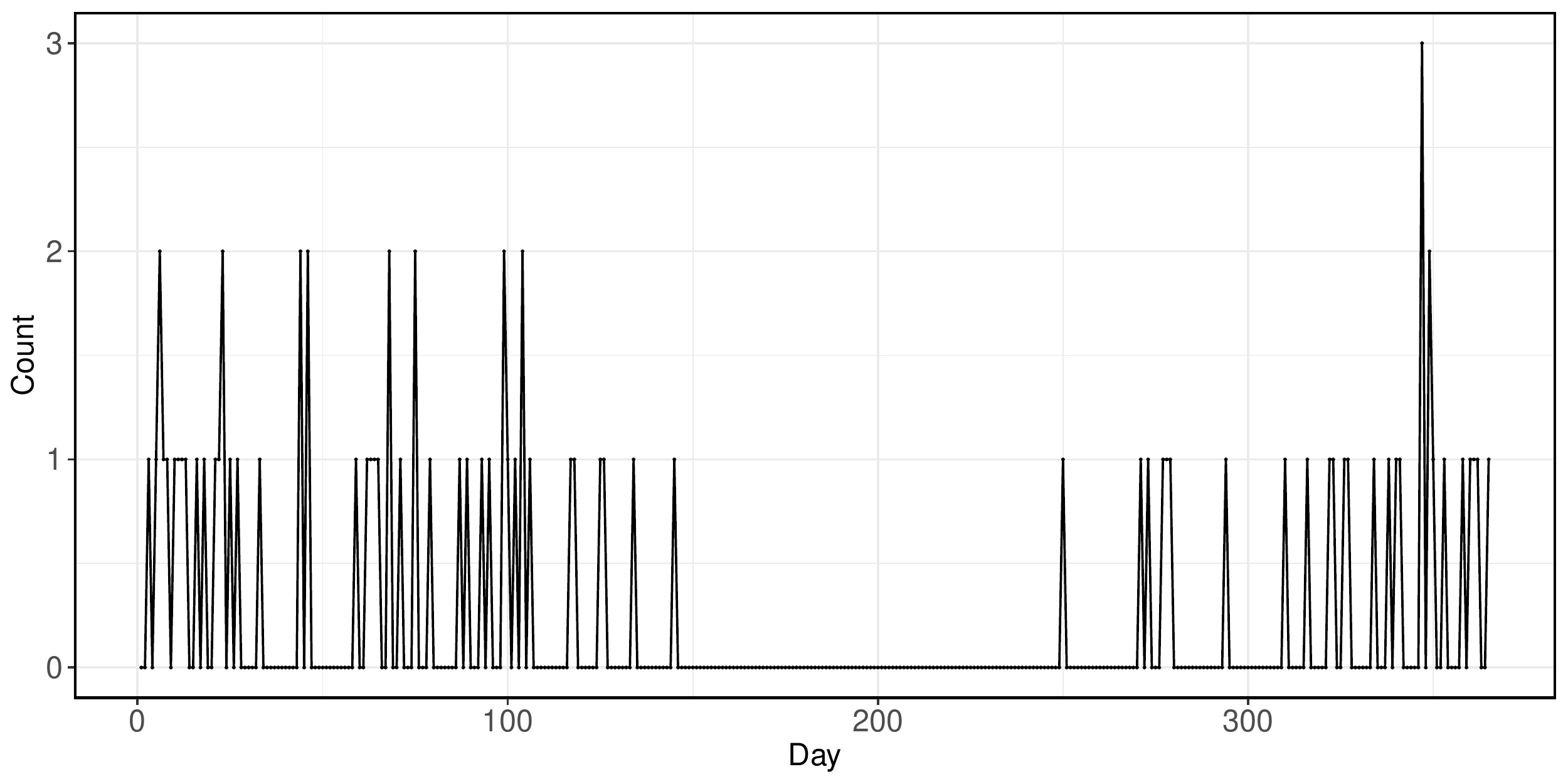}
        \caption{temporal counts}
        \label{subfig:PP_actual_temporal_2020}
    \end{subfigure}%
    \caption{Spatial and temporal predictions (plots: a, b) based on our point process model and actual realisations (plots: c, d) for the year 2020. Shadows in (b) bound the 95$\%$ confidence intervals. The unit in spatial plots is $\text{metre}^{-2}$; the unit in temporal plots is $\text{day}^{-1}$.}
    \label{fig:PP_prediction}
\end{figure}

Similar to \cite{Lu2021ISI}, we separate the fire data into two sets: the data on the period 2004--2019 and the data on the year 2020, for fitting and predicting respectively. Performing the model selection discussed above on the fitting data, we obtain corresponding optimal prediction models for all four house types and find that wind chill alone is mostly sufficient to obtain accurate risk predictions, except for house type~4, where the interaction term provides improved performance. Finally, the optimal temporal risk intensity functions obtained for the four house types read as follows:
\begin{equation}
\begin{split}
    \lambda_{1}(t)=&\exp(\theta_{1,1}+\theta_{1,2}\cos(\frac{2\pi}{365}t)+\theta_{1,3}\sin(\frac{2\pi}{365}t)+\theta_{1,4}\cos(\frac{4\pi}{365}t)+\theta_{1,5}\sin(\frac{4\pi}{365}t)\\
    &+\theta_{1,6}\cos(\frac{6\pi}{365}t)+\theta_{1,7}\sin(\frac{6\pi}{365}t)+\theta_{1,8}\cos(\frac{7\pi}{365}t)+\theta_{1,9}\sin(\frac{8\pi}{365}t)\\
    &+\theta_{1,10}V_{\tau,3}(t)+\theta_{1,11}V^{2}_{\tau,3}(t)),\\
    \lambda_{2}(t)=&\exp(\theta_{2,1}+\theta_{2,2}\cos(\frac{2\pi}{365}t)+\theta_{2,3}\sin(\frac{2\pi}{365}t)+\theta_{2,4}\cos(\frac{4\pi}{365}t)+\theta_{2,5}\sin(\frac{4\pi}{365}t)\\
    &+\theta_{2,6}\cos(\frac{6\pi}{365}t)+\theta_{2,7}\sin(\frac{6\pi}{365}t)+\theta_{2,8}V_{\tau,3}(t)+\theta_{2,9}V^{2}_{\tau,3}(t)+\theta_{2,10}V^{3}_{\tau,3}(t)\\
    &+\theta_{2,11}V^{4}_{\tau,3}(t)),\\
    \lambda_{3}(t)=&\exp(\theta_{3,1}+\theta_{3,2}\cos(\frac{2\pi}{365}t)+\theta_{3,3}\sin(\frac{2\pi}{365}t)+\theta_{3,4}\cos(\frac{4\pi}{365}t)+\theta_{3,5}\sin(\frac{4\pi}{365}t)\\
    &+\theta_{3,6}\cos(\frac{6\pi}{365}t)+\theta_{3,7}\sin(\frac{6\pi}{365}t)+\theta_{3,8}V_{\tau,3}(t)),\\
    \lambda_{4}(t)=&\exp(\theta_{4,1}+\theta_{4,2}\cos(\frac{2\pi}{365}t)+\theta_{4,3}\sin(\frac{2\pi}{365}t)+\theta_{4,4}\cos(\frac{4\pi}{365}t)+\theta_{4,5}\sin(\frac{4\pi}{365}t)\\
    &+\theta_{4,6}\cos(\frac{6\pi}{365}t)+\theta_{4,7}\sin(\frac{6\pi}{365}t)+\theta_{4,8}\cos(\frac{7\pi}{365}t)+\theta_{4,9}\sin(\frac{8\pi}{365}t)\\
    &+\theta_{4,10}V_{\tau,3}(t)+\theta_{4,11}V^{2}_{\tau,3}(t)+\theta_{4,12}V^{3}_{\tau,3}(t)+\theta_{4,13}V_{\tau,1}(t)V_{\tau,3}(t)),
\end{split}
\label{e:model}
\end{equation}
where $t$ refers to a specific day during 2004--2020, thus $t\in [0, 6205]$. We provide the estimates of all model parameters (i.e.\ $\theta$)  in Table~\ref{tab:parameter_estimates}. Additionally, we also plot the spatial and temporal predictions and the actual realisations of year 2020 in Figure \ref{fig:PP_prediction}. Overall, both the spatial and temporal predictions capture the correct trends. Specifically, in the spatial domain, our model learns the urbanity features of the Twente region which are highly relevant for the fire data. Since the spatial prediction depends on the distribution of houses of different types which can only be accessed as actual data, our model displays similar spatial patterns for different years. In the temporal domain, our model captures the periodic pattern on an annual basis and adds weather variable dependent information, which explains the noisier aspect of Figure~\ref{fig:PP_prediction}(b) compared to Figure~\ref{fig:PP_prediction}(a). A comparison with the areal unit model and the $\rho$-tuning experiments will be specifically discussed in Section \ref{sec:disc}. In summary, given appropriate building information, our prediction model can detect areas with higher risks of chimney fires in Twente and, additionally, estimate the risks for specific days based on the weather forecast.

\footnotetext{Smoothed by a Gaussian kernel with a standard deviation of 1,000 metres.}

\subsection{Confidence Interval}
\label{subsec:confidence_interval}
For risk prediction problems, it is important to quantify the uncertainty of the estimates. 
Baddeley et al. \cite{Baddeley2014LRPP} demonstrated asymptotic normality for their logistic regression estimators for stationary point processes in space when the domain increases. Recently, Choiruddin et al. \cite{Choiruddin2018increasingDomainCI} proved
asymptotic normality in an increasing-domain setting for spatial point process models that involve explanatory variables. In a companion paper, we study consistency and asymptotic normality for the logistic regression estimators for spatio-temporal Poisson point processes in an infill asymptotics setting. Specifically, under appropriate conditions, the maximiser $\hat \theta$ of (\ref{e:sLR}), under true value $\theta_{o}$, is approximately normally distributed with mean $\theta_{o}$ and covariance matrix $G$ given by the
inverse of the Godambe matrix \cite{Godambe2010GodambeMatrix}. A plug-in estimator for $G$ is
\begin{equation}
    \hat{G}= \left(\int_{T}\int_{W}\frac{\lambda_{\hat{\theta}}(u,t)\rho(u,t)}{\lambda_{\hat{\theta}}(u,t)+\rho(u,t)}C_{p}(u,t)C_{q}(u,t)dudt \right)_{p,q=1}^{m},
    \label{e:Godambe}
\end{equation}
assuming we have $m$ parameters in $\theta$. Approximate confidence intervals for the model parameters are readily
obtained. 

For chimney fire modeling, the approximate 95$\%$ confidence intervals of the model parameters are listed in 
Table~\ref{tab:parameter_estimates}. The Delta method \cite{VerHoef2012deltaMethod} can then be used to caclulate approximate confidence intervals for predicted fire intensities. We visualize the temporal risk predictions for the year 2020 in Figure~\ref{fig:PP_prediction}(b).

\begin{table}[tbh]
    \scriptsize
    \centering
    \begin{tabular}{|c|c|c|c|c|c|}
        \hline
        \textbf{Parameter} & \textbf{Estimate (CI)} &
        \textbf{Parameter} & \textbf{Estimate (CI)} &
        \textbf{Parameter} & \textbf{Estimate (CI)}\\
        \hline
        $\theta_{1,1}$ & -1.22e1($\pm$3.82e-1) & $\theta_{1,2}$ & -2.78e-1($\pm$ 4.81e-1) & $\theta_{1,3}$ & -1.44e-1($\pm$ 2.84e-1)\\
        \hline
        $\theta_{1,4}$ & -1.63e-1($\pm$ 2.64e-1) & $\theta_{1,5}$ & 7.99e-2($\pm$ 3.04e-1) & $\theta_{1,6}$ & -9.21e-3($\pm$ 2.49e-1)\\
        \hline
        $\theta_{1,7}$ & -2.33e-2($\pm$ 2.62e-1) & $\theta_{1,8}$ & 3.05e-1($\pm$ 2.19e-1) & $\theta_{1,9}$ & 1.59e-2($\pm$ 2.20e-1)\\ 
        \hline
        $\theta_{1,10}$ & -7.42e-2($\pm$ 3.56e-2) & $\theta_{1,11}$ & -6.12e-3($\pm$ 3.31e-3) & $\theta_{2,1}$ & -1.30e1($\pm$ 2.31e-1)\\ 
        \hline
        $\theta_{2,2}$ & 6.40e-2($\pm$ 2.74e-1) & $\theta_{2,3}$ & 1.42e-2($\pm$ 1.56e-1) & $\theta_{2,4}$ & -3.14e-2($\pm$ 1.64e-1)\\ 
        \hline
        $\theta_{2,5}$ & 1.69e-1($\pm$ 1.60e-1) & $\theta_{2,6}$ & 1.48e-1($\pm$ 1.24e-1) & $\theta_{2,7}$ & -4.54e-2($\pm$ 1.27e-1)\\ 
        \hline
        $\theta_{2,8}$ & -6.81e-2($\pm$ 2.64e-2) & $\theta_{2,9}$ & 2.33e-3($\pm$ 3.65e-3) & $\theta_{2,10}$ & -8.50e-6($\pm$ 2.47e-4)\\ 
        \hline
        $\theta_{2,11}$ & -2.76e-5($\pm$ 1.93e-5) & $\theta_{3,1}$ & -1.43e1($\pm$ 9.48e-1) & $\theta_{3,2}$ & 1.57e0($\pm$ 1.60e0)\\ 
        \hline
        $\theta_{3,3}$ & 2.25e-1($\pm$ 6.00e-1) & $\theta_{3,4}$ & -1.19e0($\pm$ 1.09e0) & $\theta_{3,5}$ & -3.19e-1($\pm$ 7.20e-1)\\ 
        \hline
        $\theta_{3,6}$ & 6.32e-1($\pm$ 6.06e-1) & $\theta_{3,7}$ & 1.65e-1($\pm$ 5.39e-1) & $\theta_{3,8}$ & -1.12e-1($\pm$ 5.25e-2)\\ 
        \hline
        $\theta_{4,1}$ & -1.39e1($\pm$ 2.41e-1) & $\theta_{4,2}$ & 1.49e-1($\pm$ 3.05e-1) & $\theta_{4,3}$ & 6.11e-2($\pm$ 1.76e-1)\\ 
        \hline
        $\theta_{4,4}$ & -2.06e-1($\pm$ 1.83e-1) & $\theta_{4,5}$ & 9.00e-2($\pm$ 1.95e-1) & $\theta_{4,6}$ & 2.77e-2($\pm$ 1.61e-1)\\ 
        \hline
        $\theta_{4,7}$ & 7.11e-2($\pm$ 1.72e-1) & $\theta_{4,8}$ & 1.82e-1($\pm$ 1.36e-1) & $\theta_{4,9}$ & -4.84e-2($\pm$ 1.38e-1)\\ 
        \hline
        $\theta_{4,10}$ & -9.38e-2($\pm$ 4.06e-2) & $\theta_{4,11}$ & -7.92e-4($\pm$ 1.99e-3) & $\theta_{4,12}$ & -2.35e-4($\pm$ 1.62e-4)\\ 
        \hline
        $\theta_{4,13}$ & 2.99e-3($\pm$ 2.10e-3) & & & &\\ 
        \hline
    \end{tabular}
    \caption{Parameter estimates for a Poisson point process model defined by the intensity functions of (\ref{e:model}) and their 95$\%$ confidence intervals (CI). The `e' denotes a base of 10.} 
    \label{tab:parameter_estimates}
\end{table}

\section{Model Validation}
\label{sec:mdvd}
In this section, we validate the independence assumption underlying Poisson point processes by analyzing the second-order properties. We also validate the first-order structure of our model by visualizing the residuals.

\subsection{Second-order Property Analysis}
To assess the validity of the Poisson assumption, we perform a second-order analysis on the chimney fire data using two summary statistics: the pair correlation function and the $K$-function (see, e.g. \cite{Lieshout2019SpatialStatistics}). The pair correlation function can be used to detect inter-point interaction for specific distances, whereas the $K$-function is a cumulative statistic that can be used to test the existence of point interactions in data.

\subsubsection{The Pair Correlation Function}
Consider a point process $X$ defined on a space-time domain $W\times T \subset \mathbb{R}^{2}\times\mathbb{R}$ that is locally finite. The second-order product density $\lambda^{(2)}(x,y)$ of the pair $(x,y)$, where $x,y\in W\times T$, is the infinitesimal probability that $X$ places points at $x$ and $y$. The pair correlation function for the pair $(x,y)$ is then defined as
\begin{equation}
    g(x,y)=\frac{\lambda^{2}(x,y)}{\lambda(x)\lambda(y)},
    \label{e:pcf}
\end{equation}
where $\lambda(x)$ and $\lambda(y)$ denote the intensities at $x$ and $y$, and we use the convention that $a/0=0$ for all $a\geq0$. For purposes of visualisation, we consider the pair correlation functions for the projections in space and time domain separately. Under appropriate stationarity and isotropy assumptions, in $d$ dimensions (spatially, $d=2$; temporally, $d=1$), (\ref{e:pcf}) is a function of $r=|| x- y||$ only\footnote{Spatially, $||x-y||$ is the distance between two points; temporally, $||x-y||$ is the time interval between two points.} and can be estimated by
\begin{equation}
    \hat{g}(r)=\frac{1}{s_{d}r^{d-1}}\sum_{x,y \in X}^{\neq}\frac{k_{b}(r-||x-y||)}{\lambda(x)\lambda(y)|W\cap W_{x-y}|},
    \label{e:hat_pcf}
\end{equation}
where $s_{d}$ denotes the surface area of the unit sphere in $\mathbb{R}^{d}$,
$\sum^{\neq}$ denotes the summation over all pairs of distinct points, $k_{b}(\cdot)$ is an one-dimensional smoothing kernel with bandwidth $b$, $1/|W \cap W_{x-y}|$ is the spatial edge correction factor \cite{Ohser1981edgeCorrection} (temporally, $1/|T\cap T_{x-y}|$), and $|A|$ is the volume of $A \subset \mathbb{R}^{d}$. If there is no interaction in the point pattern, the pair correlation function is $1$ at any distance. A value of $\hat{g}(r)$ larger than $1$ suggests clustering in the sense that points are more likely to occur from one another at distance $r$, whereas a value smaller than $1$ suggests inhibition that points tend to mutually avoid each other at this distance. Moreover, if the intensity function $\lambda(x)$ is unknown, a plug-in estimator can be used.

\begin{figure}[h]
    \centering
    \begin{subfigure}[]{0.48\textwidth}
        \centering
        \includegraphics[scale=0.36]{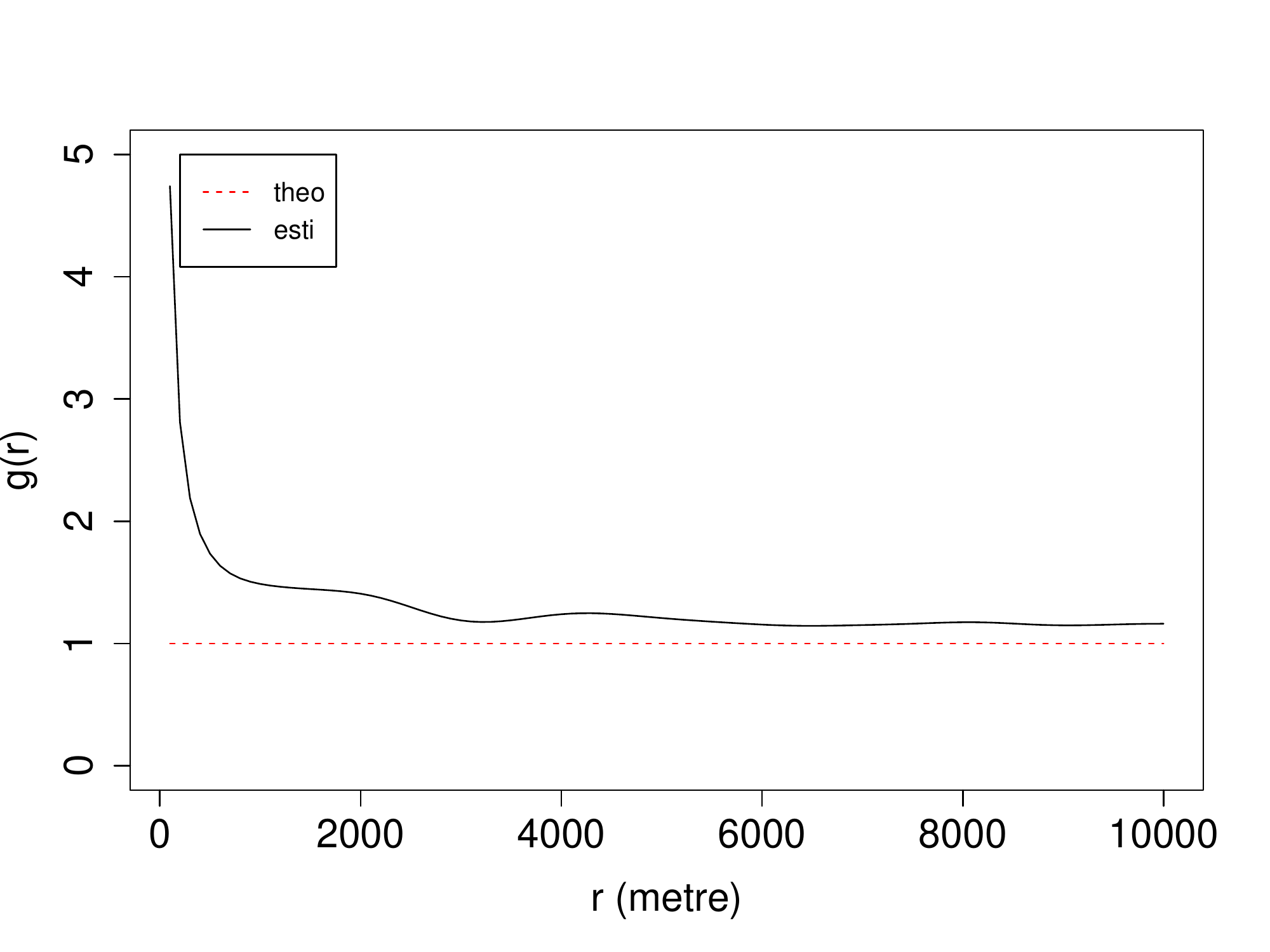}
        \caption{spatial result}
        \label{subfig:pcf_spatial}
    \end{subfigure}%
    \begin{subfigure}[]{0.48\textwidth}
        \centering
        \includegraphics[scale=0.36]{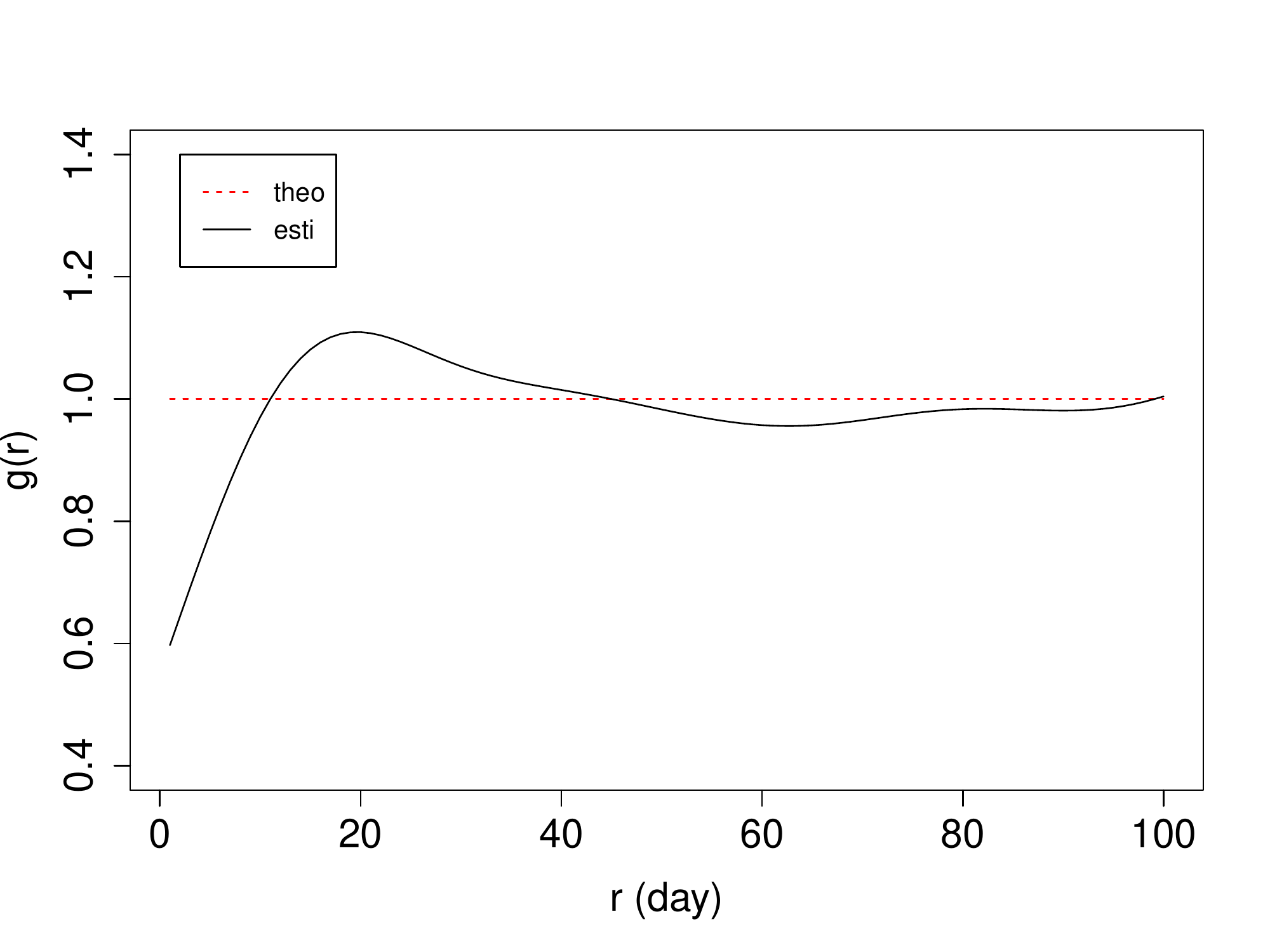}
        \caption{temporal result}
        \label{subfig:pcf_temporal}
    \end{subfigure}%
    \caption{Estimated (solid black line) and theoretical (dashed red line) pair correlation functions for the spatial and temporal projections (cf.\ Figure~\ref{fig:maps}(b) and \ref{fig:temporal_distribution}) of chimney fire incidents during 2004--2020.}
    \label{fig:pcf_results}
\end{figure}

In practice, we employ (\ref{e:hat_pcf}) with a Gaussian smoothing kernel as $k_b$ and plug in the fitted intensity functions based on data  of all $17$ years. We set the possible interaction intervals for the spatial and temporal domain to $10,000$ metres and $100$ days, respectively, and the smoothing bandwidths to $500$ metres and $10$ days. The estimated pair correlation functions are plotted in Figure \ref{fig:pcf_results}. Overall, both the spatial and temporal results show a curve that converges to $1$. At smaller range, there is some attraction between the points spatially. Temporally, possibly because of the precision of the observations (daily basis), there is some repulsion and a weak peak at an interval of about twenty days. The latter might be due to an underestimate of the intensity function on a specific day.

\subsubsection{The $K$-function}
Since the pair correlation function shows some interactions at small distances, we perform a joint space-time $K$-function test on the chimney fire data. The $K$-function, also known as Ripley's reduced second moment function, of a stationary point process $X$ is defined as the expected number of other points within a given distance $r$ of an arbitrary point divided by the intensity. In an inhomogeneous and space-time version, it takes the form of 
\begin{equation}
    K_{inhom}(r,v)=\frac{1}{|B|}E\left\{\sum_{x\in X\cap B}\sum_{y\in X;y\neq x}\frac{1(||x(u)-y(u)||\leq r,||x(t)-y(t)||\leq v)}{\lambda(x)\lambda(y)}\right\},
    \label{e:Kinhom}
\end{equation}
where $B$ denotes any Borel set, $x$ and $y$ are distinctive points with $x(u),y(u)$ and $x(t),y(t)$ representing their spatial and temporal positions, and $1(\cdot)$ is the indicator function. Under appropriate weak stationarity assumptions \cite{Gabriel2009K}, a point-wise unbiased estimator of the $K$-function is
\begin{equation}
    \hat{K}_{inhom}(r,v)=\sum_{x\in X}\sum_{y\in X;y\neq x}\frac{1(||x(u)-y(u)||\leq r,||x(t)-y(t)||\leq v)}{\lambda(x)\lambda(y)|(W\times T)\cap (W\times T)_{x-y}|},
    \label{e:hatK}
\end{equation}
with notation as in the previous section. If no interactions exist, $K_{inhom}(r,v)=2\pi r^{2}v$. Moreover, when the intensity function $\lambda(x)$ is unknown, a plug-in estimator can be used.

\begin{figure}[h]
    \centering
    \begin{subfigure}[]{0.48\textwidth}
        \centering
        \includegraphics[scale=0.36]{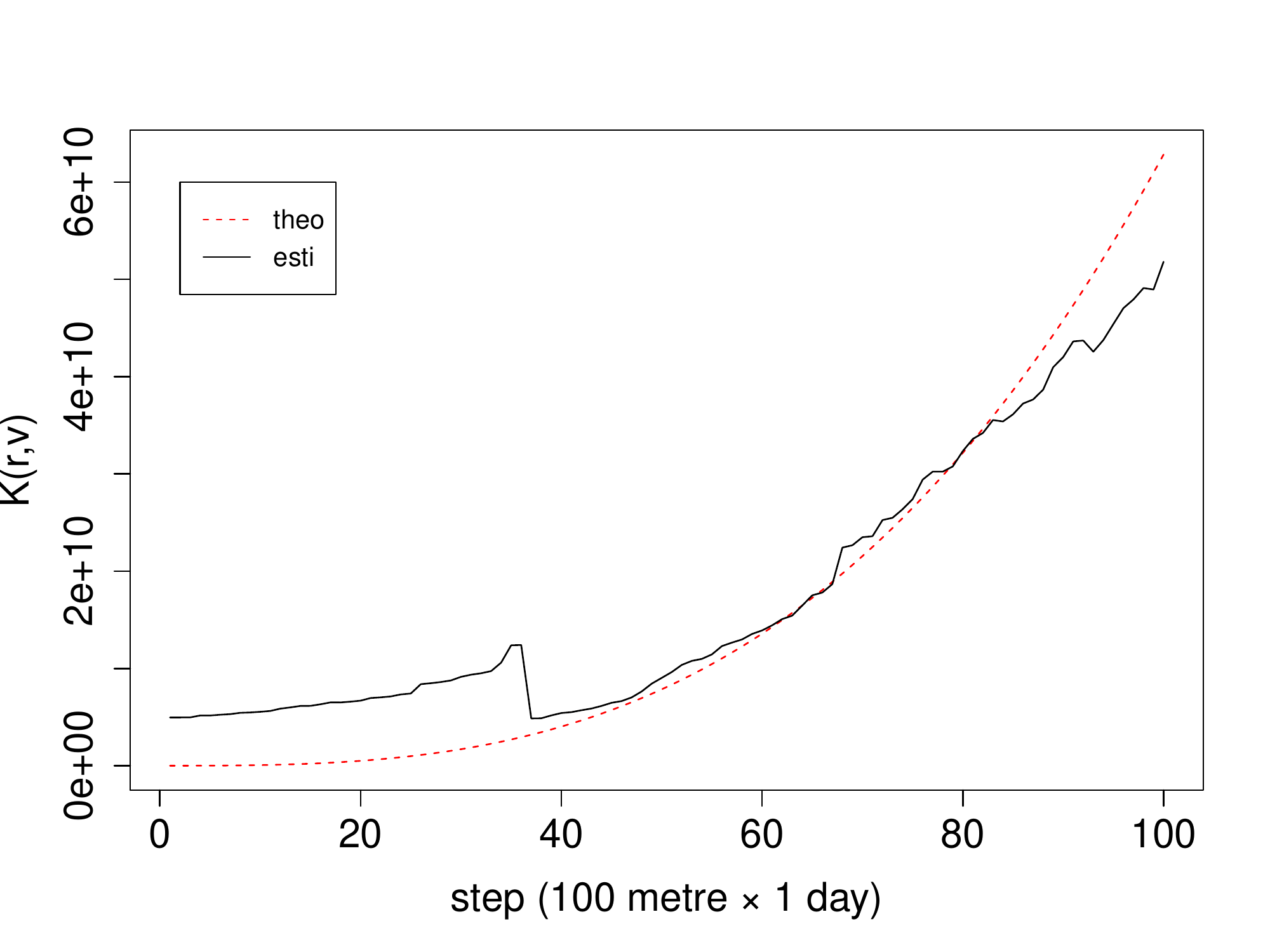}
        \caption{inhomogeneous K function}
        \label{subfig:K_function}
    \end{subfigure}%
    \begin{subfigure}[]{0.48\textwidth}
        \centering
        \includegraphics[scale=0.36]{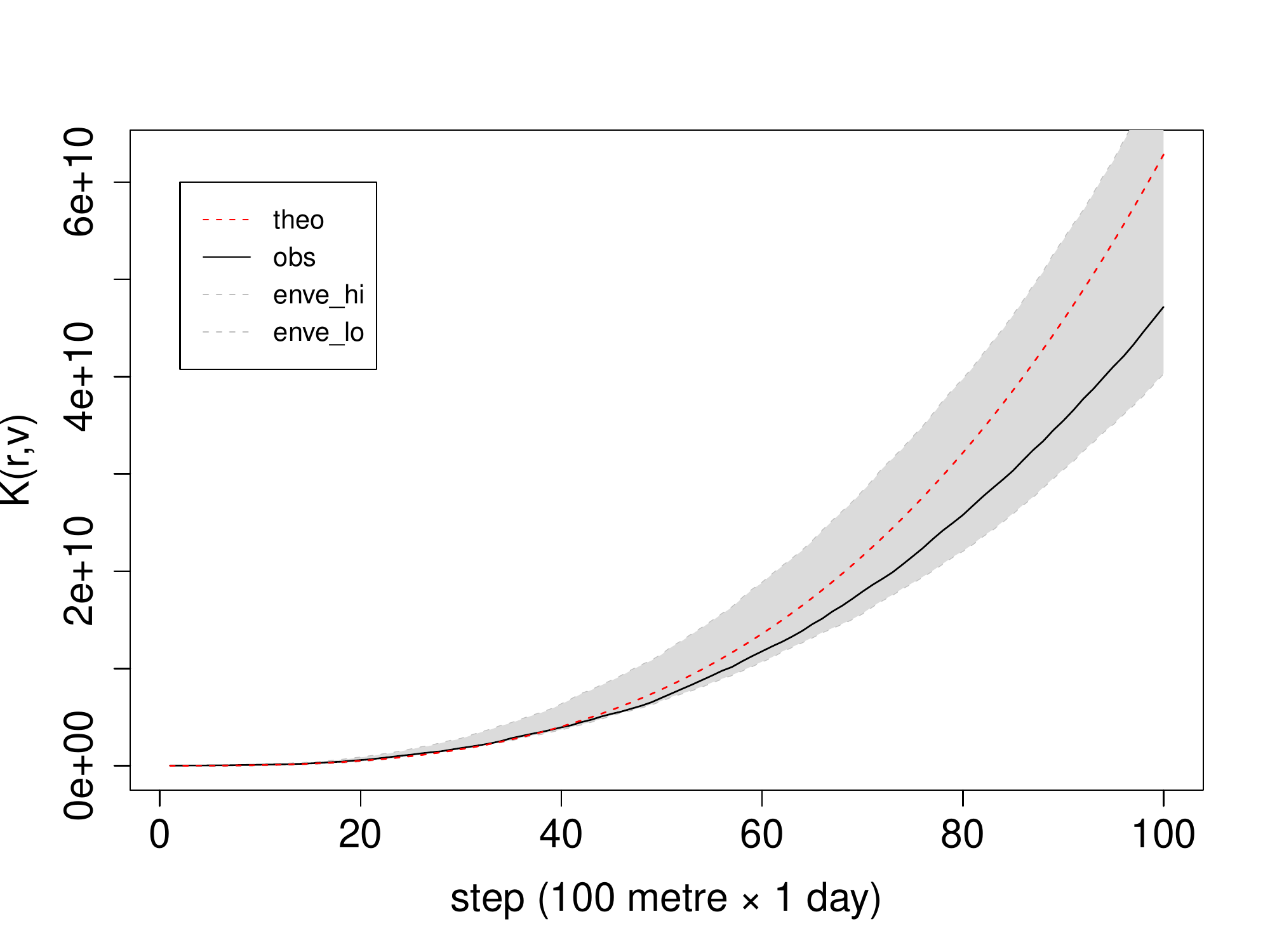}
        \caption{Monte Carlo envelope}
        \label{subfig:K_function_test}
    \end{subfigure}%
    \caption{Estimated (solid black line) and theoretical (dashed red line) inhomogeneous $K$-functions for the space-time pattern of chimney fire incidents during 2004--2020 (a) and the Monte Carlo envelope over 99 simulations of the fitted model as well as the empirical $K$-function (b).}
    \label{fig:K_results}
\end{figure}

We compute the inhomogeneous $K$-function over data from all $17$ years. To obtain an elegant view of it, we set the testing sequences of spatial and temporal distances to $10,000$ metres and $100$ days and divide both of them into $100$ step size pairs of $100$ metres and $1$ day. We calculate the $K$-function using the fitted intensity functions to estimate $\lambda(\cdot)$ in (\ref{e:hatK}) and plot the result in Figure~\ref{fig:K_results}(a). Same with the pair correlation function, we find some evidence of clustering. To test whether the interaction is statistically significant, we implement a Monte Carlo test. We generate $99$ realisations of a Poisson point process with the fitted intensity functions, and compute their $K$-functions  and that of the actual chimney fire data using kernel estimators to estimate $\lambda(\cdot)$ in (\ref{e:hatK}). We plot the local envelopes in Figure~\ref{fig:K_results}(b). As the empirical $K$-function of the actual chimney fire data lies completely within the envelope, there is no reason to look beyond a Poisson model.

\subsection{Residual Analysis}
\begin{figure}[h]
    \centering
    \begin{subfigure}[t]{0.32\textwidth}
        \centering
        \includegraphics[scale=0.276]{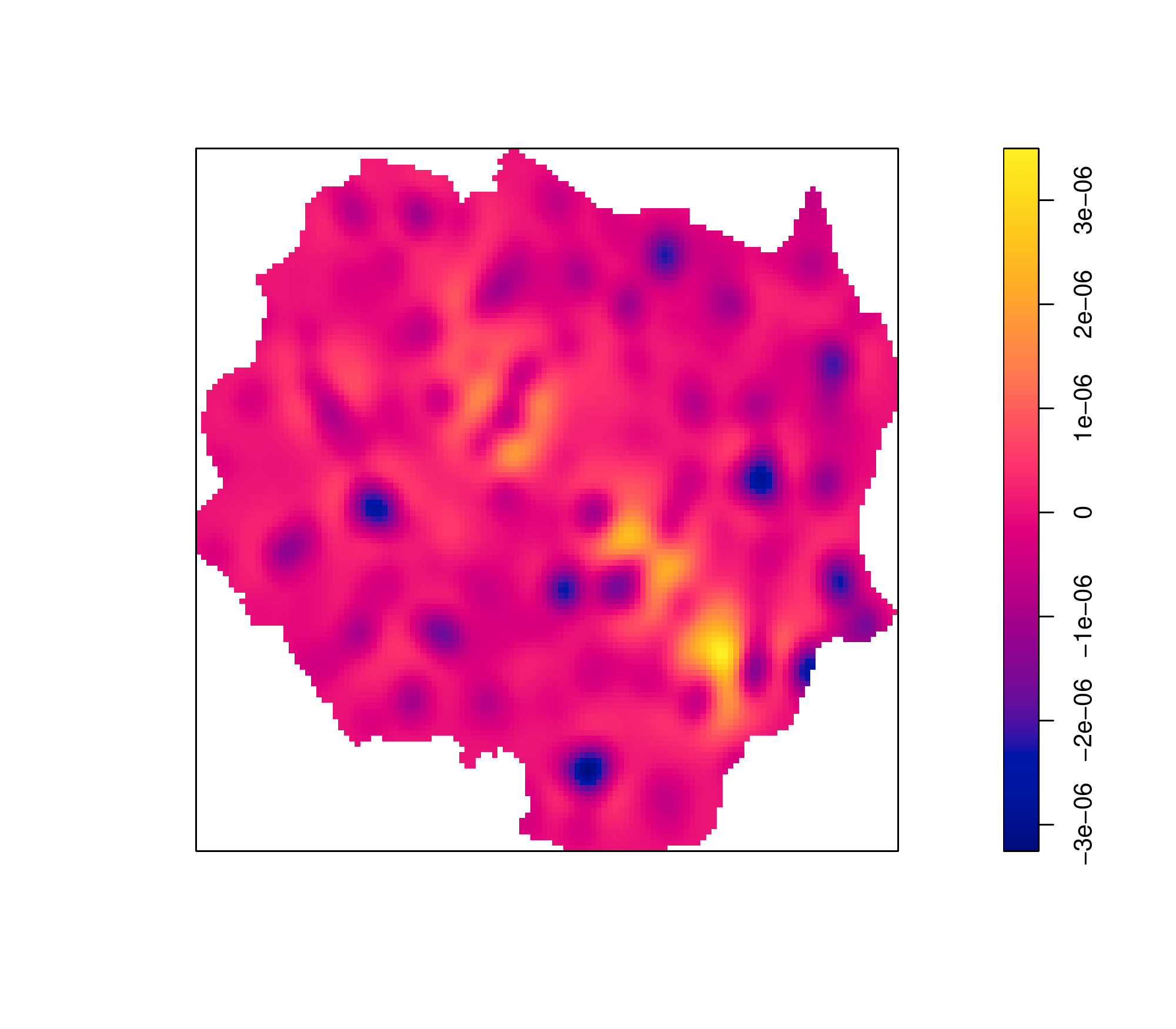}
        \caption{spatial residuals}
        \label{subfig:spatial_residuals}
    \end{subfigure}%
    \begin{subfigure}[t]{0.66\textwidth}
        \centering
        \includegraphics[scale=0.34]{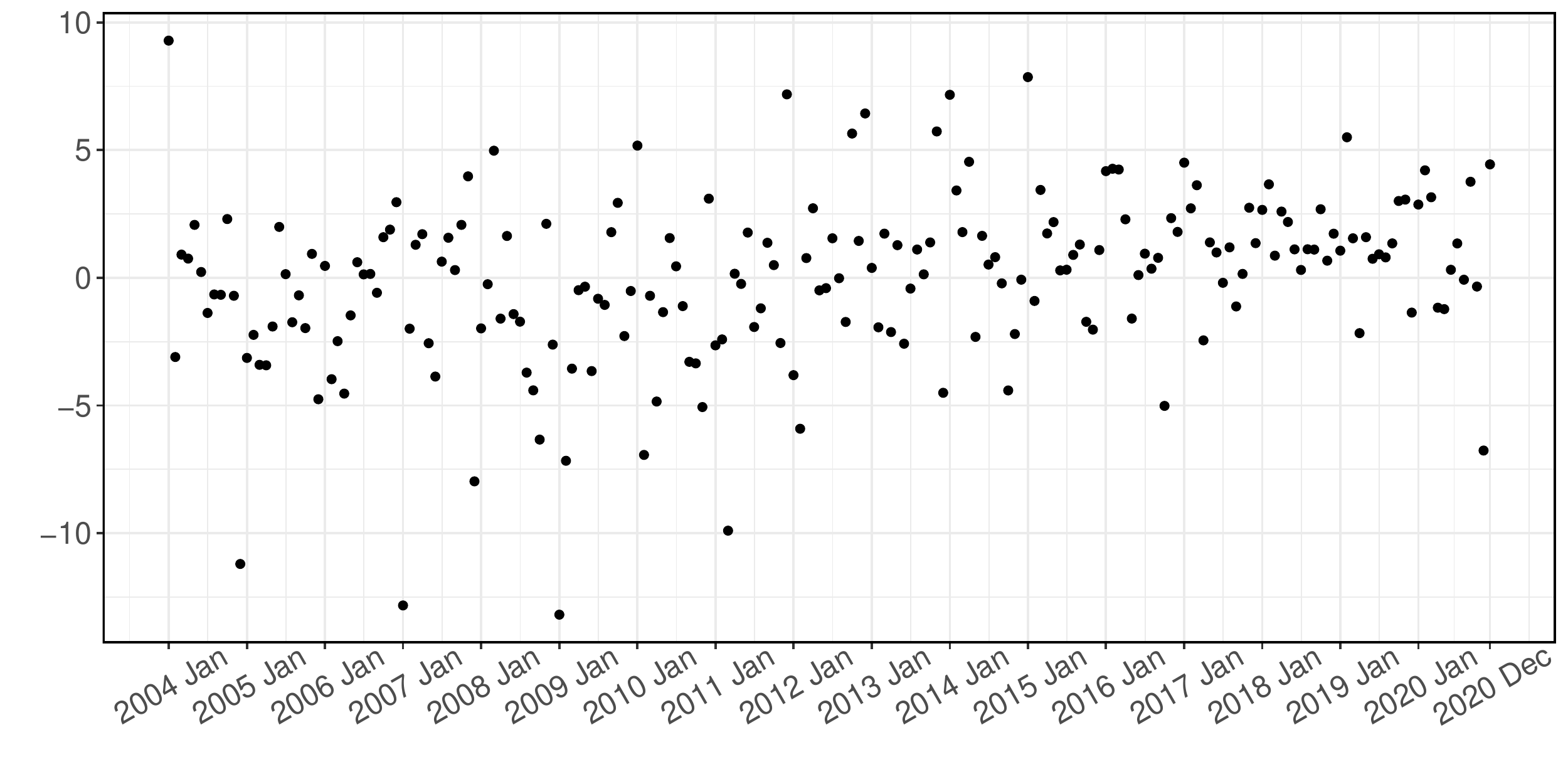}
        \caption{temporal residuals (monthly counts)}
        \label{subfig:temporal_residuals}
    \end{subfigure}%
    \caption{Spatial (a) and temporal (b) residuals based on the chimney fire data during 2004--2020. The unit in the spatial plot is $\text{metre}^{-2}$; the unit in the temporal plot is $\text{month}^{-1}$.}
    \label{fig:residual_study}
\end{figure}

To further verify the validity of our point process model, we perform a residual analysis to check whether typical patterns still exist after the modeling of Poisson point process with specific form defined in (\ref{e:intensity}), (\ref{e:lambda_k}) and (\ref{e:model}). We fit the model using the data of all $17$ years and compute spatial and temporal residuals separately. As suggested in \cite{Baddeley2005spatialResiduals}, the spatial residuals are defined as the difference between the smoothed fitted fire risks and the smoothed actual realisations using a Gaussian kernel with a standard deviation of $1,000$ metres. Temporally, we define the residuals as the difference between predicted and actual monthly counts. The results of the residual analysis are plotted in Figure \ref{fig:residual_study}. In the spatial domain, no specific pattern is shown, except that our point process model overestimates for big city centres (e.g.\ Enschede and Hengelo) whereas underestimates a bit for small city centres. It is reasonable, as our model merely depends the spatial distribution of chimney fires on the density of corresponding house types and we observe a saturation pattern of chimney fire occurrences when the density of a house type increases (cf.\ Figure \ref{fig:variable_relations}). A possible solution for that is to apply a piece-wise function instead of the linear function to model the relation between chimney fire occurrences and the density of a house type. In the temporal domain, no specific pattern is evident. 

\section{Discussion}
\label{sec:disc}
In this section, we discuss the data-driven modeling process designed for chimney fire prediction, which can be generalized to similar fire types. We also compare the areal unit model in \cite{Lu2021ISI} with our point process model. Additionally, we present the influence of tuning $\rho$ to important regions and times in the logistic regression estimation.

\subsection{Modelling Process}
\label{subsec:modelling_process}
In this paper, we design a two-step data-driven modeling process for chimney fire prediction. 

In the first step, we combine machine learning and statistics, where the former is utilized to select the most important environmental variables and the latter appropriately approximates and explains the relations between explanatory variables and fire incidence. We employ parametric functions to model such relations and fit them using the logistic regression approach, so that the influence of a variable on chimney fires can be analyzed by its coefficient in the function model. This combination offers significant advantages. From the machine learning side: i) compared to naive statistics used for the selection of explanatory variables, random forests can detect both linear and non-linear correlation between a candidate variable and chimney fire occurrences; ii) random forests also consider the dependence among candidate variables non-parametrically; iii) by applying the conditional permutation importance technique instead of the traditional one, the bias towards correlated variables is suppressed as well. From the statistics side: i) classic statistical methods for regression modeling with specific mathematical expressions are more interpretable and based on natural assumptions; ii) statistical methods also allow for significance testing and uncertainty quantification.

In the second step, we use summary statistics, the pair correlation function and the $K$-function, to detect point interactions. We also perform a residual analysis to validate the model structure. Although, in our study, a Poisson point process model is already proved sufficient for chimney fire prediction, a hierarchical modeling where random effects are incorporated to deal with latent effects, might be more appropriate for different fire types.

\subsection{Model Comparison}
\label{subsec:model_comparison}
\begin{figure}[h]
    \centering
    \begin{subfigure}[t]{0.32\textwidth}
        \centering
        \includegraphics[scale=0.286]{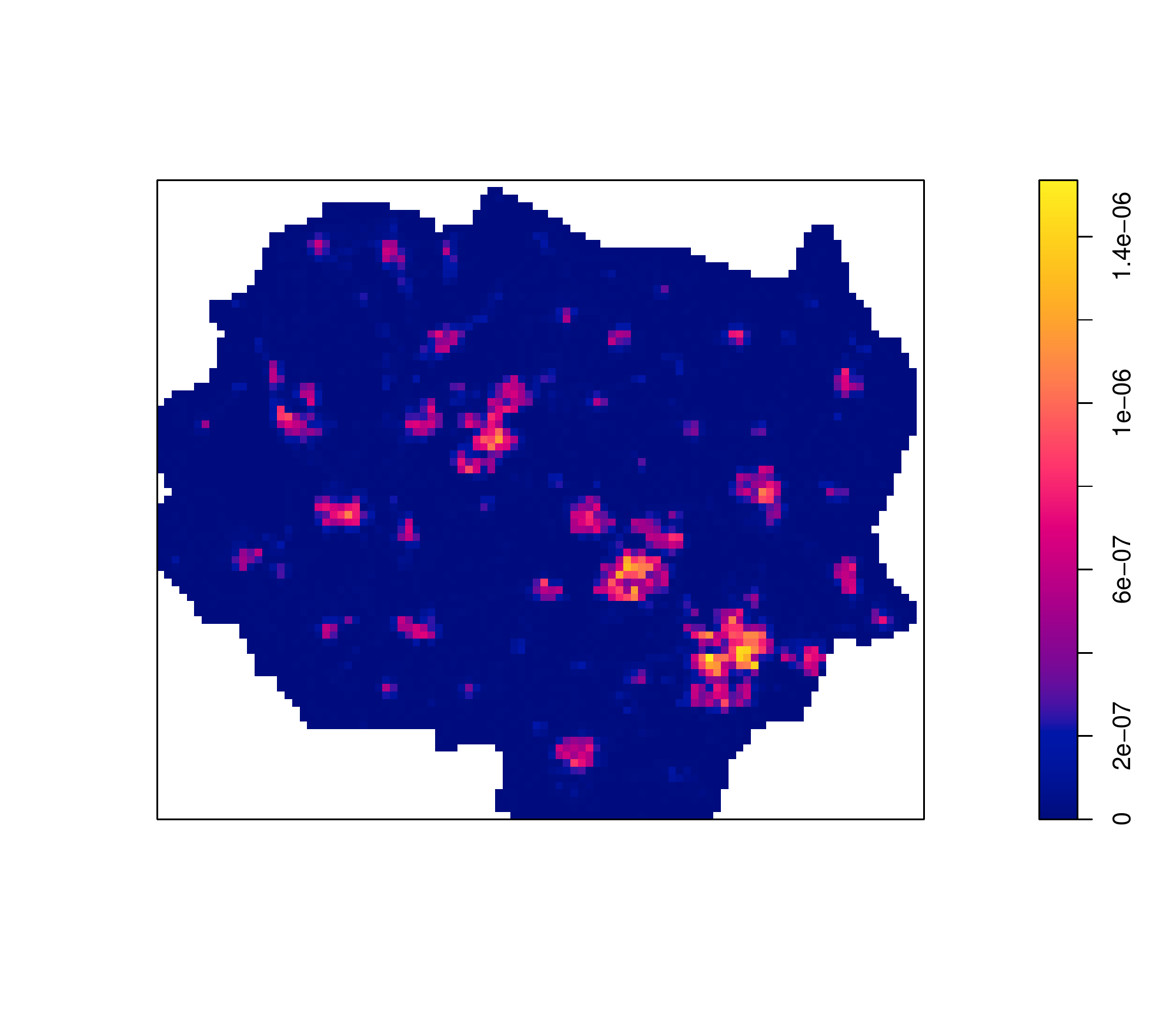}
        \caption{spatial predictions}
        \label{subfig:AU_pred_spatial_2020}
    \end{subfigure}%
    \begin{subfigure}[t]{0.66\textwidth}
        \centering
        \includegraphics[scale=0.34]{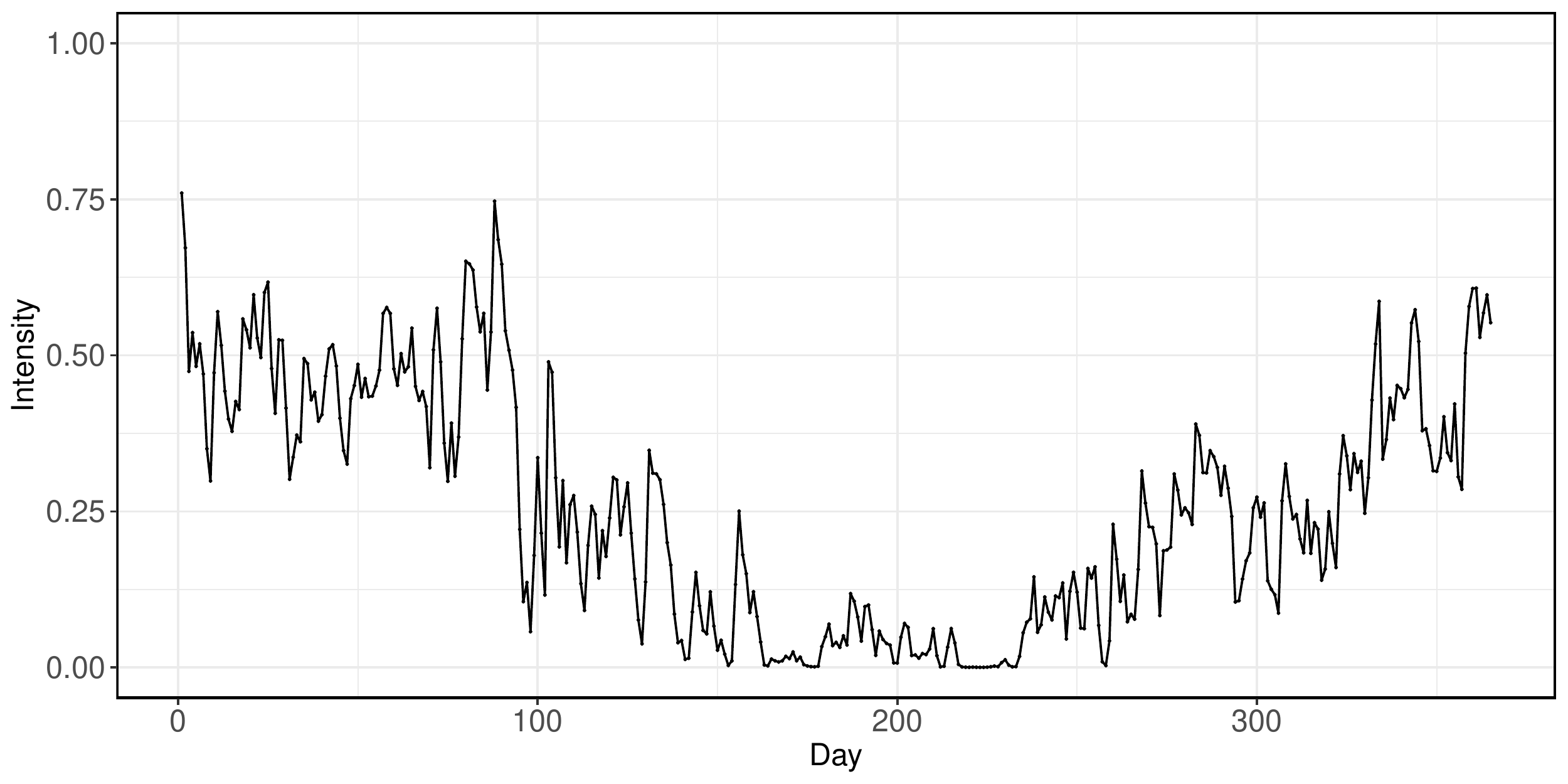}
        \caption{temporal predictions}
        \label{subfig:AU_pred_temporal_2020}
    \end{subfigure}%
    \caption{Spatial (a) and temporal (b) predictions based on the areal unit model in \cite{Lu2021ISI} for the year 2020. The unit in the spatial plot is $\text{metre}^{-2}$; the unit in the temporal plot is $\text{day}^{-1}$.}
    \label{fig:AU_prediction}
\end{figure}

To compare the performance of the areal unit model in \cite{Lu2021ISI} and our point process model, we plot spatial and temporal predictions for the year 2020 for the former model in Figure~\ref{fig:AU_prediction}. Since the spatial predictions from this areal unit model are computed for 500$m\times$ 500$m$ area boxes, to comply with the scaling level of point process model, we average the predictions by the volume of these area boxes. Generally, comparing Figure~\ref{fig:AU_prediction} to Figure~\ref{fig:PP_prediction}, we can find that both the areal unit model and the point process model capture similar spatial and temporal patterns from the fire data. However, the spatial predictions from the point process model behave smoother than those from the areal unit model, as we apply the kernel smoothing in the former to obtain the density of a house type at a location, which helps establish spatially continuous predictions. The temporal predictions from the two models are almost identical for every day, because our prediction model (cf.\ (\ref{e:intensity}), (\ref{e:lambda_k}) and (\ref{e:model})) mainly learns temporal information and merely distributes temporal predictions to spatial locations according to the densities of certain house types.

More specifically, in terms of the total chimney fire risks predicted for the year 2020, the areal unit model predicts $92.86$ and the point process model predicts $92.95$. The  observed count of $81$ lies in the 95$\%$ confidence intervals of both models. However, the areal unit model overestimates many more fire risks for certain area boxes of big city centres (e.g.\ Enschede). Such phenomena not only result from the different smoothing strategies the two models employ, but also derive from the fact that we fit the fire data over the whole Twente region, which means that detailed information between the number of houses of a type and chimney fire occurrences can be lost. An additional explanation is offered by the saturation effect suggested by Figure~\ref{fig:variable_relations}. From a computational perspective, the logistic regression approach employed for our point process model enables to capture spatial information efficiently and obtains more reasonable spatial predictions.

\subsection{$\rho$ Tuning in the Logistic Regression Estimation}
\label{subsec:rho_tuning}
The role of the dummy point process $D$ is to estimate the integral in (\ref{e:CM}). In \cite{Baddeley2014LRPP}, a rule of thumb was proposed that $\rho(u,t)$ should be at least four times $\lambda(u,t)$; additionally, a data-driven selection of $\rho$ was suggested to obtain relatively precise estimate under a limited number of dummy points.

\begin{figure}[h]
    \centering
    \begin{subfigure}[t]{0.32\textwidth}
        \centering
        \includegraphics[scale=0.276]{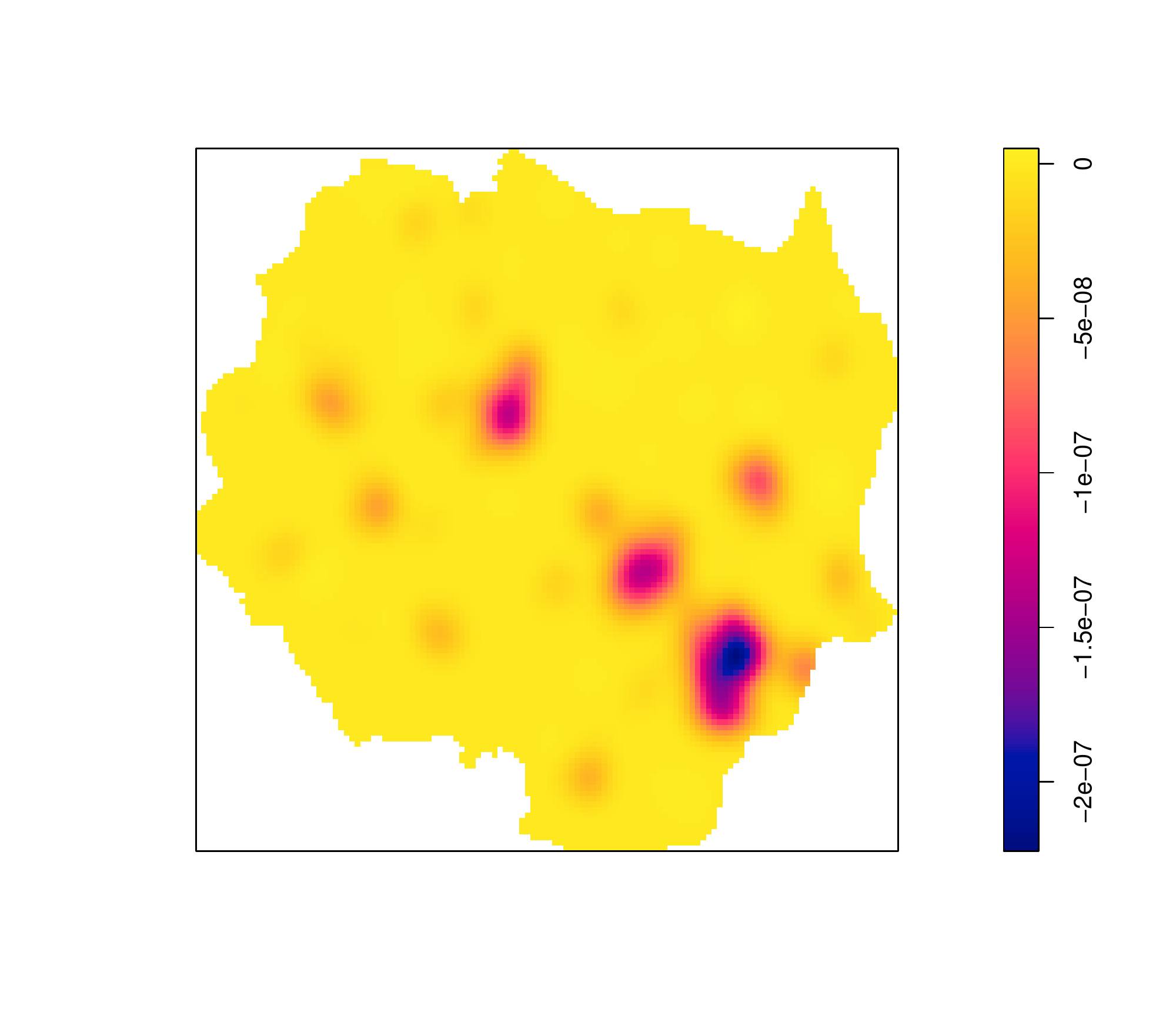}
        \caption{spatial difference}
        \label{subfig:rho_tuning_spatial}
    \end{subfigure}%
    \begin{subfigure}[t]{0.66\textwidth}
        \centering
        \includegraphics[scale=0.34]{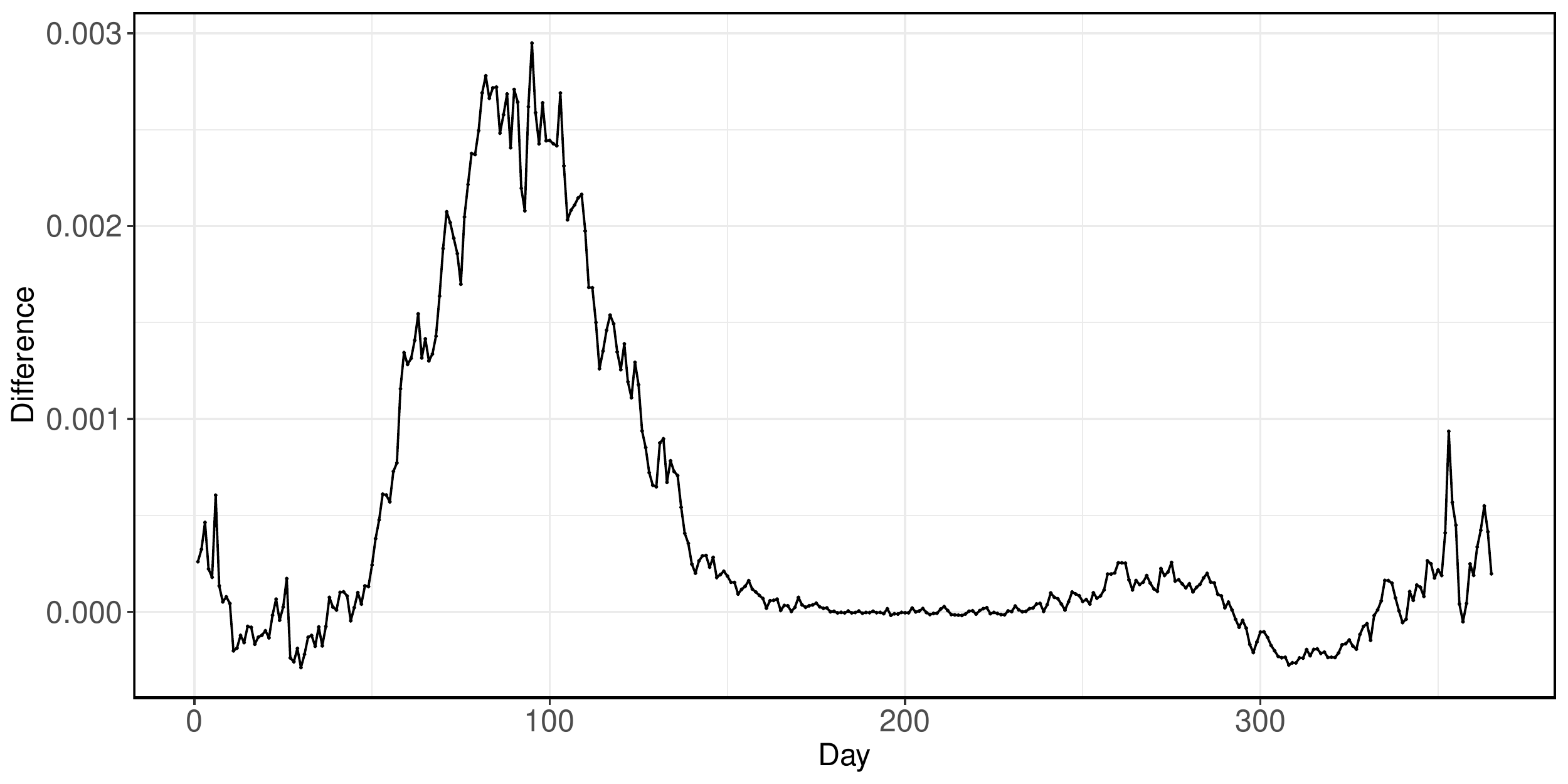}
        \caption{temporal difference}
        \label{subfig:rho_tuning_temporal}
    \end{subfigure}%
    \caption{Spatial (a) and temporal (b) difference of the fitted intensities using tuned $\rho$ compared to uniform $\rho$ with an equal expected number of dummy points in the logistic regression estimation (averaged on 60 random runs). The unit in the spatial plot is $\text{metre}^{-2}$; the unit in the temporal plot is $\text{day}^{-1}$.}
    \label{fig:rho_tuning}
\end{figure}

Based on such considerations, in (\ref{e:rho}), we tune $\rho$ to concentrate on urban areas and winter. Here, we  perform two experiments to assess the influence of $\rho$ tuning in the spatial and temporal domains separately by fitting our prediction model with and without $\rho$ tuning on the data of all $17$ years and plotting the difference of the fitted values. We also set the expectations of the number of dummy points as equal both with and without $\rho$ tuning. In addition, since the realisations of the dummy point process generated in different runs contain randomness, we perform each experiment $60$ times and plot the averaged spatial and temporal differences. Figure~\ref{fig:rho_tuning}(a) shows that, spatially, a concentration on urban areas reduces the bias caused by the saturation effect shown in Figure~\ref{fig:variable_relations}.  Figure~\ref{fig:rho_tuning}(b) shows that, temporally, with a concentration on winter seasons, the fitted risk intensities increase in March and April, which are known as the weather tipping seasons and are found more likely to catch chimney fires in \cite{School2018Thesis}. To summarize, tuning $\rho$ to important regions and times is helpful to estimate the fire risk intensities more efficiently and to capture important underlying patterns.

\section{Conclusion}
\label{sec:conclusion}
In this paper, we proposed a data-driven modeling process for chimney fire risk prediction, which can be generalized to studies on other fire types. Firstly, we applied random forests and permutation importance techniques to identify the important explanatory variables non-parametrically from a large number of environmental variables. From a practical perspective, our results indicate that pre-war detached or semi-detached houses run a higher risk of chimney fires and therefore public awareness campaigns should preferentially
target the owners of such houses. Secondly, based on the observed relations between the selected variables and chimney fire occurrences, we designed a Poisson point process model to predict fire risk to learn the observed spatio-temporal point pattern directly. We applied the logistic regression estimation for model fitting and provided asymptotic confidence intervals. Additionally, to validate our model assumption, we performed a second-order property analysis and a residual study to test point interactions in the fire data. Last but not least, we reviewed our modeling process, compared the areal unit model and the point process model and discussed the $\rho$ tuning required for the logistic regression estimation.

For future work, the first interesting direction would be to extend the data-driven modeling process of chimney fire risk prediction to similar fire types. Additionally, we are investigating the asymtotics of the logistic regression estimation under the framework of infill asymptotics using limit results for U-statistics \cite{Reitzner2013CLT}.

\section*{Acknowledgements}
This work was funded by the Dutch Research Council (NWO) for the project `Data Driven Risk Management for Fire Services' (18004). We thank the user committee for their valuable input. We also thank Emiel Borggreve, Niels Peters and Etienne Mulder from the Twente Fire Brigade for their help with data cleaning and pre-processing.

\bibliography{main.bbl}

\newpage

\begin{center}
{\Large\bf Supplementary Material}
\end{center}

\appendix

\section{Extra variable importance experiments}
\begin{figure}[h]
    \centering
    \begin{subfigure}[t]{0.48\textwidth}
        \centering
        \includegraphics[scale=0.32]{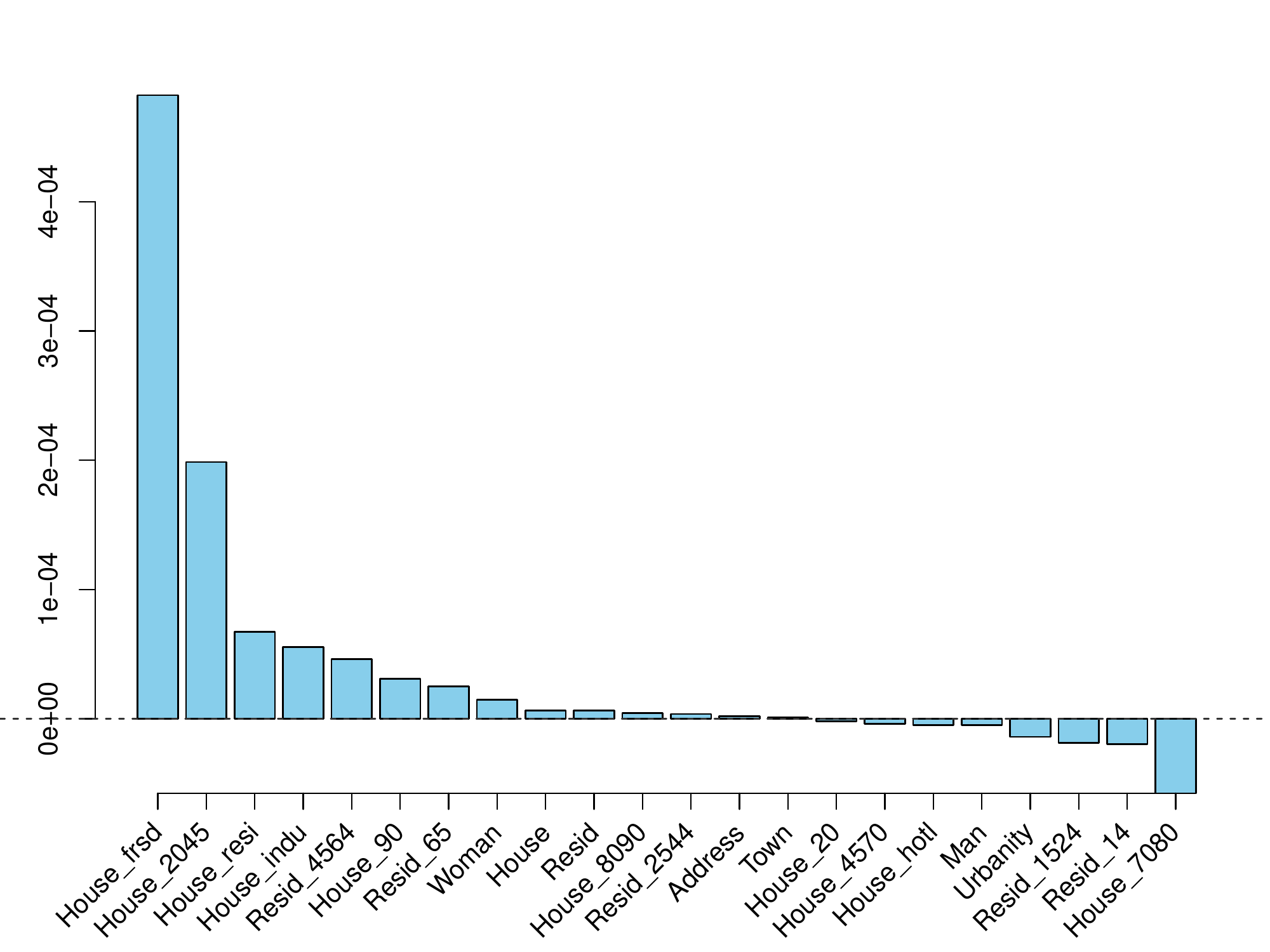}
        \caption{spatial, mtry=15}
        \label{subfig:VItests_spatial_mtry15}
    \end{subfigure}%
    \begin{subfigure}[t]{0.48\textwidth}
        \centering
        \includegraphics[scale=0.32]{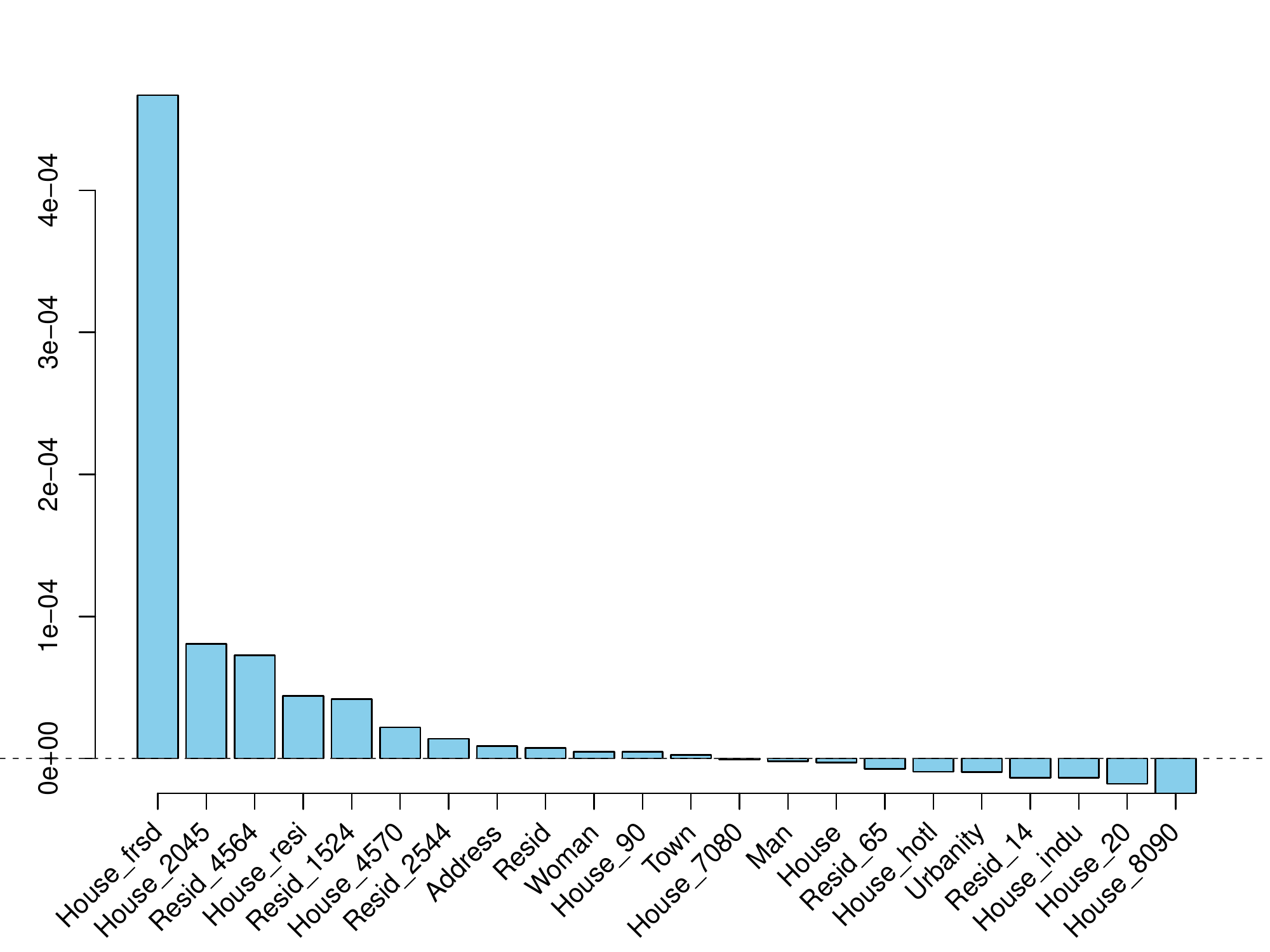}
        \caption{spatial, mtry=22}
        \label{subfig:VItests_spatial_mtry22}
    \end{subfigure}%
    \medskip
    
    \begin{subfigure}[t]{0.48\textwidth}
        \centering
        \includegraphics[scale=0.32]{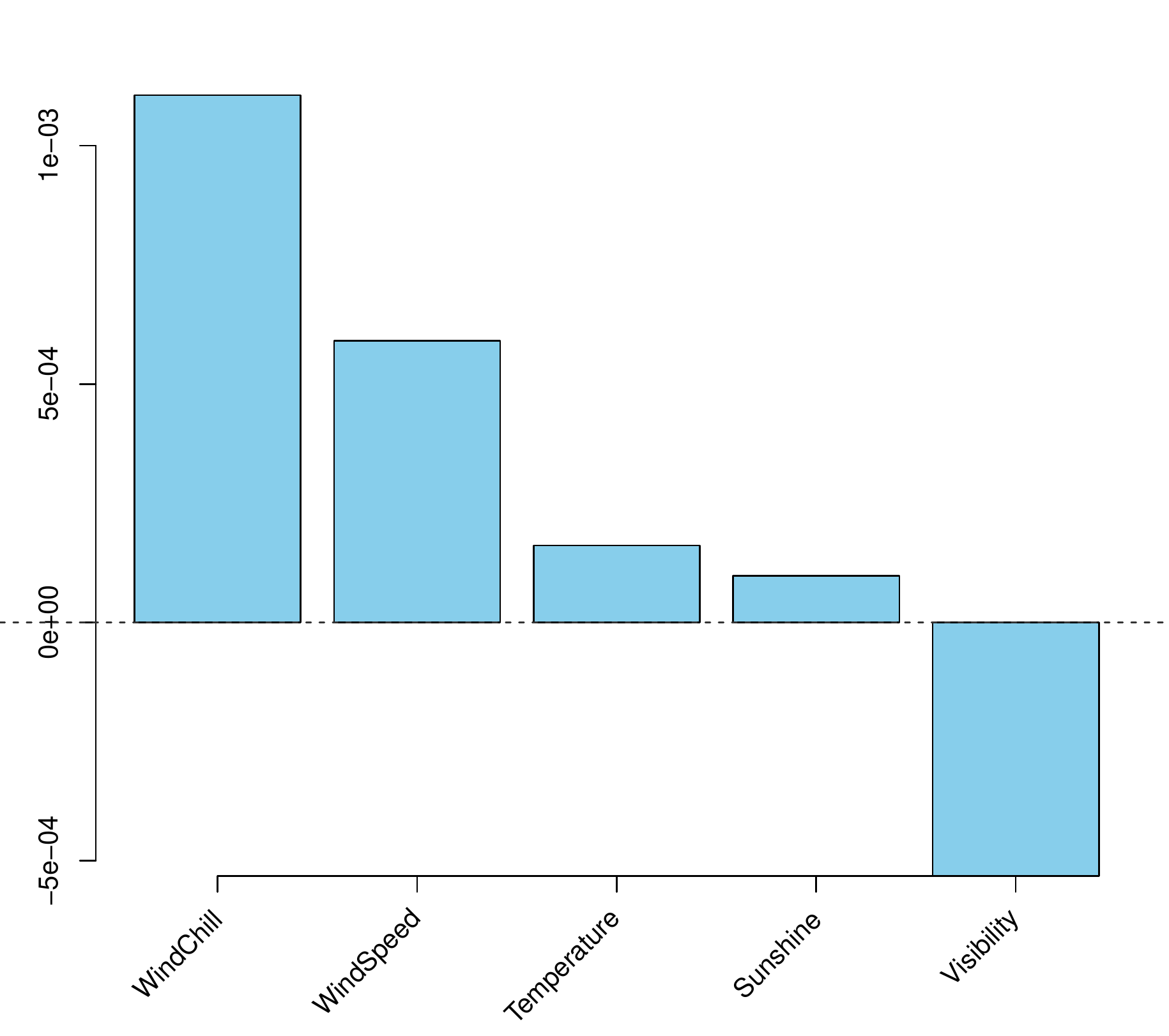}
        \caption{temporal, mtry=3}
        \label{subfig:VItests_temporal_mtry3}
    \end{subfigure}%
    \begin{subfigure}[t]{0.48\textwidth}
        \centering
        \includegraphics[scale=0.32]{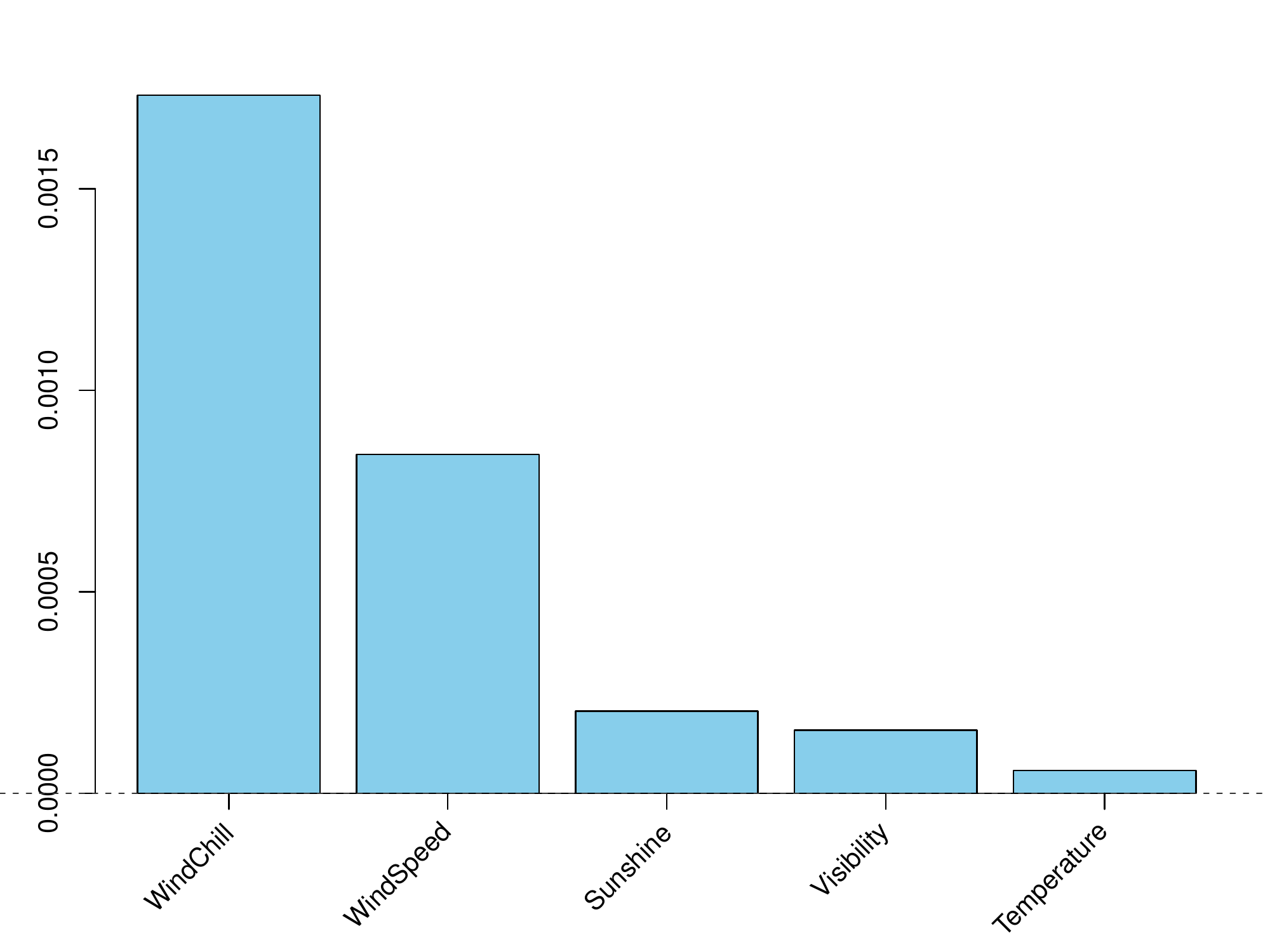}
        \caption{temporal, mtry=5}
        \label{subfig:VItests_temporal_mtry5}
    \end{subfigure}%
    \caption{Tests on the influence of the hyper parameter -- mtry -- in the selection of explanatory variables.}
    \label{fig:VItests_mtry}
\end{figure}

\begin{figure}[h]
    \centering
    \begin{subfigure}[t]{0.48\textwidth}
        \centering
        \includegraphics[scale=0.32]{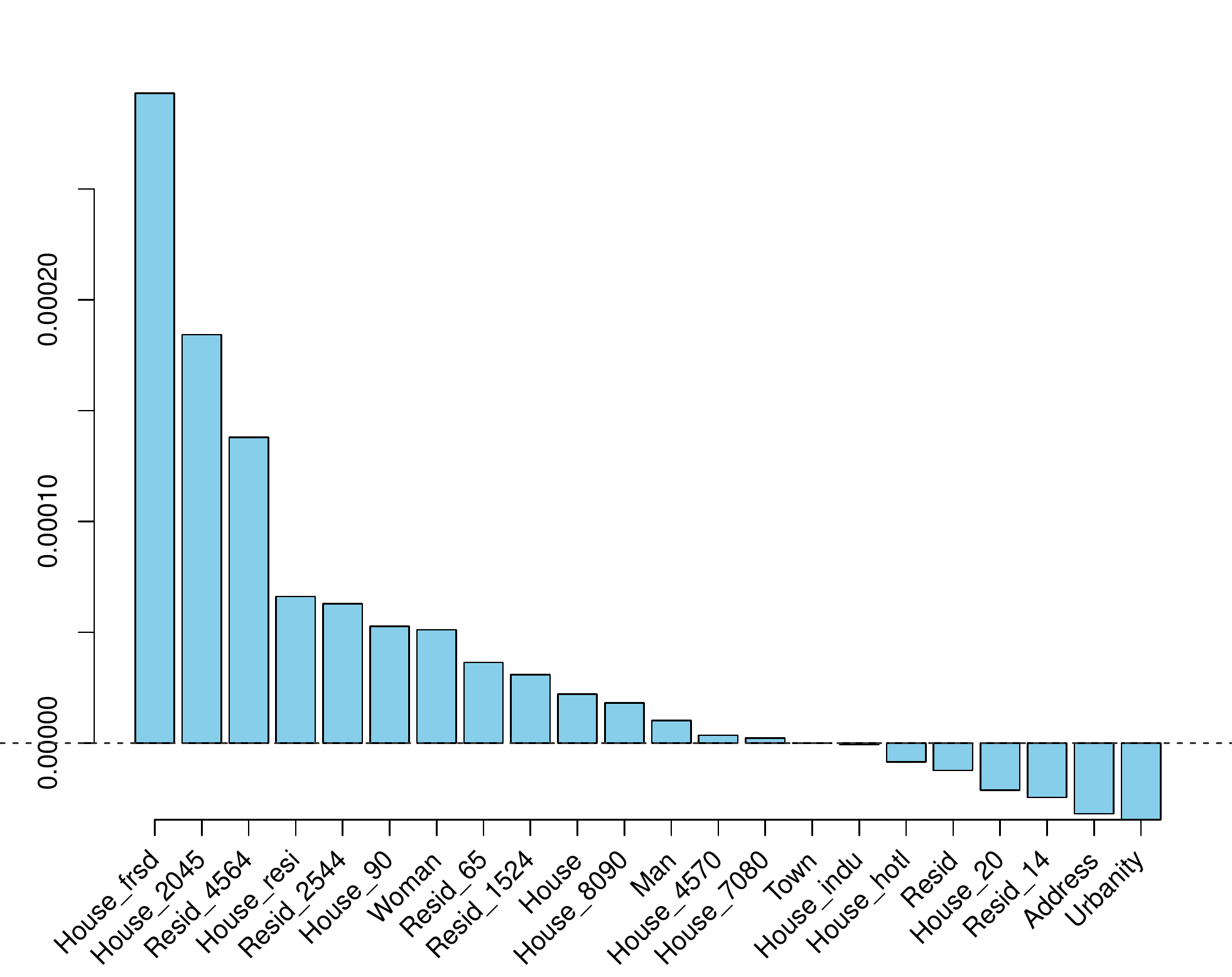}
        \caption{spatial, random seed=30}
        \label{subfig:VItests_spatial_randomseed30}
    \end{subfigure}%
    \begin{subfigure}[t]{0.48\textwidth}
        \centering
        \includegraphics[scale=0.32]{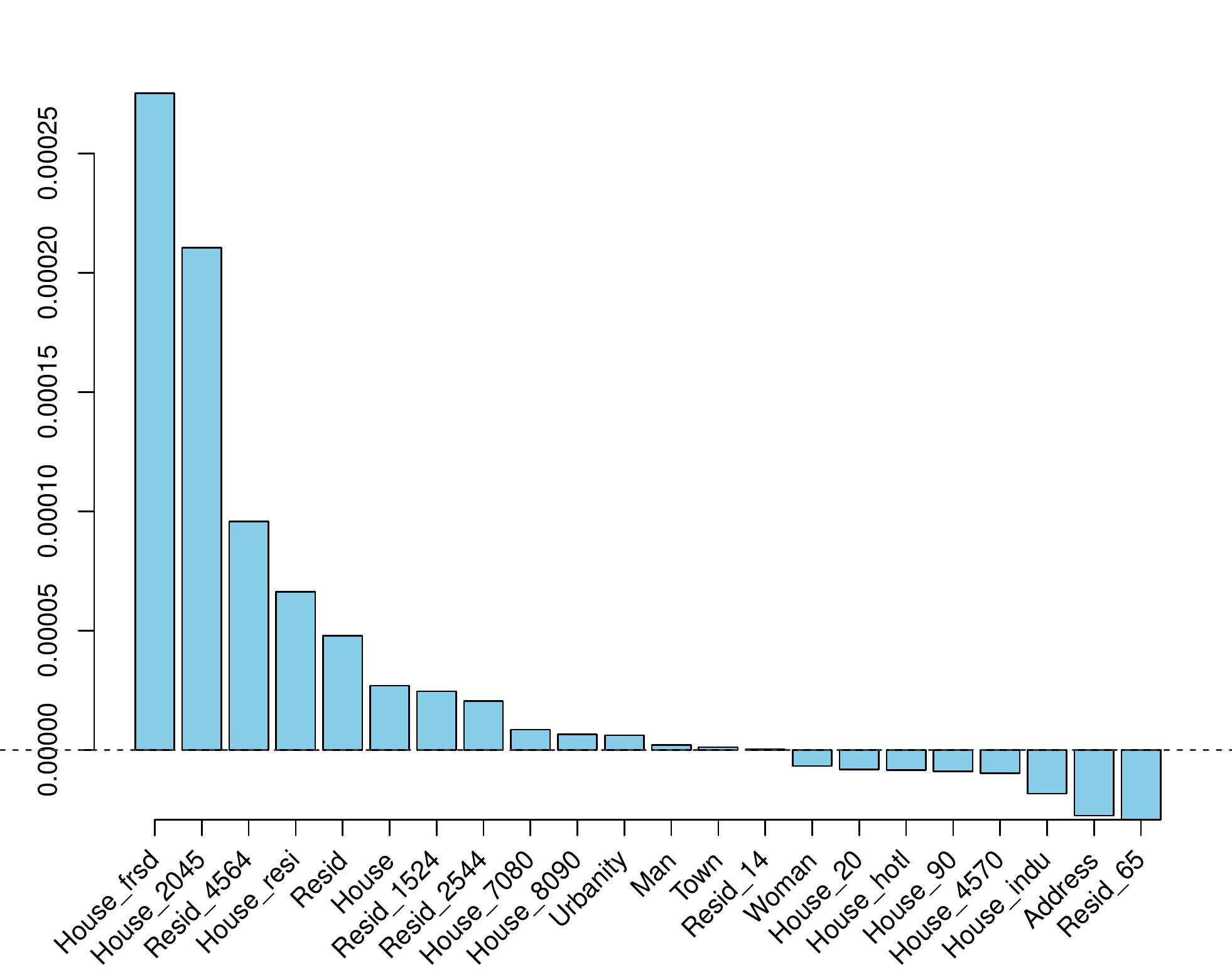}
        \caption{spatial, random seed=200}
        \label{subfig:VItests_spatial_randomseed200}
    \end{subfigure}%
    \medskip
    
    \begin{subfigure}[t]{0.48\textwidth}
        \centering
        \includegraphics[scale=0.32]{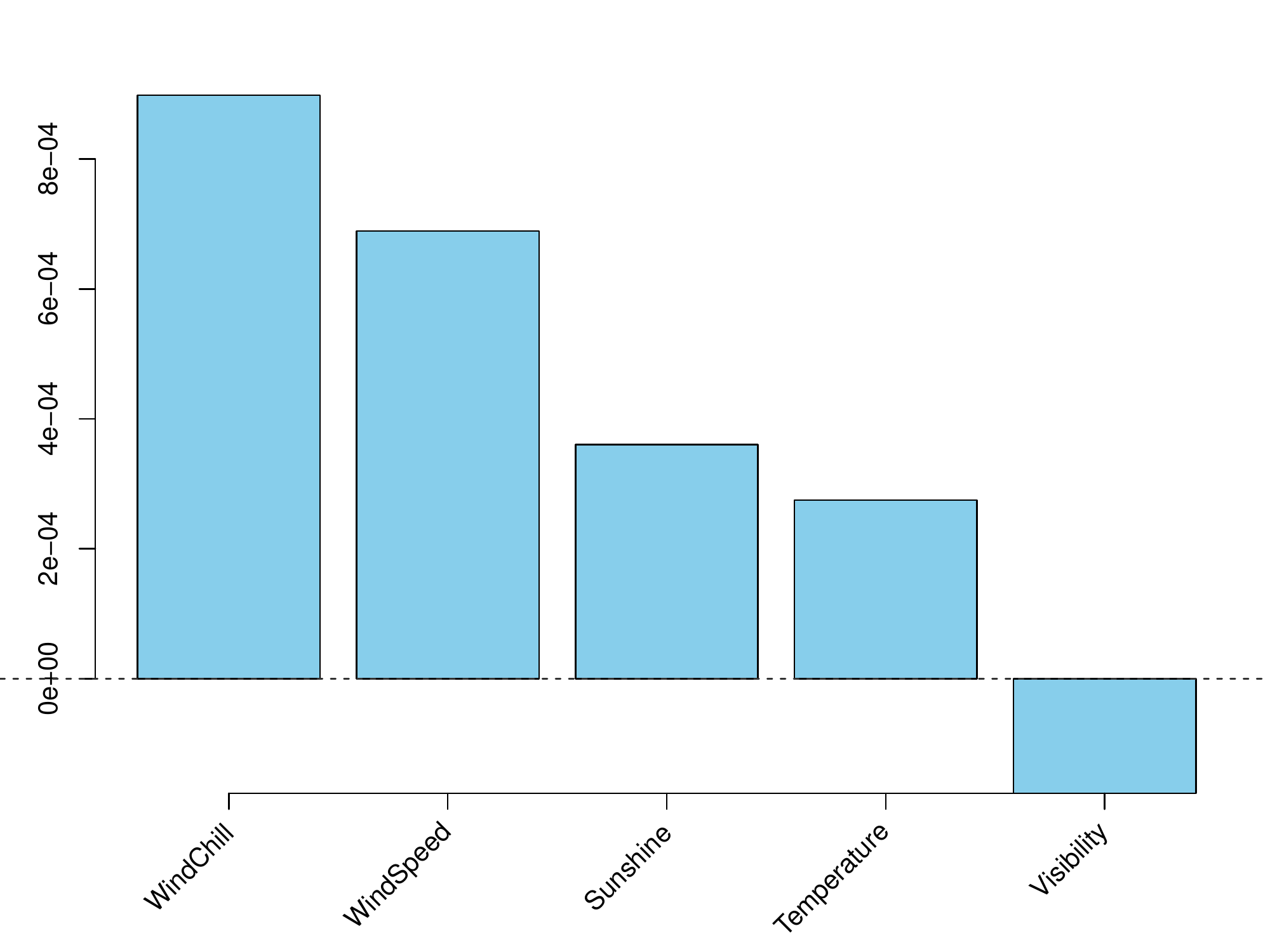}
        \caption{temporal, random seed=30}
        \label{subfig:VItests_temporal_randomseed30}
    \end{subfigure}%
    \begin{subfigure}[t]{0.48\textwidth}
        \centering
        \includegraphics[scale=0.32]{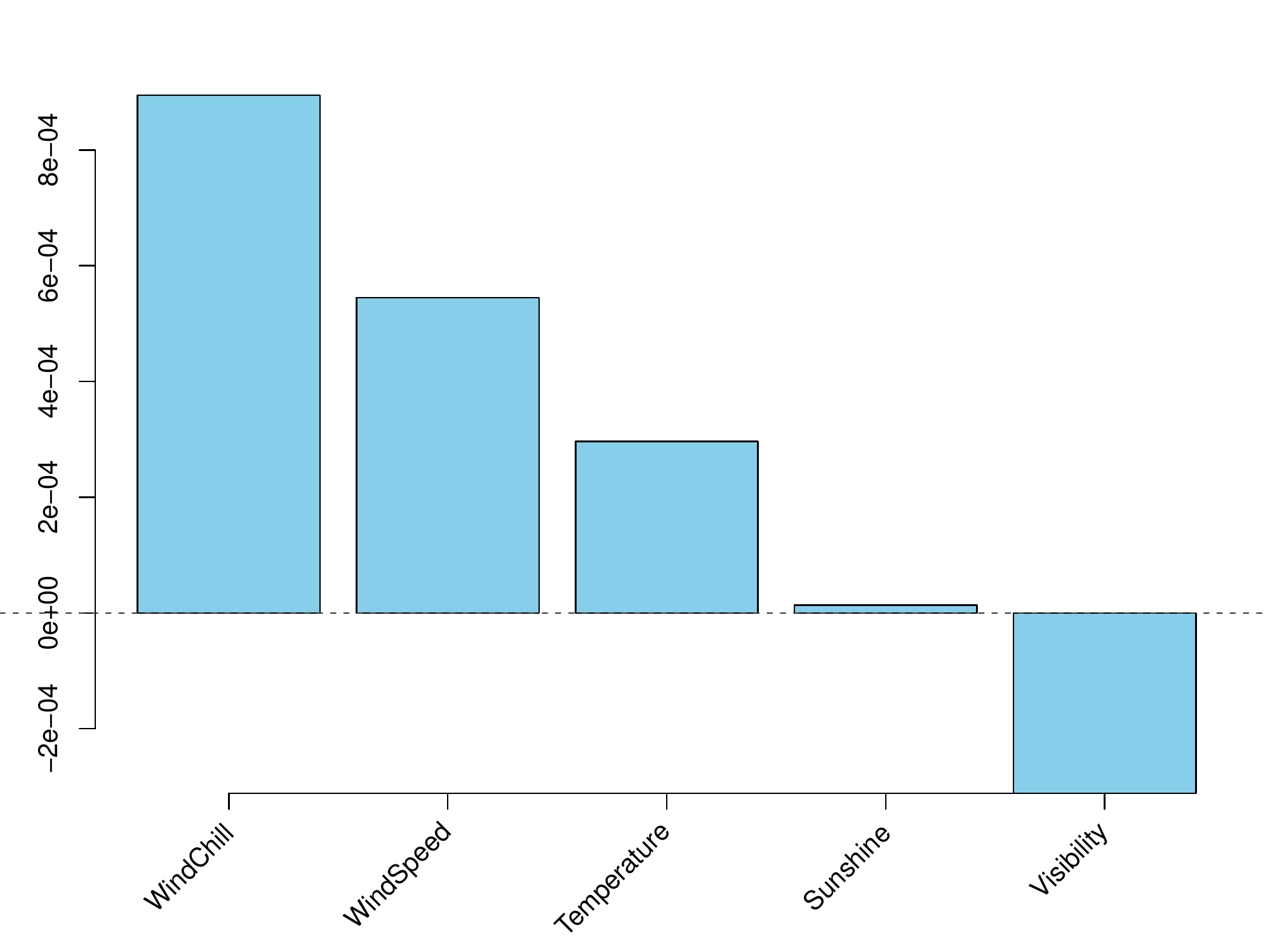}
        \caption{temporal, random seed=200}
        \label{subfig:VItests_temporal_randomseed200}
    \end{subfigure}%
    \caption{Tests on the influence of the hyper parameter -- random seed -- in the selection of explanatory variables.}
    \label{fig:VItests_randomseed}
\end{figure}

\begin{figure}[ht]
    \centering
    \begin{subfigure}[t]{0.48\textwidth}
        \centering
        \includegraphics[scale=0.32]{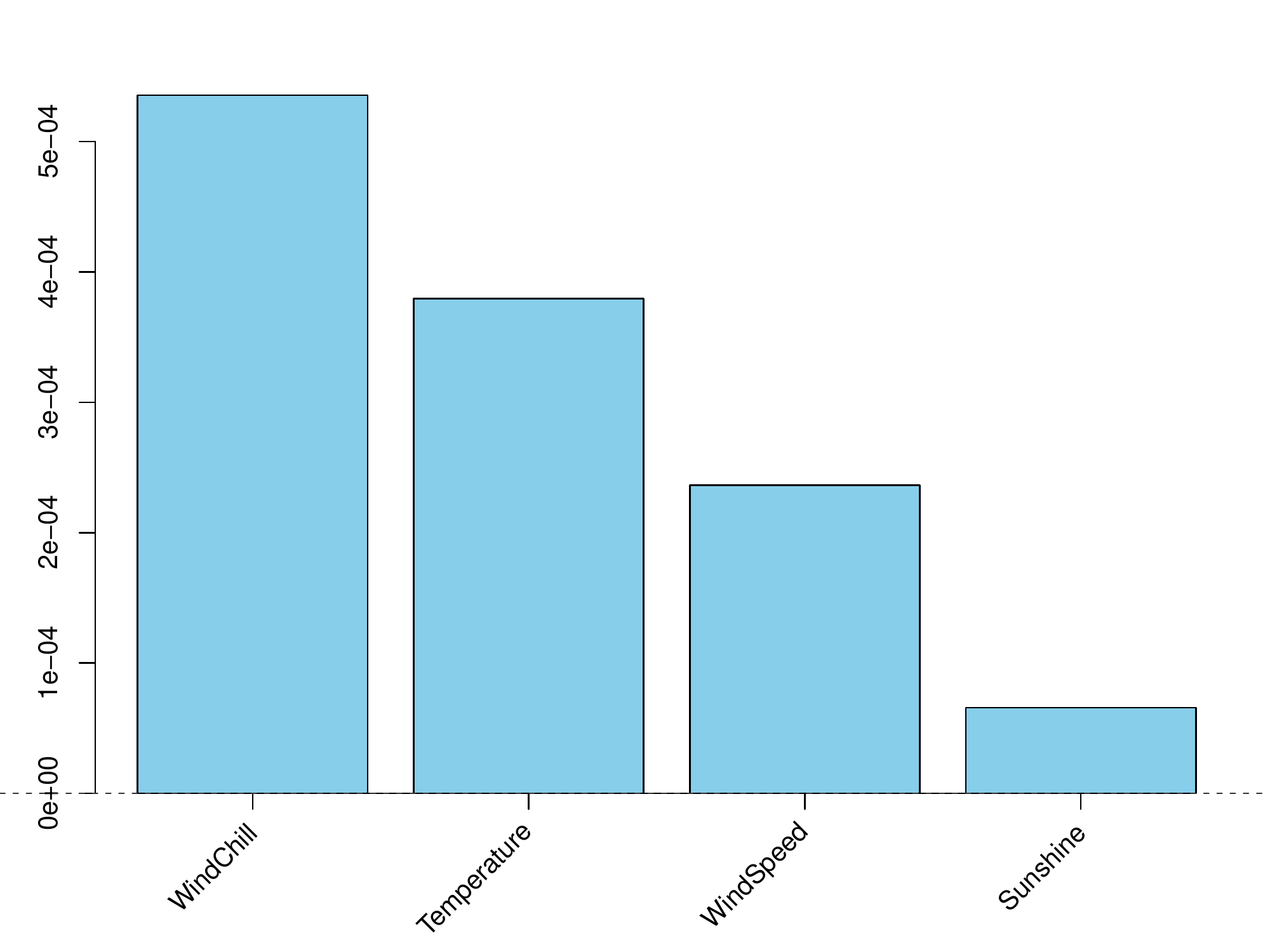}
        \caption{Heino}
        \label{subfig:VItests_Heino}
    \end{subfigure}%
    \begin{subfigure}[t]{0.48\textwidth}
        \centering
        \includegraphics[scale=0.32]{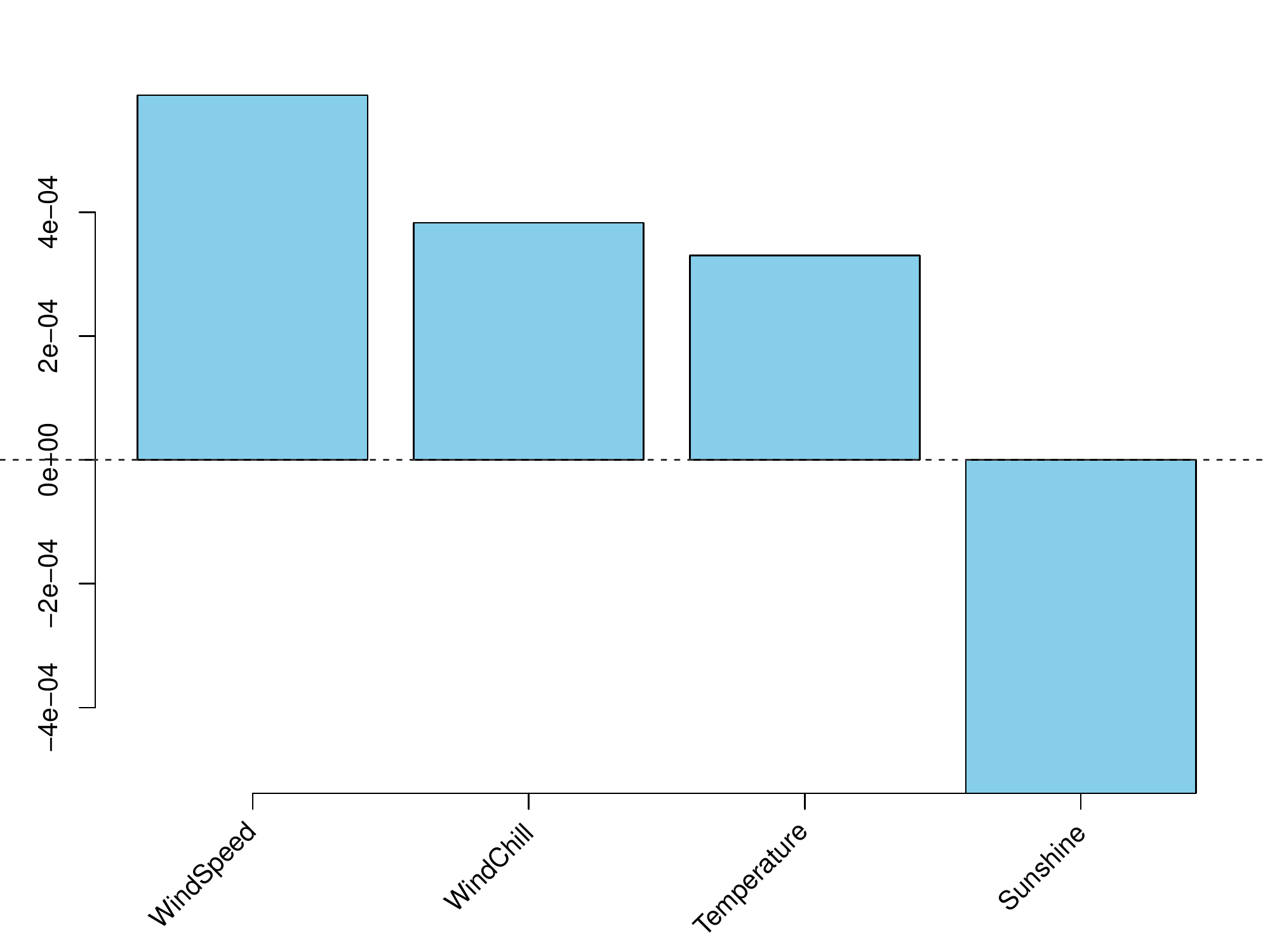}
        \caption{Hupsel}
        \label{subfig:VItests_Hupsel}
    \end{subfigure}%
    \caption{Tests on the influence of the weather variation in the selection of explanatory variables. Note that the variable, visibility, is not accessible in the weather stations, Heino and Hupsel.}
    \label{fig:VItests_weather_variation}
\end{figure}

In the selection of explanatory variables using random forests and conditional permutation techniques, to test the influence of the number of randomly sampled input variables at each tree node (i.e.\ \textit{mtry}) and different random seed settings, we perform extra experiments to construct the random forests with multiple configurations and plot the variable importance results in Figure~\ref{fig:VItests_mtry} and Figure~\ref{fig:VItests_randomseed}. The results show that both the \textit{mtry} and random seed settings may influence the detailed importance scores (the increase of the prediction error) of some variables, but they will not influence the detection of the most significant variables. We also perform an importance analysis on the temporal data from two neighbouring weather stations to assess the influence of small variations in weather among different parts of the Twente region. The results are shown in Figure~\ref{fig:VItests_weather_variation} and indicate that those small variations can also be ignored.

\end{document}